# Anharmonic Molecular Mechanics: *Ab Initio* Based Morse Parameterisations for the Popular MM3 Force Field

R. J. Shannon[1,2]*, B. Hornung[1], D. P. Tew[1], D. R.Glowacki[1,3]*

[1] *School of Chemistry, Cantock's Close, University of Bristol, Bristol, BS8 1TS, UK*

[2]*Department of Mechanical Engineering, Stanford University, 452 Escondido Mall, Stanford, CA 94305, USA;*

[3]*Department of Computer Science, University of Bristol, Bristol BS8 1UB, UK*

Abstract

Methodologies for creating reactive potential energy surfaces from molecular mechanics force-fields are becoming increasingly popular. To date, molecular mechanics force-fields in biochemistry and small molecule organic chemistry tend to use harmonic expressions to treat bonding stretches, which is a poor approximation in reactive and non-equilibirum molecular dynamics simulations since bonds are often displaced significantly from their equilibrium positions. For such applications there is need for a better treatment of anharmonicity. In this contribution, Morse bonding potentials have been extensively parameterized for the atom types in the MM3 force field of Allinger and co-workers using high level CCSD(T)(F12*) energies. To our knowledge this is amongst the first instances of a comprehensive parametrization of Morse potentials in a popular organic chemistry force field. In the context of molecular dynamics simulations, this data will: (1) facilitate the fitting of reactive potential energy surfaces using empirical valence bond approaches, and (2) enable more accurate treatments of energy transfer.

## 1. Introduction

In the field of biochemistry and small molecule organic chemistry, Molecular mechanics force fields have offered a computationally efficient way of calculating energies and forces of large-scale molecular systems for many years. There are numerous force fields available [1-5] all of which have been extensively parameterised against experimental and/or theoretical data and these force fields have been utilised successfully within molecular dynamics simulations. In their basic form, molecular mechanics force fields require a connectivity specification (this is not true for force fields from other fields such as solid state physics [6-8]) and are non-reactive by definition (although modified approaches such as the ReaxFF[9] force-field do allow bonding breaking/making to occur and the recently developed QMDFF does provide a black box framework for fitting system specific potentials which are dissociative[10]). However, there are deficiencies to the harmonic bond representation used in popular force fields. For example, in recent years an increasing number of approaches like the empirical valence bond (EVB) approach have been used to create reactive potential energy surfaces utilising a basis of unreactive molecular mechanics force fields. [11-16] In such reactive simulations, bond breaking and bond forming processes lead to scenarios where chemical bonds are substantially displaced from their equilibrium bonding distance, highlighting serious deficiencies in many commonly used force fields. Moreover, it has been known for some time that an accurate treatment of anharmonicity plays a crucial role in accurately treating inter and intra-molecular energy transfer rates.[17-19]

Typically, bonding terms used in molecular mechanics force fields are polynomial expansions (up to fourth order), but this approximation breaks down for elongated bonds, e.g., during reactive events or in cases where molecules have significant internal energy. The primary motivation for treating bonds as harmonic rather than anharmonic in the past was one of computational efficiency, owing to the fact that force evaluations for a harmonic bond term are significantly cheaper than for a Morse type oscillator. With modern computational facilities however, such considerations are less important (although the additional computational cost may still remain an important consideration for very large systems or timescales), and routine molecular dynamics studies are perfectly feasible with a force field comprising of Morse oscillators rather then the less physically realistic harmonic oscillator.

One commonly used force field is the MM3 force field of Allinger and co-workers, [20-22] which we have used extensively over the years to model reactive dynamics and transient ultrafast infrared spectroscopy in a range of both strongly coupled and weakly coupled solvents.[11, 17] This force-field has gone through several iterations from MM1 to MM3, as well as the closely related MMFF, all of which have been characterised on a large set of small hydrocarbon species over a long series of publications. In this work we replace the standard harmonic bonding terms in the MM3 force field with a Morse oscillator fit to *ab initio* calculations using explicitly correlated coupled-cluster singles, doubles and perturbative triples theory, CCSD(T)(F12*). [23]

The aim of this article is to provide a comprehensive Morse parameterisation of the popular MM3 force field such that researchers wishing to include anharmonicity in their simulations can utilise this new Morse MM3 force field with trivial effort. The Morse MM3 force field described herein, has been designed such that it tends to the results of the original, MM3 force field at geometries near equilibrium but gives a more physical description of the energy at large bond lengths. Section 2 in this paper details the methodology for obtaining the Morse parameters and presents the new force field along with some analysis. Section 3 demonstrates an EVB fitting application for this new force field through comparison with the original MM3 force field. Finally in section 4 we present some conclusions.

## 2. Morse fitting

Dissociation energies for the MM3 bonding entries were obtained as follows. First a test set of molecules was created and MM3 atoms types were assigned to each species. This test set primarily included molecules which were used to parameterise the existing MM3 parameters as described in various publications,[24-61] however it was necessary to add a substantial number of additional molecules in order to cover a larger selection of the MM3 bonding types. This test set is listed in the online supplementary information. It comprises 254 species and covers all but 29 of the 213 MM3 bonding types. Most of the remaining bond types for which dissociation energies have not been obtained were either found to be incomplete, in the sense that a full set of MM3 parameters was not present for a species involving this type of bond, or high level energy calculations were problematic for the given bond type due to molecular size. For example the MM3 types corresponding to Ferrocene have not been included.

Having defined the test set, the geometry of each species was initially optimised using the Tinker molecular dynamics package and the MM3 parameter set.[62] Starting from the optimised geometry, any given bond was then subject to a total of 7 displacements from the equilibrium geometry, covering the range $r - 0.3$ to $r + 0.6$, where in this case $r$ is the (MM3) optimised nuclear separation for the given bond in Å.

At each of these displaced geometries CCSD(T)(F12*) / aug-cc-pVDZ energies were computed using the Molpro software [63] to construct an *ab initio*, anharmonic potential for a given bond. For second row atoms and above, ecp2 core potentials in combination with the aug-cc-pVDZ-PP basis sets were used. Previous experience shows that the near basis set limit CCSD(T) harmonic and anharmonic potential energy surface parameters delivered by the CCSD(T)(F12*) method are typically within 1-2 % of experimentally derived values. [64] We fit Morse parameters $(D_e, \alpha, r_e \text{ and } C)$ in Eq. 1 to each of the *ab initio* potentials via least squares fit.

$$V(r) = D_e(1 - e^{-\alpha(r-r_e)})^2 - C \qquad \text{(Eq. 1)}$$

In order to assess the consistency between the fitted *ab initio* Morse potentials and the harmonic parameters already present in the MM3 force field, we can derive harmonic force constants ($k_\alpha = 2\alpha^2 D_e$) from the fitted Morse curves.

Fig 1 compares these $k_\alpha$ with the force constants already present in the MM3 force field $k_h$ . The two values of $k$ are plotted for a selection of the bonds involving MM3 type 1 (sp3 Carbon) and for the bonds involving MM3 type 3 (carbonyl carbon) in Figure 1.

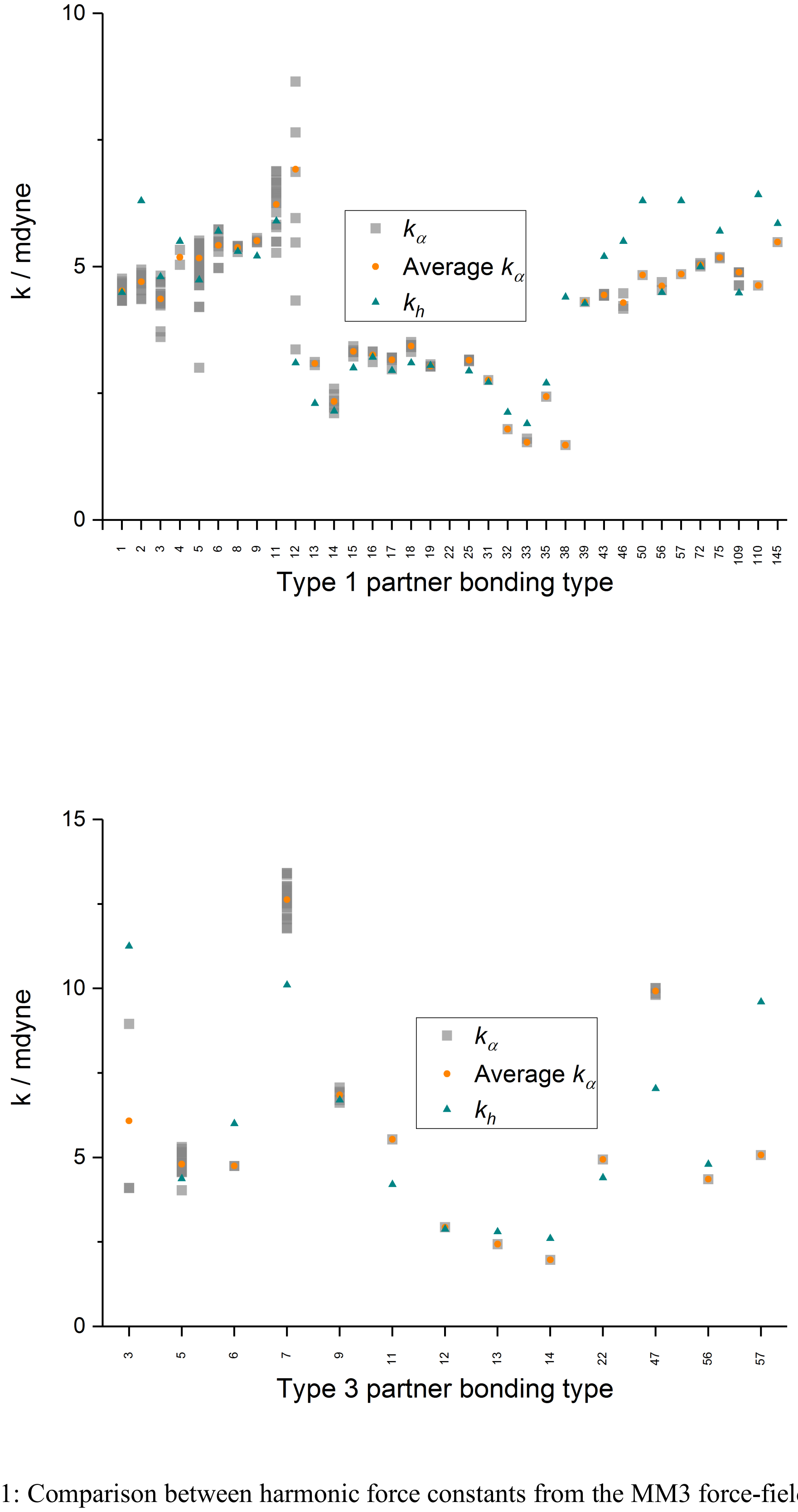

Figure 1: Comparison between harmonic force constants from the MM3 force-field and those from the Morse potential fits. Grey squares give spread of $k_\alpha$ values from fits, orange circles give average $k_\alpha$ and cyan triangles give $k_h$ from the MM3 parameter file. MM3 types for the two atoms in a bond are shown on the x-axis. The upper panel displays results for a portion of the MM3 type 1 bonds and the lower panel displays results for the MM3 type 3 bonds.

For both panels in Figure 1 the trends in the different $k_\alpha$ and $k_h$ values agree well but there are significant discrepancies, particularly for the type 3 bonds. The primary reason for this appears to be due to the wide variation in molecular structures, which feature a single bond type. For example, both 1,2 benzoquinone and glyoxal involve a bond between two type 3 atoms, and yet the chemical environment surrounding these bonds is clearly different. This gives rise to different *ab initio* Morse potentials, as demonstrated in Figure 2. For comparison, Figure 2 shows a harmonic curve based upon the original MM3 data; the two Morse curves deviate significantly from the original harmonic potential. Owing to the organic growth of the MM3 force field over the years, we were unable to obtain the definitive list of test molecules used to parameterise the original data. As such the test set used here likely deviates from the original MM3 test set and this could account for some of the discrepancies.

Figure 3 shows a similar comparison of Morse curves for 3 examples of the bonding between MM3 types 3 and 5. Here the agreement is much improved both between the different Morse curves and also between the Morse curves and the original harmonic potential.

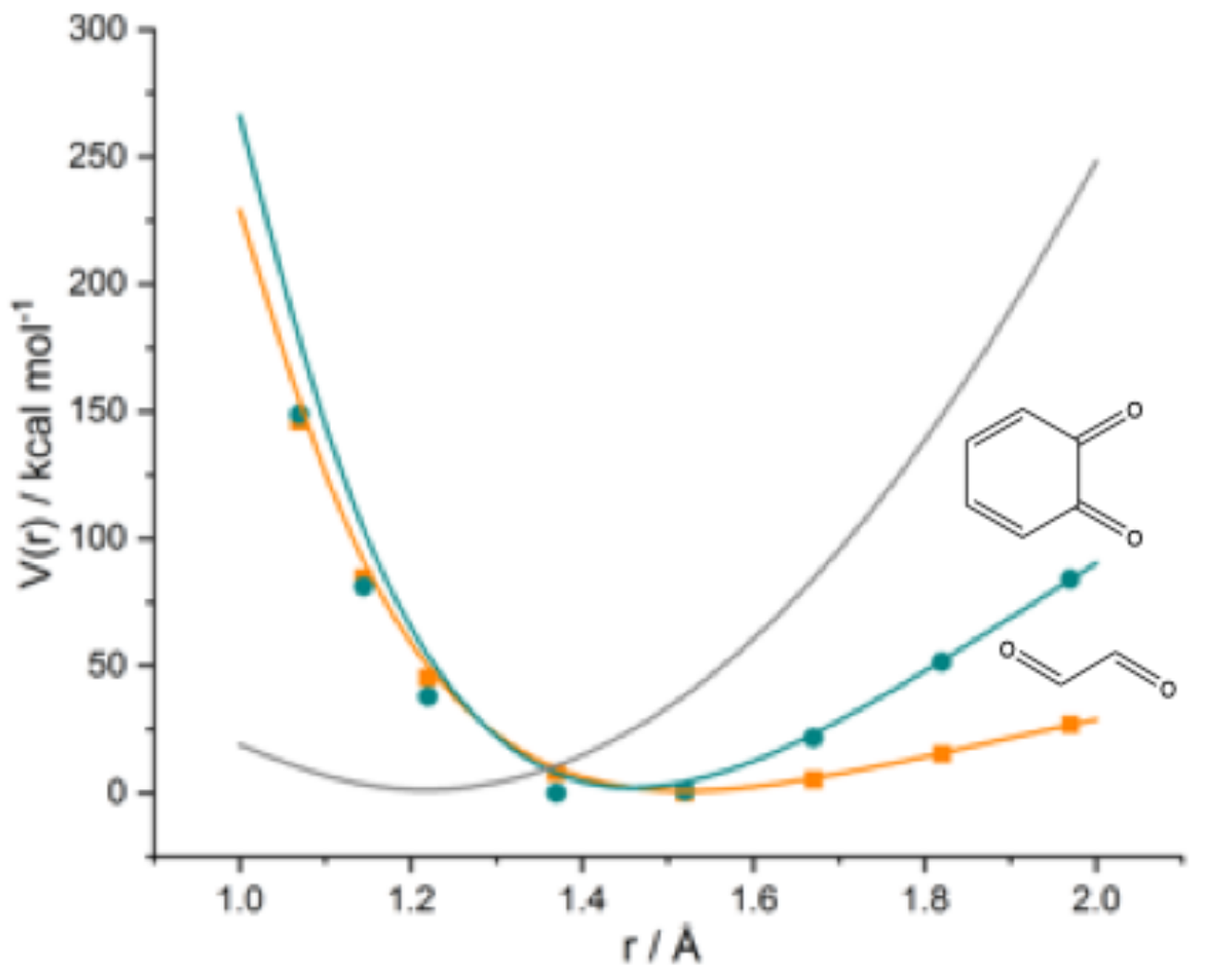

Figure 2: *Ab initio* Morse potentials for 1,2 benzoquinone (orange) and glyoxal (cyan). In each case a bond between two MM3 type 3 atoms (C CSP2 CARBONYL) is displayed. The points correspond to the CCSD(T)(F12*) / aug-cc-pVDZ energies and the lines correspond to the fitted Morse potentials. The grey curve corresponds to the harmonic potential from the original MM3 data.

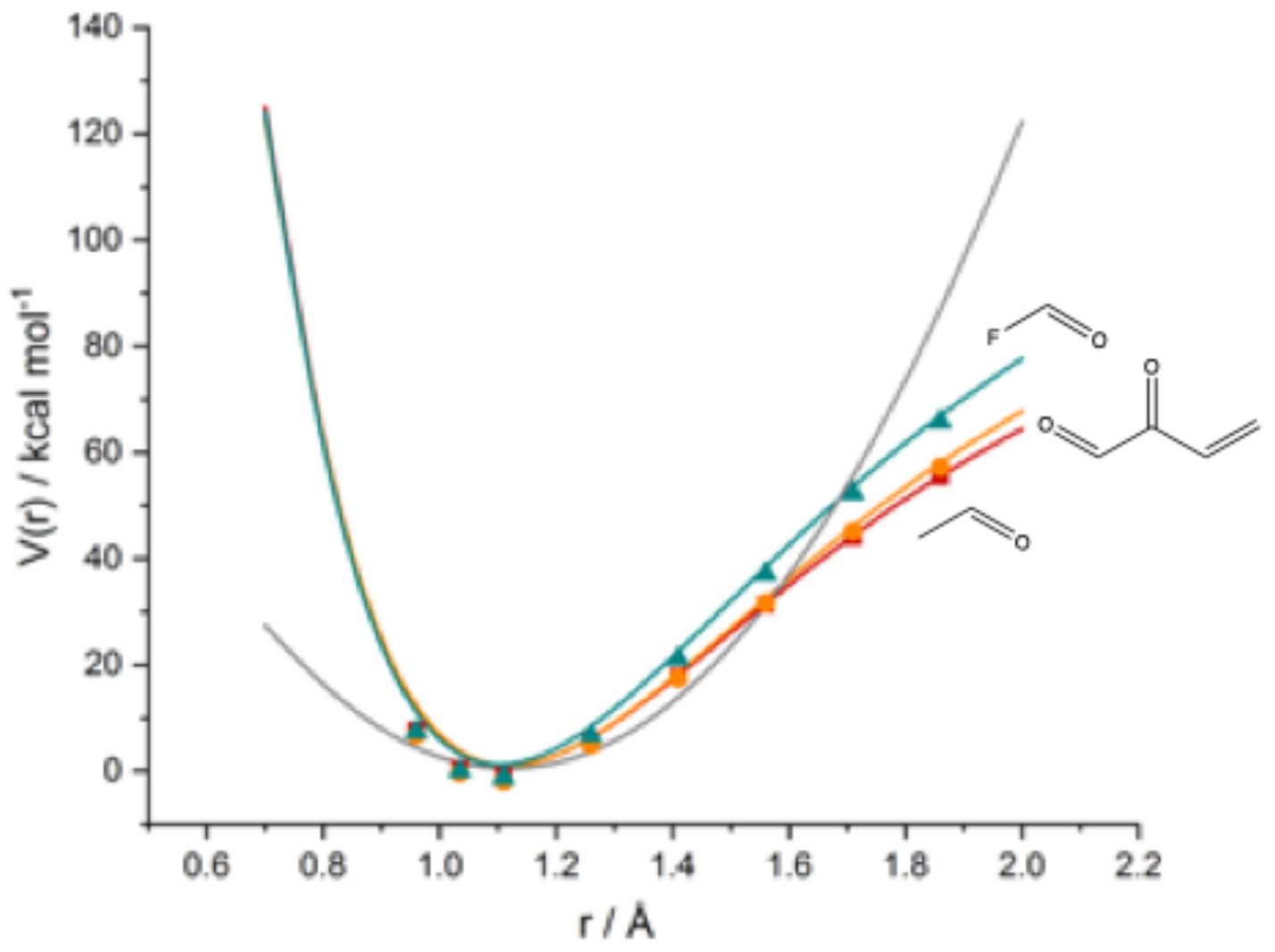

Figure 3: *Ab initio* Morse potentials for formyl flouride (orange), 2-Oxo-3-butenal and formaldehyde (red). In this case the bonds are between one MM3 type 3 atom (C CSP2 CARBONYL) and one MM3 type 5 atom (H EXCEPT ON N,O,S). The points correspond to the CCSD(T)(F12*) / aug-cc-pVDZ energies and the lines correspond to the fitted Morse potentials. The grey curve corresponds to the harmonic potential from the original MM3 data.

To generate the new Morse MM3 force field we retain the $r_h$ and $k_h$ values from the original MM3 dataset so that the Morse MM3 force field is a minimal perturbation of the original MM3 force field. Given the reasonable agreement between $k_\alpha$ and $k_h$, we decided to only transfer the $D_e$ values and determine α ($\alpha_h$) from Eq.2: [65]

$$\alpha_h = \sqrt{\frac{k_h}{2D_e}} \qquad \text{(Eq. 2)}$$

We emphasise that the $D_e$ parameters are determined such that the Morse potential reproduces CCSD(T)(F12*)/cc-pVDZ-F12 energies at relatively small bond stretches and should not be used as reliable bond dissociation values. The full list of new MM3 parameters is given in Table 1.

| **Atom Type 1 (MM3 type number)** | **Atom Type 2 (MM3 type number)** | $k_h$ / mdyne Å$^{-1}$ | $r_h$ / Å | $D_e$ / mdyne Å | $k_\alpha$/ mdyne Å$^{-1}$ | $\alpha_{calc}$/ Å$^{-1}$ |
|---|---|---|---|---|---|---|
| C CSP3 ALKANE (1) | C CSP3 ALKANE (1) | 4.490 | 1.525 | 1.130 | 5.069 | 1.409 |
| C CSP3 ALKANE (1) | C CSP2 ALKENE (2) | 6.300 | 1.499 | 1.242 | 5.529 | 1.593 |
| C CSP3 ALKANE (1) | C CSP2 CARBONYL (3) | 4.800 | 1.509 | 0.994 | 4.892 | 1.554 |
| C CSP3 ALKANE (1) | C CSP ALKYNE (4) | 5.500 | 1.470 | 0.995 | 5.183 | 1.663 |
| C CSP3 ALKANE (1) | H EXCEPT ON N,O,S (5) | 4.740 | 1.112 | 0.854 | 5.168 | 1.666 |
| C CSP3 ALKANE (1) | O C-O-H, C-O-C, O-O (6) | 5.700 | 1.413 | 0.851 | 5.409 | 1.830 |
| C CSP3 ALKANE (1) | N NSP3 (8) | 5.300 | 1.448 | 1.140 | 5.670 | 1.525 |
| C CSP3 ALKANE (1) | N NSP2 (9) | 5.210 | 1.446 | 0.968 | 5.510 | 1.640 |
| C CSP3 ALKANE (1) | F FLUORIDE (11) | 5.900 | 1.390 | 0.989 | 6.227 | 1.727 |
| C CSP3 ALKANE (1) | Cl CHLORIDE (12) | 3.100 | 1.791 | 0.629 | 6.043 | 1.569 |
| C CSP3 ALKANE (1) | Br BROMIDE (13) | 2.300 | 1.944 | 0.708 | 3.087 | 1.275 |
| C CSP3 ALKANE (1) | I IODIDE (14) | 2.150 | 2.166 | 0.635 | 2.337 | 1.301 |
| C CSP3 ALKANE (1) | S =-S- SULFIDE (15) | 3.000 | 1.805 | 0.651 | 3.325 | 1.518 |
| C CSP3 ALKANE (1) | S+ >S+ SULFONIUM (16) | 3.213 | 1.816 | 0.666 | 3.254 | 1.554 |
| C CSP3 ALKANE (1) | S >S=O SULFOXIDE (17) | 2.950 | 1.800 | 0.541 | 3.160 | 1.651 |
| C CSP3 ALKANE (1) | S >SO2 SULFONE (18) | 3.100 | 1.772 | 0.674 | 3.435 | 1.516 |
| C CSP3 ALKANE (1) | Si SILANE (19) | 3.050 | 1.876 | 0.812 | 3.040 | 1.370 |
| C CSP3 ALKANE (1) | C CYCLOPROPANE (22) | 5.000 | 1.511 | 0.919 | 4.784 | 1.649 |
| C CSP3 ALKANE (1) | P >P- PHOSPHINE (25) | 2.940 | 1.843 | 0.702 | 3.152 | 1.448 |
| C CSP3 ALKANE (1) | B >B- TRIGONAL (26) | 4.501 | 1.577 | 0.862 | 3.781 | 1.616 |
| C CSP3 ALKANE (1) | Ge GERMANIUM (31) | 2.720 | 1.949 | 0.744 | 2.761 | 1.352 |
| C CSP3 ALKANE (1) | Sn TIN (32) | 2.124 | 2.147 | 0.657 | 2.139 | 1.272 |
| C CSP3 ALKANE (1) | Pb LEAD (IV) (33) | 1.900 | 2.242 | 0.553 | 1.793 | 1.311 |
| C CSP3 ALKANE (1) | Se SELENIUM (34) | 2.680 | 1.948 | 0.692 | 2.823 | 1.391 |
| C CSP3 ALKANE (1) | Te TELLURIUM (35) | 2.700 | 2.140 | 0.625 | 2.445 | 1.469 |
| C CSP3 ALKANE (1) | N -N=C-/PYR (DELOCLZD) (37) | 5.000 | 1.434 | 0.756 | 5.207 | 1.819 |
| C CSP3 ALKANE (1) | N+ NSP3 AMMONIUM (39) | 4.274 | 1.511 | 0.810 | 4.422 | 1.624 |
| C CSP3 ALKANE (1) | N NSP2 PYRROLE (40) | 4.230 | 1.490 | 0.997 | 5.472 | 1.456 |
| C CSP3 ALKANE (1) | O OSP2 FURAN (41) | 5.400 | 1.420 | 0.782 | 5.411 | 1.858 |
| C CSP3 ALKANE (1) | N =N-O AZOXY (LOCAL) (43) | 5.200 | 1.483 | 0.532 | 4.442 | 2.210 |
| C CSP3 ALKANE (1) | N NITRO (46) | 5.500 | 1.495 | 0.629 | 4.288 | 2.091 |
| C CSP3 ALKANE (1) | C BENZENE (LOCALIZED) (50) | 6.300 | 1.499 | 0.951 | 4.832 | 1.820 |
| C CSP3 ALKANE (1) | C CSP3 CYCLOBUTANE (56) | 4.490 | 1.525 | 0.881 | 4.615 | 1.596 |
| C CSP3 ALKANE (1) | C CSP2 CYCLOBUTENE (57) | 6.300 | 1.499 | 0.924 | 4.851 | 1.847 |
| C CSP3 ALKANE (1) | N =N- IMINE (LOCALZD) (72) | 5.000 | 1.439 | 0.718 | 5.035 | 1.866 |
| C CSP3 ALKANE (1) | O O-H, O-C (CARBOXYL) (75) | 5.700 | 1.413 | 0.860 | 5.176 | 1.820 |
| C CSP3 ALKANE (1) | N =-N= AZOXY (LOCAL) (109) | 4.480 | 1.453 | 0.711 | 4.887 | 1.775 |
| C CSP3 ALKANE (1) | N+ -N(+)= IMMINIUM (110) | 6.420 | 1.470 | 0.821 | 4.627 | 1.978 |
| C CSP3 ALKANE (1) | O >N-OH HYDROXYAMINE (145) | 5.850 | 1.368 | 0.751 | 5.483 | 1.973 |
| C CSP3 ALKANE (1) | N >N-OH HYDROXYAMINE (146) | 4.800 | 1.414 | 0.796 | 5.201 | 1.736 |
| C CSP3 ALKANE (1) | N NSP3 HYDRAZINE (150) | 3.800 | 1.443 | 0.773 | 5.179 | 1.567 |
| C CSP3 ALKANE (1) | S >SO2 SULFONAMIDE (154) | 3.104 | 1.776 | 0.743 | 3.681 | 1.445 |
| C CSP3 ALKANE (1) | N NSP3 SULFONAMIDE (155) | 4.454 | 1.454 | 0.885 | 5.155 | 1.587 |
| C CSP3 ALKANE (1) | O O-P=O PHOSPHATE (159) | 4.400 | 1.424 | 0.895 | 5.224 | 1.568 |
| C CSP2 ALKENE (2) | C CSP2 ALKENE (2) | 7.500 | 1.332 | 1.602 | 9.671 | 1.530 |
| C CSP2 ALKENE (2) | C CSP2 CARBONYL (3) | 8.500 | 1.354 | 0.779 | 4.868 | 2.336 |
| C CSP2 ALKENE (2) | C CSP ALKYNE (4) | 11.200 | 1.312 | 1.030 | 6.061 | 2.332 |
| C CSP2 ALKENE (2) | H EXCEPT ON N,O,S (5) | 5.150 | 1.101 | 0.912 | 5.475 | 1.680 |
| C CSP2 ALKENE (2) | O C-O-H, C-O-C, O-O (6) | 6.000 | 1.354 | 0.912 | 6.567 | 1.813 |
| C CSP2 ALKENE (2) | N NSP3 (8) | 6.320 | 1.369 | 1.039 | 6.578 | 1.744 |
| C CSP2 ALKENE (2) | N NSP2 (9) | 5.960 | 1.410 | 1.054 | 6.214 | 1.681 |
| C CSP2 ALKENE (2) | F FLUORIDE (11) | 5.500 | 1.354 | 0.985 | 6.384 | 1.671 |
| C CSP2 ALKENE (2) | Cl CHLORIDE (12) | 2.800 | 1.727 | 1.219 | 3.443 | 1.071 |
| C CSP2 ALKENE (2) | Br BROMIDE (13) | 2.500 | 1.890 | 0.727 | 3.336 | 1.311 |
| C CSP2 ALKENE (2) | I IODIDE (14) | 2.480 | 2.075 | 0.681 | 2.784 | 1.349 |
| C CSP2 ALKENE (2) | S >SO2 SULFONE (18) | 2.800 | 1.772 | 0.662 | 3.350 | 1.455 |
| C CSP2 ALKENE (2) | Si SILANE (19) | 3.000 | 1.854 | 0.836 | 3.124 | 1.339 |
| C CSP2 ALKENE (2) | C CYCLOPROPANE (22) | 5.700 | 1.460 | 0.955 | 5.146 | 1.728 |
| C CSP2 ALKENE (2) | P >P- PHOSPHINE (25) | 2.910 | 1.828 | 0.723 | 3.230 | 1.419 |
| C CSP2 ALKENE (2) | B >B- TRIGONAL (26) | 3.460 | 1.550 | 0.905 | 4.054 | 1.383 |
| C CSP2 ALKENE (2) | Ge GERMANIUM (31) | 3.580 | 1.935 | 0.745 | 2.822 | 1.550 |
| C CSP2 ALKENE (2) | D DEUTERIUM (36) | 5.150 | 1.096 | 0.912 | 5.512 | 1.680 |
| C CSP2 ALKENE (2) | N -N=C-/PYR (DELOCLZD) (37) | 9.000 | 1.271 | 1.586 | 10.187 | 1.685 |
| C CSP2 ALKENE (2) | C CSP2 CYCLOPROPENE (38) | 9.600 | 1.336 | 0.884 | 5.572 | 2.330 |
| C CSP2 ALKENE (2) | N+ NSP3 AMMONIUM (39) | 11.090 | 1.260 | 1.039 | 8.383 | 2.310 |
| C CSP2 ALKENE (2) | N NSP2 PYRROLE (40) | 11.090 | 1.266 | 1.791 | 10.620 | 1.759 |
| C CSP2 ALKENE (2) | O OSP2 FURAN (41) | 12.200 | 1.218 | 1.041 | 7.421 | 2.420 |
| C CSP2 ALKENE (2) | S SSP2 THIOPHENE (42) | 7.171 | 1.537 | 1.408 | 5.197 | 1.596 |
| C CSP2 ALKENE (2) | N NITRO (46) | 5.050 | 1.473 | 0.653 | 4.624 | 1.966 |
| C CSP2 ALKENE (2) | C BENZENE (LOCALIZED) (50) | 5.280 | 1.434 | 0.996 | 5.337 | 1.628 |
| C CSP2 ALKENE (2) | C CSP2 CYCLOBUTENE (57) | 7.500 | 1.333 | 0.925 | 5.579 | 2.013 |
| C CSP2 ALKENE (2) | N =N- IMINE (LOCALZD) (72) | 9.000 | 1.270 | 1.281 | 11.272 | 1.875 |
| C CSP2 ALKENE (2) | C =C=O KETENE (106) | 11.900 | 1.311 | 1.127 | 8.914 | 2.298 |
| C CSP2 ALKENE (2) | N =N-OH OXIME (108) | 8.702 | 1.284 | 1.086 | 11.358 | 2.001 |
| C CSP2 ALKENE (2) | N+ -N(+)= PYRIDINIUM (111) | 8.300 | 1.274 | 1.521 | 9.081 | 1.652 |
| C CSP2 ALKENE (2) | N =N-O AXOXY (DELOC) (143) | 3.800 | 0.827 | 0.317 | 3.745 | 2.448 |
| C CSP2 ALKENE (2) | N =-N= AZOXY (DELOC) (144) | 5.200 | 1.395 | 0.656 | 7.176 | 1.990 |
| C CSP2 CARBONYL (3) | C CSP2 CARBONYL (3) | 11.250 | 1.217 | 0.896 | 4.082 | 2.505 |
| C CSP2 CARBONYL (3) | H EXCEPT ON N,O,S (5) | 4.370 | 1.118 | 0.732 | 4.861 | 1.728 |
| C CSP2 CARBONYL (3) | O C-O-H, C-O-C, O-O (6) | 6.000 | 1.354 | 0.740 | 5.482 | 2.014 |
| C CSP2 CARBONYL (3) | O O=C CARBONYL (7) | 10.100 | 1.208 | 1.501 | 12.653 | 1.834 |
| C CSP2 CARBONYL (3) | N NSP2 (9) | 6.700 | 1.377 | 0.984 | 6.778 | 1.845 |
| C CSP2 CARBONYL (3) | F FLUORIDE (11) | 4.200 | 1.381 | 0.852 | 5.532 | 1.570 |
| C CSP2 CARBONYL (3) | Cl CHLORIDE (12) | 2.880 | 1.816 | 0.842 | 2.935 | 1.308 |
| C CSP2 CARBONYL (3) | Br BROMIDE (13) | 2.800 | 1.990 | 0.617 | 2.436 | 1.506 |
| C CSP2 CARBONYL (3) | I IODIDE (14) | 2.600 | 2.228 | 0.568 | 1.970 | 1.513 |
| C CSP2 CARBONYL (3) | C CYCLOPROPANE (22) | 4.400 | 1.447 | 0.803 | 4.938 | 1.656 |
| C CSP2 CARBONYL (3) | O CARBOXYLATE ION (47) | 7.035 | 1.276 | 1.130 | 9.920 | 1.764 |
| C CSP2 CARBONYL (3) | C CSP3 CYCLOBUTANE (56) | 4.800 | 1.509 | 0.728 | 4.357 | 1.816 |
| C CSP2 CARBONYL (3) | C CSP2 CYCLOBUTENE (57) | 9.600 | 1.351 | 0.751 | 5.069 | 2.528 |
| C CSP2 CARBONYL (3) | O O-H, O-C (CARBOXYL) (75) | 6.000 | 1.354 | 0.833 | 6.401 | 1.897 |
| C CSP2 CARBONYL (3) | O O=C-C=O (76) | 10.800 | 1.209 | 1.232 | 12.009 | 2.094 |
| C CSP2 CARBONYL (3) | O O=C-O-H (ACID) (77) | 9.800 | 1.214 | 1.597 | 13.498 | 1.752 |
| C CSP2 CARBONYL (3) | O O=C-O-C (ESTER) (78) | 9.800 | 1.214 | 1.532 | 12.741 | 1.788 |
| C CSP2 CARBONYL (3) | O O=C-X (HALIDE) (80) | 11.650 | 1.204 | 1.625 | 13.872 | 1.893 |
| C CSP2 CARBONYL (3) | O O=C-C=C< (81) | 9.640 | 1.208 | 1.255 | 11.118 | 1.960 |
| C CSP2 CARBONYL (3) | O O=C-O-C=O (82) | 10.600 | 1.198 | 1.638 | 13.967 | 1.799 |
| C CSP2 CARBONYL (3) | O O=C(C=C)(O-C=O) (102) | 9.600 | 1.204 | 1.581 | 13.275 | 1.742 |
| C CSP2 CARBONYL (3) | O O=C(C=O)(C=C<) (120) | 10.800 | 1.209 | 1.137 | 11.549 | 2.179 |
| C CSP2 CARBONYL (3) | O =-O- ANHYDRIDE (LOCL) (148) | 4.300 | 1.405 | 0.749 | 5.027 | 1.694 |
| C CSP ALKYNE (4) | C CSP ALKYNE (4) | 15.250 | 1.210 | 2.203 | 16.345 | 1.860 |
| C CSP ALKYNE (4) | N NSP (10) | 17.330 | 1.158 | 1.959 | 18.727 | 2.103 |
| C CSP ALKYNE (4) | C CSP2 CYCLOBUTENE (57) | 11.200 | 1.312 | 1.079 | 5.756 | 2.279 |
| C CSP ALKYNE (4) | H H-C ACETYLENE (124) | 5.970 | 1.080 | 1.079 | 6.265 | 1.663 |
| H EXCEPT ON N,O,S (5) | S+ >S+ SULFONIUM (16) | 3.800 | 1.346 | 0.769 | 3.945 | 1.572 |
| H EXCEPT ON N,O,S (5) | S >S=O SULFOXIDE (17) | 3.170 | 1.372 | 0.518 | 3.313 | 1.749 |
| H EXCEPT ON N,O,S (5) | S >SO2 SULFONE (18) | 3.800 | 1.346 | 0.644 | 3.649 | 1.718 |
| H EXCEPT ON N,O,S (5) | Si SILANE (19) | 2.650 | 1.483 | 0.777 | 2.955 | 1.306 |
| H EXCEPT ON N,O,S (5) | C CYCLOPROPANE (22) | 5.080 | 1.086 | 0.877 | 5.493 | 1.702 |
| H EXCEPT ON N,O,S (5) | P >P- PHOSPHINE (25) | 3.065 | 1.420 | 0.667 | 3.328 | 1.516 |
| H EXCEPT ON N,O,S (5) | Ge GERMANIUM (31) | 2.550 | 1.529 | 0.689 | 2.626 | 1.361 |
| H EXCEPT ON N,O,S (5) | Sn TIN (32) | 2.229 | 1.696 | 0.662 | 2.127 | 1.297 |
| H EXCEPT ON N,O,S (5) | Pb LEAD (IV) (33) | 1.894 | 1.775 | 0.549 | 1.725 | 1.314 |
| H EXCEPT ON N,O,S (5) | Se SELENIUM (34) | 3.170 | 1.472 | 0.693 | 3.457 | 1.513 |
| H EXCEPT ON N,O,S (5) | Te TELLURIUM (35) | 2.850 | 1.670 | 0.636 | 2.736 | 1.497 |
| H EXCEPT ON N,O,S (5) | C CSP2 CYCLOPROPENE (38) | 4.600 | 1.072 | 0.913 | 5.879 | 1.587 |
| H EXCEPT ON N,O,S (5) | C BENZENE (LOCALIZED) (50) | 5.150 | 1.101 | 0.903 | 5.436 | 1.689 |
| H EXCEPT ON N,O,S (5) | C CSP3 CYCLOBUTANE (56) | 4.740 | 1.112 | 0.834 | 5.170 | 1.685 |
| H EXCEPT ON N,O,S (5) | C CSP2 CYCLOBUTENE (57) | 5.150 | 1.101 | 0.886 | 5.468 | 1.705 |
| H EXCEPT ON N,O,S (5) | P >P=O PHOSPHATE (153) | 3.280 | 1.398 | 0.719 | 3.608 | 1.510 |
| O C-O-H, C-O-C, O-O (6) | O C-O-H, C-O-C, O-O (6) | 3.950 | 1.448 | 0.401 | 4.485 | 2.219 |
| O C-O-H, C-O-C, O-O (6) | Si SILANE (19) | 5.050 | 1.636 | 0.978 | 5.148 | 1.607 |
| O C-O-H, C-O-C, O-O (6) | H =-OH ALCOHOL (21) | 7.630 | 0.947 | 0.914 | 8.386 | 2.043 |
| O C-O-H, C-O-C, O-O (6) | H COOH CARBOXYL (24) | 7.150 | 0.972 | 0.940 | 7.854 | 1.951 |
| O C-O-H, C-O-C, O-O (6) | P >P- PHOSPHINE (25) | 2.900 | 1.615 | 0.781 | 4.390 | 1.363 |
| O C-O-H, C-O-C, O-O (6) | B >B- TRIGONAL (26) | 4.619 | 1.362 | 1.042 | 5.816 | 1.488 |
| O C-O-H, C-O-C, O-O (6) | C CSP3 CYCLOBUTANE (56) | 2.800 | 1.415 | 0.818 | 5.316 | 1.308 |
| O C-O-H, C-O-C, O-O (6) | C CSP2 CYCLOBUTENE (57) | 6.000 | 1.355 | 0.892 | 6.863 | 1.834 |
| O C-O-H, C-O-C, O-O (6) | H H-O ENOL/PHENOL (73) | 7.200 | 0.972 | 0.891 | 8.216 | 2.010 |
| O C-O-H, C-O-C, O-O (6) | P >P=O PHOSPHATE (153) | 5.300 | 1.599 | 1.398 | 9.924 | 1.377 |
| O O=C CARBONYL (7) | S >S=O SULFOXIDE (17) | 7.100 | 1.487 | 0.633 | 1.845 | 2.369 |
| O O=C CARBONYL (7) | S >SO2 SULFONE (18) | 9.420 | 1.442 | 1.180 | 9.402 | 1.998 |
| O O=C CARBONYL (7) | N NITRO (46) | 7.500 | 1.223 | 1.079 | 10.556 | 1.864 |
| O O=C CARBONYL (7) | C C=O CYCLOBUTANONE (58) | 10.150 | 1.202 | 1.558 | 13.099 | 1.805 |
| O O=C CARBONYL (7) | C C=O CYCLOPROPANONE (67) | 11.420 | 1.196 | 1.539 | 13.191 | 1.926 |
| O O=C CARBONYL (7) | C =C=O KETENE (106) | 10.500 | 1.165 | 1.667 | 15.588 | 1.775 |
| O O=C CARBONYL (7) | P >P=O PHOSPHATE (153) | 8.900 | 1.487 | 1.341 | 10.415 | 1.822 |
| O O=C CARBONYL (7) | S >SO2 SULFONAMIDE (154) | 8.677 | 1.463 | 1.181 | 9.937 | 1.916 |
| N NSP3 (8) | H NH AMINE/IMINE (23) | 6.420 | 1.015 | 0.885 | 7.128 | 1.904 |
| N NSP3 (8) | C BENZENE (LOCALIZED) (50) | 6.320 | 1.378 | 0.965 | 6.262 | 1.810 |
| N NSP3 (8) | C CSP3 CYCLOBUTANE (56) | 5.300 | 1.448 | 0.846 | 5.270 | 1.769 |
| N NSP2 (9) | S >SO2 SULFONE (18) | 6.100 | 1.660 | 0.899 | 5.108 | 1.842 |
| N NSP2 (9) | H H-N-C=O AMIDE (28) | 6.770 | 1.028 | 1.048 | 7.356 | 1.797 |
| S =-S- SULFIDE (15) | S =-S- SULFIDE (15) | 2.620 | 2.019 | 0.429 | 2.890 | 1.747 |
| S =-S- SULFIDE (15) | H SH THIOL (44) | 3.870 | 1.342 | 0.905 | 24.559 | 1.462 |
| Si SILANE (19) | Si SILANE (19) | 1.650 | 2.324 | 0.672 | 1.918 | 1.108 |
| Si SILANE (19) | C CYCLOPROPANE (22) | 3.500 | 1.837 | 0.839 | 3.203 | 1.444 |
| Si SILANE (19) | C CSP3 CYCLOBUTANE (56) | 1.300 | 1.881 | 0.767 | 2.897 | 0.921 |
| H =-OH ALCOHOL (21) | O >N-OH HYDROXYAMINE (145) | 7.500 | 0.974 | 0.824 | 8.418 | 2.134 |
| H =-OH ALCOHOL (21) | O O-P=O PHOSPHATE (159) | 7.780 | 0.948 | 1.002 | 8.287 | 1.970 |
| C CYCLOPROPANE (22) | C CYCLOPROPANE (22) | 5.000 | 1.485 | 1.397 | 5.221 | 1.338 |
| C CYCLOPROPANE (22) | Ge GERMANIUM (31) | 2.700 | 1.911 | 0.752 | 2.835 | 1.340 |
| C CYCLOPROPANE (22) | C CSP2 CYCLOPROPENE (38) | 4.400 | 1.488 | 1.492 | 5.878 | 1.214 |
| C CYCLOPROPANE (22) | N NITRO (46) | 4.350 | 1.478 | 0.610 | 4.214 | 1.889 |
| C CYCLOPROPANE (22) | C CSP3 CYCLOBUTANE (56) | 4.400 | 1.505 | 0.872 | 4.717 | 1.588 |
| H NH AMINE/IMINE (23) | N NSP2 PYRROLE (40) | 6.500 | 1.030 | 1.027 | 7.238 | 1.779 |
| H NH AMINE/IMINE (23) | N =N-O AZOXY (LOCAL) (43) | 5.520 | 1.040 | 0.568 | 6.073 | 2.205 |
| H NH AMINE/IMINE (23) | N =N- IMINE (LOCALZD) (72) | 5.970 | 1.019 | 0.735 | 6.585 | 2.015 |
| H NH AMINE/IMINE (23) | N =-N= AZOXY (LOCAL) (109) | 5.950 | 1.028 | 0.718 | 6.883 | 2.035 |
| H NH AMINE/IMINE (23) | N >N-OH HYDROXYAMINE (146) | 6.150 | 1.021 | 0.771 | 6.846 | 1.998 |
| H NH AMINE/IMINE (23) | N NSP3 HYDRAZINE (150) | 6.360 | 1.021 | 0.774 | 6.814 | 2.027 |
| H NH AMINE/IMINE (23) | N NSP3 SULFONAMIDE (155) | 6.378 | 1.020 | 0.957 | 7.023 | 1.825 |
| H COOH CARBOXYL (24) | O O-H, O-C (CARBOXYL) (75) | 7.150 | 0.974 | 0.934 | 7.996 | 1.956 |
| Ge GERMANIUM (31) | Ge GERMANIUM (31) | 1.450 | 2.404 | 0.542 | 2.038 | 1.157 |
| Pb LEAD (IV) (33) | Pb LEAD (IV) (33) | 2.050 | 1.944 | 0.509 | 0.891 | 1.419 |
| C CSP2 CYCLOPROPENE (38) | C CSP2 CYCLOPROPENE (38) | 9.600 | 1.303 | 0.613 | 3.431 | 2.797 |
| N+ NSP3 AMMONIUM (39) | N NSP2 PYRROLE (40) | 11.000 | 1.230 | 0.597 | 5.133 | 3.034 |
| N+ NSP3 AMMONIUM (39) | H AMMONIUM (48) | 6.140 | 1.053 | 0.861 | 6.329 | 1.888 |
| O OSP2 FURAN (41) | H H-O ENOL/PHENOL (73) | 7.200 | 0.960 | 0.883 | 8.460 | 2.019 |
| O OSP2 FURAN (41) | N =N-OH OXIME (108) | 4.320 | 1.404 | 0.533 | 4.792 | 2.014 |
| N =N-O AZOXY (LOCAL) (43) | O AMINE OXIDE OXYGEN (69) | 8.800 | 1.269 | 0.816 | 8.425 | 2.322 |
| N =N-O AZOXY (LOCAL) (43) | N =-N= AZOXY (LOCAL) (109) | 7.100 | 1.262 | 1.078 | 9.846 | 1.815 |
| C BENZENE (LOCALIZED) (50) | C BENZENE (LOCALIZED) (50) | 6.560 | 1.389 | 2.546 | 10.280 | 1.135 |
| C CSP3 CYCLOBUTANE (56) | C CSP3 CYCLOBUTANE (56) | 4.490 | 1.500 | 1.529 | 6.251 | 1.212 |
| C CSP2 CYCLOBUTENE (57) | C CSP2 CYCLOBUTENE (57) | 7.500 | 1.332 | 1.972 | 12.598 | 1.379 |
| O AMINE OXIDE OXYGEN (69) | N =N-O AXOXY (DELOC) (143) | 9.000 | 1.282 | 0.639 | 5.299 | 2.654 |
| O >N-OH HYDROXYAMINE (145) | N >N-OH HYDROXYAMINE (146) | 4.500 | 1.405 | 0.571 | 5.132 | 1.984 |
| N NSP3 HYDRAZINE (150) | N NSP3 HYDRAZINE (150) | 3.000 | 1.549 | 0.634 | 5.372 | 1.538 |
| P >P=O PHOSPHATE (153) | O O-P=O PHOSPHATE (159) | 5.700 | 1.600 | 1.046 | 7.226 | 1.651 |
| S >SO2 SULFONAMIDE (154) | N NSP3 SULFONAMIDE (155) | 3.944 | 1.697 | 0.703 | 4.727 | 1.674 |

**Table 1**: List of Bonding Interactions fitted in this work, showing the original MM3 force constants ($k_h$), the original MM3 equilibrium bond distances ($r_h$), the fitted values of $D_e$, and the corresponding values of $\alpha_{calc}$ obtained using Eq (2). The harmonic force constants $k_\alpha$ (calculated using $k_\alpha = 2\alpha^2 D_e$) are also shown to enable comparison with the values of $k_h$

Using the data from the current work, it is possible to compare how well the harmonic and Morse MM3 potentials reproduce the accurate CCSD(T)(F12*) energies. Figure 4 compares cumulative (across the whole dataset) RMSD values for both harmonic and Morse potentials relative to the *ab initio* data at different bond displacements from the harmonic equilibrium distance $r_h$. At each point the *ab initio* energy is defined as $E(r) - E(r_h)$ where $E$ denotes the CCSD(T)(F12*) energy at a given bond distance. From Figure 4 it is immediately evident that the Morse representation gives a far more realistic description of the CCSD(T)(F12*) energies, even at relatively small bond displacements

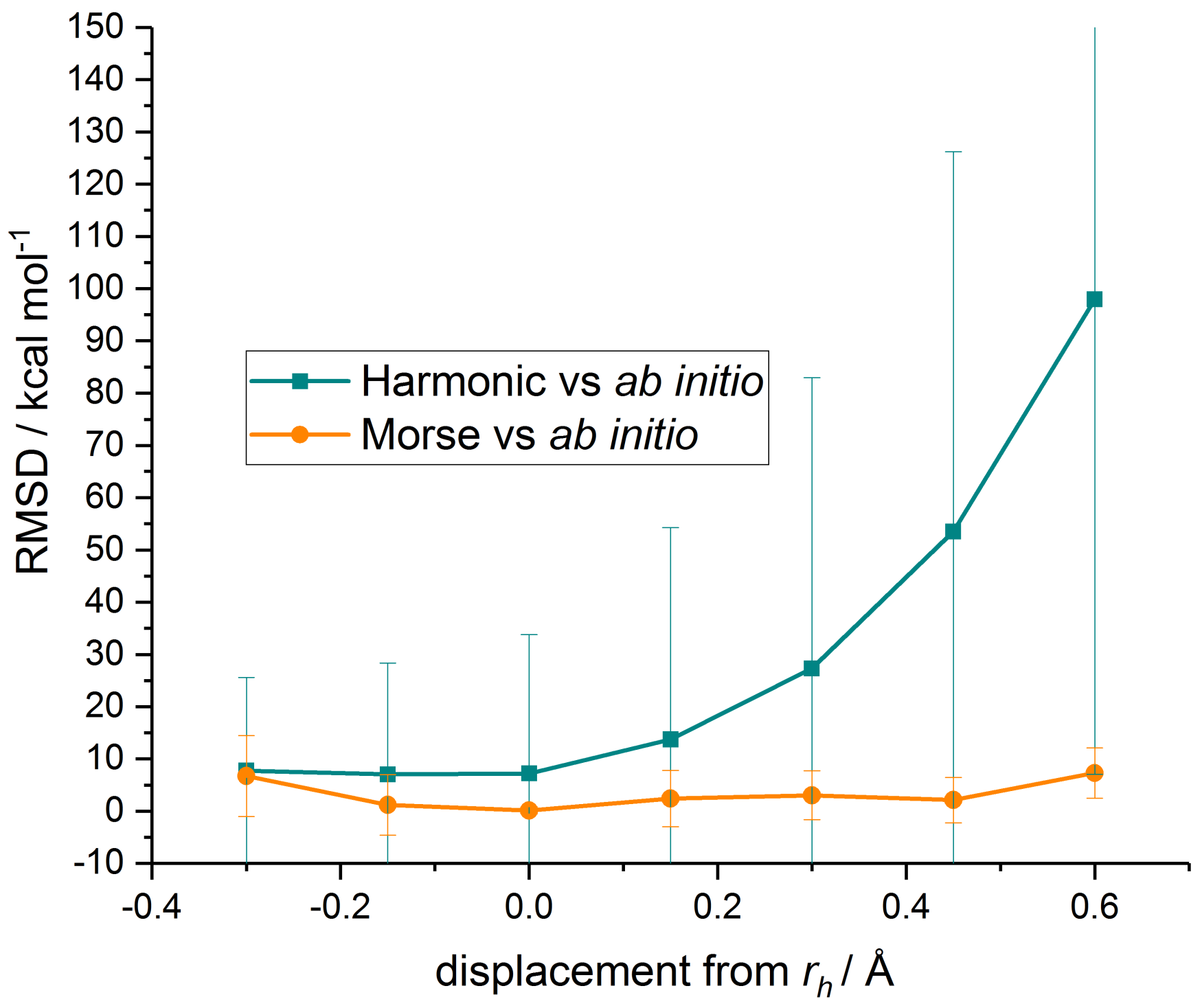

Figure 4: RMSD for the Morse and original harmonic potential energies relative to the CCSD(T)(F12*) energies as a function of the displacement from the harmonic equilibrium bond distance $r_h$. The RMSD values are averaged over all data and the error bars are at the 1σ level.

3. An EVB fitting comparison between Morse and Harmonic MM3 force fields

An accurate treatment of anharmonicity becomes particularly important when bonds are elongated from their equillibirum geometries, as occurs for example during reactive potential energy surface fitting, which highlights the difference between harmonic and anharmonic implementations of the MM3 force field. In the current work we have chosen to compare the behaviour of these two force fields by fitting a reactive potential energy surface for the $H_3CCN \Leftrightarrow CNCH_3$ reaction. As noted in the introduction, molecular mechanics organic chemistry force fields are unreactive by definition (even in the new Morse form where dissociation is possible) since there is an explicit connectivity description required. As such in order to describe chemical reaction in the current work, we construct a reactive potential energy surface using an empirical valence bonding (EVB) approach based upon molecular mechanics descriptions of reactant and product connectivities. This approach has been described in detail previously. [66-67]

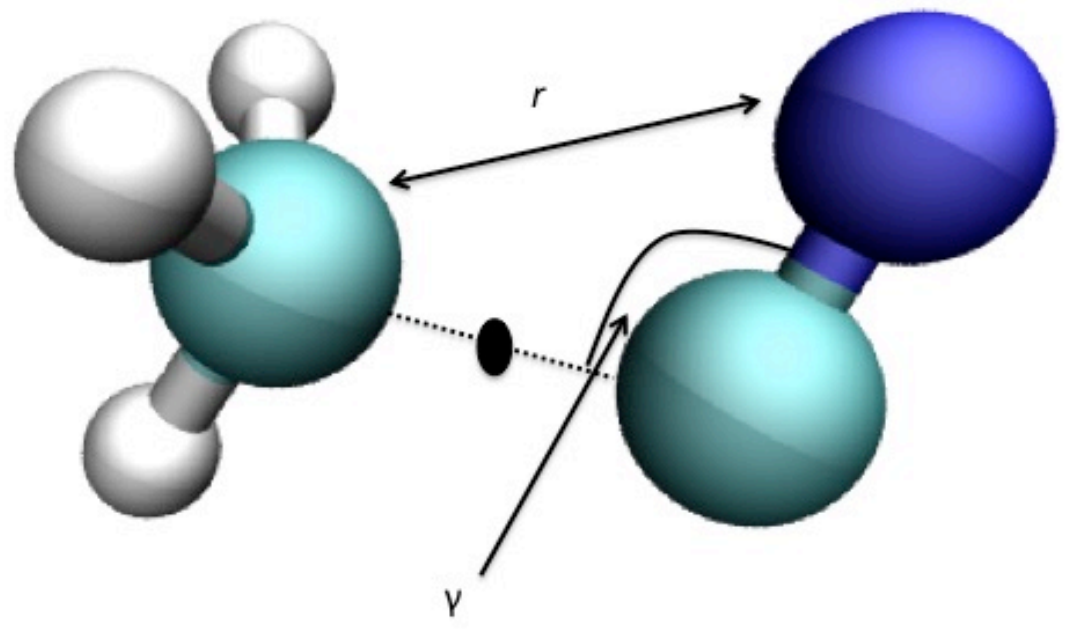

Figure 5: Jacobi coordinate system used for EVB fitting. Here $r$ is the length between the nitrogen atom and the $CH_3$ carbon in Å and $\gamma$ and the angle between the nitrogen, the central carbon and a pivot point (black circle) located 0.62 Å from the $CH_3$ carbon

In order to fully parameterise an EVB potential it is necessary to fit to accurate potential energy data for a given system. In this work we have performed a rigid scan over the Jacobi coordinates $\gamma$ and $r$, both of which are shown schematically in Figure 5.

The angle γ was varied between 3.14 and 0 radians in increments of -0.16 radian and at each fixed γ, $r$ was varied between 2.07Å and 3.07 Å in increments of 0.05 Å giving a total of 400 distinct geometries. At each of these geometries high level energies were obtained at the CCSD(T)(F12*) /aug–cc-pVDZ level of theory using the Molpro electronic structure theory code. Further details of the EVB fitting procedure along with the specific MM3 types and fitting parameters used are detailed in the online supporting information.

The difference between harmonic vs. anharmonic MM3 force fields is apparent when considering the EVB fits. Figure 6 shows the variation in goodness of fit metric $\Delta E$ (Eq.S4 in the supporting information) with iteration of the fitting algorithm. Stochastic errors are obtained with standard bootstrap sampling of the twenty-one separate fits in both cases. [68] Fig 6 clearly shows that the quality of fit is better using the Morse potentials, with a mean unsigned error of $\Delta E$ of 4.03 ± 0.78 kcal mol$^{-1}$ compared to 26.5 ± 4.4 kcal mol$^{-1}$ for the harmonic potentials. This clearly demonstrates that accurately treating anharmonicity at large inter-nuclear separations, is key to quantitatively describing reactive potential energy surfaces like that in the $CH_4$ + CN system. The new dissociation energies presented in this work are expected to be of great utility in deriving EVB and multi-state EVB potential energy surfaces for similar reactive dynamics studies.

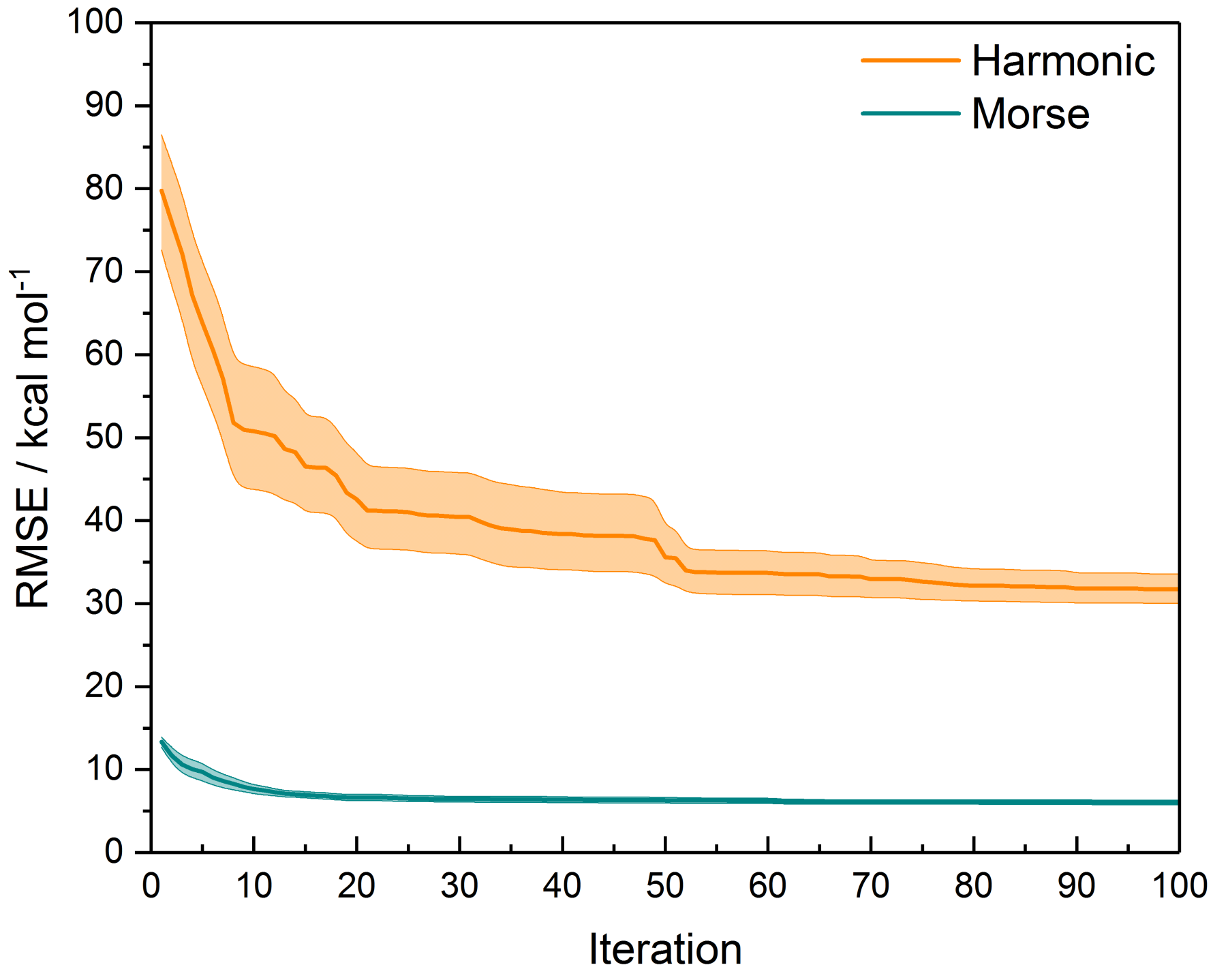

Figure 6: Root mean square errors for both Harmonic and Morse MM3 force fields as a function of fitting iteration. The confidence bands were calculated with standard bootstrap sampling of the twenty-one separate fits in both cases.

## 3 Conclusions

To our knowledge this work represents one of the first parameterisations of Morse potentials for a widely used molecular mechanics force field for organic chemistry. Morse parameters have been fit to high level CCSD(T)(F12*) potentials for all bonds stretches in a total of 254 molecular species and these potentials are in generally good agreement with the harmonic MM3 derived $\alpha_h$. In this article we have shown that this new Morse MM3 force field offers significant improvements upon the original harmonic version, particularly for fitting reactive EVB surfaces. This new Morse MM3 force field is ideally suited to molecular mechanics-based approaches for studying reaction dynamics, which should be of broad practical use to molecular modellers. In previous work[67, 69] we have discussed in detail how modeling energy transfer from vibrationally hot solutes to solvents (and corresponding transient vibrational spectra) depends critically on an accurate treatment of bond anharmonicity. Beyond the EVB fitting application discussed above, accurate treatments of energy transfer offer another application where the new Morse MM3 potential should prove valuable.

## Acknowledgements

Funding for DRG is from a Royal Society URF and the EPSRC via Programme Grant EP/P021123/1. Funding for RJS was provided by the U.S. Air Force of Scientific Research (AFOSR) under Contract No. FA9550-16-1-0051 and EPSRC Programme Grant EP/P021123/1.

Supplementary Information

1. List of species used to parameterise the Morse MM3 forcefield

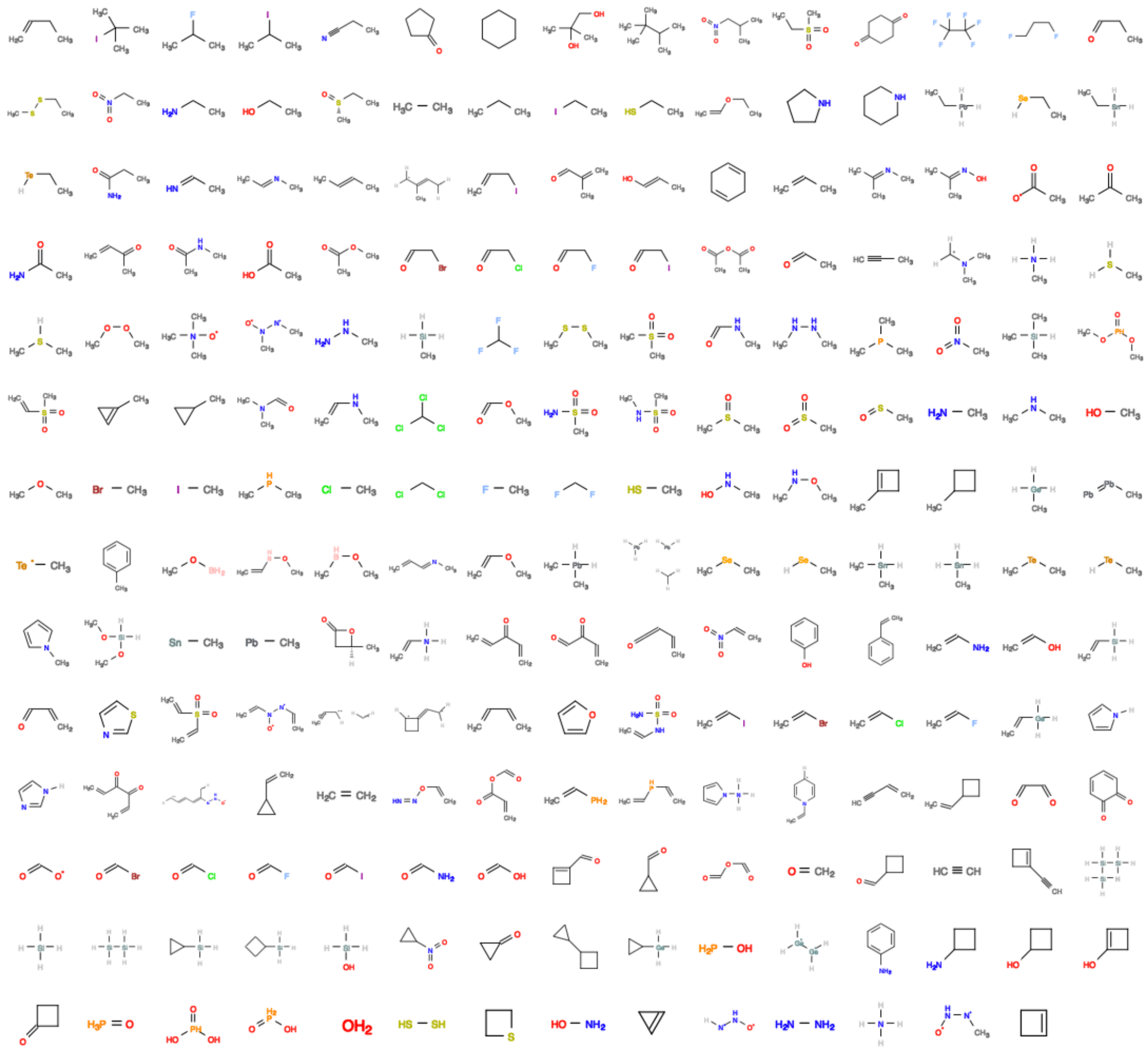

Figure S1: Structures of all compounds used to fit the new anharmonic force field. Below is the same list of compounds in SMILES format

| Type 1 | Type 2 | SMILES code | $r_h$ - 0.15 Å | $r_h$ - 0.075 Å | $r_h$ | $r_h$ +0.15 Å | $r_h$ +0.3 Å | $r_h$ +0.45 Å | $r_h$ +0.6 Å | $r_h$ +0.75 Å |
|---|---|---|---|---|---|---|---|---|---|---|
| | | | CCSD(T)(F12*) energies relative to $r_h$ / kJ mol$^{-1}$ | | | | | | | |
| 1 | 1 | CCC=C | 8.58 | 1.68 | 0.00 | 6.18 | 18.78 | 33.48 | 48.10 | 61.52 |
| 1 | 1 | CC(I)(C)C | 7.56 | 1.22 | 0.00 | 6.76 | 19.79 | 34.81 | 49.65 | 63.25 |
| 1 | 1 | CC(I)(C)C | 7.56 | 1.22 | 0.00 | 6.76 | 19.79 | 34.81 | 49.65 | 63.25 |
| 1 | 1 | CC(I)(C)C | 7.56 | 1.22 | 0.00 | 6.76 | 19.79 | 34.81 | 49.65 | 63.25 |
| 1 | 1 | CC(F)C | 9.73 | 2.05 | 0.00 | 6.10 | 19.00 | 34.13 | 49.20 | 63.07 |
| 1 | 1 | CC(F)C | 9.73 | 2.05 | 0.00 | 6.10 | 19.00 | 34.13 | 49.20 | 63.07 |
| 1 | 1 | CC(I)C | 7.65 | 1.23 | 0.00 | 6.97 | 20.26 | 35.56 | 50.67 | 64.52 |
| 1 | 1 | CC(I)C | 7.65 | 1.23 | 0.00 | 6.97 | 20.26 | 35.56 | 50.67 | 64.52 |
| 1 | 1 | CCC#N | 9.01 | 1.84 | 0.00 | 6.08 | 18.77 | 33.71 | 48.62 | 62.38 |
| 1 | 1 | O=C1CCCC1 | 15.52 | 3.65 | 0.00 | 8.89 | 29.81 | 55.48 | 81.73 | 106.28 |
| 1 | 1 | O=C1CCCC1 | 16.54 | 3.82 | 0.00 | 9.58 | 31.57 | 58.21 | 85.24 | 110.45 |
| 1 | 1 | O=C1CCCC1 | 10.41 | 2.21 | 0.00 | 7.58 | 24.60 | 46.23 | 69.68 | 93.47 |
| 1 | 1 | C1CCCCC1 | 10.82 | 2.15 | 0.00 | 8.99 | 28.79 | 54.08 | 81.40 | 108.46 |
| 1 | 1 | C1CCCCC1 | 16.56 | 3.15 | 0.00 | 12.41 | 37.39 | 66.54 | 95.61 | 122.49 |
| 1 | 1 | C1CCCCC1 | 10.82 | 2.15 | 0.00 | 8.98 | 28.78 | 54.06 | 81.38 | 108.44 |
| 1 | 1 | C1CCCCC1 | 10.20 | 2.12 | 0.00 | 7.72 | 24.89 | 46.50 | 69.57 | 92.47 |
| 1 | 1 | C1CCCCC1 | 10.20 | 2.12 | 0.00 | 7.72 | 24.88 | 46.49 | 69.56 | 92.46 |
| 1 | 1 | OCC(O)(C)C | 8.33 | 1.58 | 0.00 | 6.18 | 18.52 | 32.81 | 46.95 | 59.96 |
| 1 | 1 | OCC(O)(C)C | 8.05 | 1.45 | 0.00 | 6.44 | 19.05 | 33.60 | 48.00 | 61.24 |
| 1 | 1 | OCC(O)(C)C | 8.14 | 1.43 | 0.00 | 6.66 | 19.57 | 34.38 | 48.95 | 62.26 |
| 1 | 1 | CC(C(C)(C)C)C | 7.44 | 1.25 | 0.00 | 6.51 | 19.03 | 33.51 | 47.90 | 61.19 |
| 1 | 1 | CC(C(C)(C)C)C | 7.44 | 1.25 | 0.00 | 6.50 | 19.03 | 33.51 | 47.90 | 61.19 |
| 1 | 1 | CC(C(C)(C)C)C | 7.44 | 1.27 | 0.00 | 6.45 | 18.91 | 33.30 | 47.57 | 60.70 |
| 1 | 1 | CC(C(C)(C)C)C | 7.10 | 1.12 | 0.00 | 6.60 | 19.07 | 33.40 | 47.59 | 60.66 |
| 1 | 1 | CC(C(C)(C)C)C | 7.25 | 1.18 | 0.00 | 6.55 | 19.02 | 33.38 | 47.63 | 60.79 |
| 1 | 1 | CC(C(C)(C)C)C | 7.10 | 1.12 | 0.00 | 6.60 | 19.07 | 33.40 | 47.59 | 60.66 |
| 1 | 1 | CC(CN(=O)=O)C | 7.87 | 1.38 | 0.00 | 6.56 | 19.39 | 34.27 | 49.09 | 62.78 |
| 1 | 1 | CC(CN(=O)=O)C | 7.84 | 1.36 | 0.00 | 6.60 | 19.46 | 34.37 | 49.18 | 62.85 |
| 1 | 1 | CC(CN(=O)=O)C | 7.31 | 1.15 | 0.00 | 6.78 | 19.58 | 34.28 | 48.85 | 62.30 |
| 1 | 1 | CCS(=O)(=O)C | 8.41 | 1.60 | 0.00 | 6.40 | 19.30 | 34.33 | 49.29 | 63.12 |
| 1 | 1 | O=C1CCC(=O)CC1 | 14.97 | 4.06 | 0.00 | 5.91 | 23.32 | 46.65 | 72.28 | 97.78 |
| 1 | 1 | O=C1CCC(=O)CC1 | 12.14 | 2.94 | 0.00 | 6.54 | 22.63 | 42.83 | 63.85 | 83.98 |
| 1 | 1 | FC(C(F)(F)F)(F)F | 14.25 | 3.96 | 0.00 | 3.77 | 15.50 | 30.03 | 44.78 | 58.48 |
| 1 | 1 | FCCCF | 10.84 | 2.52 | 0.00 | 5.55 | 18.21 | 33.28 | 48.37 | 62.33 |
| 1 | 1 | FCCCF | 10.81 | 2.50 | 0.00 | 5.56 | 18.23 | 33.30 | 48.39 | 62.35 |
| 1 | 1 | CCC=O | 9.72 | 2.16 | 0.00 | 5.63 | 18.03 | 32.78 | 47.59 | 61.33 |
| 1 | 1 | CCSSC | 7.82 | 1.34 | 0.00 | 6.67 | 19.55 | 34.37 | 48.99 | 62.34 |
| 1 | 1 | CCN(=O)=O | 8.34 | 1.52 | 0.00 | 6.67 | 19.89 | 35.29 | 50.64 | 64.82 |
| 1 | 1 | CCN | 7.72 | 1.24 | 0.00 | 7.06 | 20.45 | 35.85 | 51.05 | 64.98 |
| 1 | 1 | CCO | 8.27 | 1.51 | 0.00 | 6.49 | 19.24 | 33.93 | 48.41 | 61.66 |
| 1 | 1 | CC[S@@](=O)C | 9.65 | 2.10 | 0.00 | 5.79 | 18.41 | 33.38 | 48.40 | 62.32 |
| 1 | 1 | CC | 8.54 | 1.62 | 0.00 | 6.50 | 19.61 | 34.94 | 50.24 | 64.38 |
| 1 | 1 | CCC | 8.15 | 1.47 | 0.00 | 6.56 | 19.51 | 34.56 | 49.52 | 63.32 |
| 1 | 1 | CCC | 8.13 | 1.47 | 0.00 | 6.57 | 19.52 | 34.57 | 49.53 | 63.32 |
| 1 | 1 | CCI | 8.03 | 1.36 | 0.00 | 6.97 | 20.46 | 36.05 | 51.48 | 65.63 |
| 1 | 1 | CCS | 7.95 | 1.40 | 0.00 | 6.60 | 19.47 | 34.35 | 49.04 | 62.42 |
| 1 | 1 | CCOC=C | 7.66 | 1.17 | 0.00 | 7.31 | 21.04 | 36.80 | 52.36 | 66.62 |
| 1 | 1 | C1CCCN1 | 16.05 | 3.41 | 0.00 | 11.19 | 35.51 | 64.77 | 94.38 | 121.76 |
| 1 | 1 | C1CCCN1 | 14.90 | 3.25 | 0.00 | 9.86 | 31.63 | 57.92 | 84.61 | 109.41 |
| 1 | 1 | C1CCCN1 | 10.34 | 2.24 | 0.00 | 7.07 | 22.98 | 42.91 | 64.37 | 86.22 |
| 1 | 1 | C1CCCNC1 | 10.78 | 2.01 | 0.00 | 9.75 | 30.62 | 56.94 | 85.08 | 112.61 |
| 1 | 1 | C1CCCNC1 | 16.06 | 2.85 | 0.00 | 13.14 | 38.83 | 68.58 | 98.08 | 125.22 |
| 1 | 1 | C1CCCNC1 | 10.42 | 2.17 | 0.00 | 7.85 | 25.34 | 47.38 | 70.94 | 94.38 |
| 1 | 1 | O=C1CCC(=O)CC1 | 14.97 | 4.06 | 0.00 | 5.91 | 23.32 | 46.65 | 72.28 | 97.78 |
| 1 | 1 | O=C1CCC(=O)CC1 | 12.14 | 2.94 | 0.00 | 6.54 | 22.63 | 42.83 | 63.85 | 83.98 |
| 1 | 1 | OCC(O)(C)C | 8.33 | 1.58 | 0.00 | 6.18 | 18.52 | 32.81 | 46.95 | 59.96 |
| 1 | 1 | OCC(O)(C)C | 8.05 | 1.45 | 0.00 | 6.44 | 19.05 | 33.60 | 48.00 | 61.24 |
| 1 | 1 | OCC(O)(C)C | 8.14 | 1.43 | 0.00 | 6.66 | 19.57 | 34.38 | 48.95 | 62.26 |
| 1 | 1 | CCOC=C | 6.85 | 0.85 | 0.00 | 7.64 | 21.47 | 37.22 | 52.73 | 66.92 |
| 1 | 1 | CC[PbH3] | 8.58 | 1.69 | 0.00 | 6.14 | 18.75 | 33.53 | 48.28 | 61.87 |
| 1 | 1 | CC[SeH] | 7.29 | 1.10 | 0.00 | 7.10 | 20.45 | 35.86 | 51.12 | 65.17 |
| 1 | 1 | CC[SnH3] | 9.16 | 1.98 | 0.00 | 5.61 | 17.76 | 32.16 | 46.60 | 59.95 |
| 1 | 1 | CC[TeH] | 7.62 | 1.26 | 0.00 | 6.79 | 19.87 | 35.04 | 50.13 | 64.05 |
| 1 | 1 | CCC(=O)N | 10.57 | 2.51 | 0.00 | 5.22 | 17.45 | 32.17 | 47.01 | 60.82 |
| 1 | 2 | CC=N | 9.70 | 2.09 | 0.00 | 5.79 | 18.11 | 32.43 | 46.54 | 59.41 |
| 1 | 2 | C/N=C/C | 9.23 | 1.79 | 0.00 | 6.63 | 20.02 | 35.52 | 50.82 | 64.77 |
| 1 | 2 | C/C=C/C | 7.78 | 1.20 | 0.00 | 7.40 | 21.31 | 37.31 | 53.13 | 67.67 |
| 1 | 2 | C/C=C/C | 7.78 | 1.20 | 0.00 | 7.40 | 21.31 | 37.31 | 53.13 | 67.66 |
| 1 | 2 | CCC=C | 7.42 | 1.07 | 0.00 | 7.41 | 21.12 | 36.79 | 52.24 | 66.41 |
| 1 | 2 | [CH2]/C=C(\C)/[CH2] | 7.37 | 1.07 | 0.00 | 7.35 | 21.02 | 36.69 | 52.21 | 66.49 |
| 1 | 2 | ICC=C | 5.20 | 0.09 | 0.00 | 8.82 | 23.47 | 39.75 | 55.55 | 69.84 |
| 1 | 2 | CC(=C)C=O | 7.67 | 1.16 | 0.00 | 7.44 | 21.40 | 37.46 | 53.41 | 68.11 |
| 1 | 2 | C/C=C/O | 8.07 | 1.30 | 0.00 | 7.39 | 21.48 | 37.77 | 53.95 | 68.84 |
| 1 | 2 | C1=CCC=CC1 | 14.33 | 2.81 | 0.00 | 13.20 | 42.56 | 80.18 | 120.50 | 159.75 |
| 1 | 2 | C1=CCC=CC1 | 14.33 | 2.82 | 0.00 | 13.20 | 42.56 | 80.18 | 120.50 | 159.75 |
| 1 | 2 | C1=CCC=CC1 | 13.16 | 2.77 | 0.00 | 11.83 | 39.86 | 76.21 | 112.64 | 142.44 |
| 1 | 2 | C1=CCC=CC1 | 13.16 | 2.77 | 0.00 | 11.83 | 39.87 | 76.21 | 112.64 | 142.45 |
| 1 | 2 | CC=C | 7.64 | 1.13 | 0.00 | 7.49 | 21.46 | 37.49 | 53.33 | 67.86 |
| 1 | 2 | CC=N | 9.70 | 2.09 | 0.00 | 5.79 | 18.11 | 32.43 | 46.54 | 59.41 |
| 1 | 2 | CN=C(C)C | 8.57 | 1.68 | 0.00 | 5.98 | 17.97 | 31.74 | 45.24 | 57.56 |
| 1 | 2 | CN=C(C)C | 9.35 | 1.89 | 0.00 | 6.28 | 19.21 | 34.21 | 49.01 | 62.52 |
| 1 | 2 | ON=C(C)C | 8.14 | 1.38 | 0.00 | 6.93 | 20.23 | 35.49 | 50.55 | 64.36 |
| 1 | 2 | ON=C(C)C | 9.66 | 1.96 | 0.00 | 6.56 | 20.13 | 36.01 | 51.81 | 66.34 |
| 1 | 2 | CN=C(C)C | 9.57 | 1.99 | 0.00 | 6.14 | 18.99 | 33.96 | 48.75 | 62.25 |
| 1 | 2 | CN=C(C)C | 9.29 | 1.98 | 0.00 | 5.61 | 17.44 | 31.15 | 44.65 | 56.99 |
| 1 | 3 | CC(=O)[O] | 6.81 | 1.20 | 0.00 | 5.30 | 15.27 | 26.32 | 36.89 | 46.41 |
| 1 | 3 | CC(=O)C | 10.55 | 2.52 | 0.00 | 4.84 | 16.12 | 29.37 | 42.49 | 54.50 |
| 1 | 3 | CC(=O)C | 10.55 | 2.52 | 0.00 | 4.84 | 16.12 | 29.37 | 42.49 | 54.50 |
| 1 | 3 | CC(=O)N | 9.77 | 2.13 | 0.00 | 5.61 | 17.66 | 31.68 | 45.49 | 58.11 |
| 1 | 3 | O=C1CCCC1 | 11.51 | 2.73 | 0.00 | 6.40 | 21.84 | 41.43 | 62.56 | 84.09 |
| 1 | 3 | O=C1CCC(=O)CC1 | 18.03 | 3.86 | 0.00 | 10.53 | 32.70 | 58.25 | 83.34 | 106.29 |
| 1 | 3 | O=C1CCC(=O)CC1 | 11.90 | 2.84 | 0.00 | 6.97 | 24.28 | 46.90 | 71.67 | 96.78 |
| 1 | 3 | CC(=O)C=C | 10.18 | 2.38 | 0.00 | 4.94 | 16.22 | 29.47 | 42.62 | 54.71 |
| 1 | 3 | CNC(=O)C | 9.65 | 2.10 | 0.00 | 5.58 | 17.52 | 31.40 | 45.08 | 57.56 |
| 1 | 3 | CC(=O)O | 12.67 | 3.18 | 0.00 | 5.08 | 17.78 | 33.02 | 48.26 | 62.32 |
| 1 | 3 | CCC=O | 8.20 | 1.49 | 0.00 | 6.24 | 18.33 | 32.03 | 45.34 | 57.38 |
| 1 | 3 | COC(=O)C | 13.64 | 3.64 | 0.00 | 4.29 | 16.32 | 30.96 | 45.68 | 59.24 |
| 1 | 3 | BrCC=O | 9.07 | 1.89 | 0.00 | 5.60 | 17.21 | 30.55 | 43.57 | 55.39 |
| 1 | 3 | ClCC=O | 9.90 | 2.43 | 0.00 | 4.09 | 13.58 | 24.65 | 35.97 | 47.08 |
| 1 | 3 | FCC=O | 8.74 | 1.74 | 0.00 | 5.84 | 17.62 | 31.05 | 44.12 | 55.94 |
| 1 | 3 | ICC=O | 9.33 | 2.00 | 0.00 | 5.49 | 17.08 | 30.44 | 43.51 | 55.40 |
| 1 | 3 | CC(=O)OC(=O)C | 9.42 | 1.89 | 0.00 | 6.31 | 19.20 | 34.06 | 48.66 | 61.95 |
| 1 | 3 | CC(=O)OC(=O)C | 9.42 | 1.89 | 0.00 | 6.31 | 19.19 | 34.06 | 48.65 | 61.94 |
| 1 | 3 | CC=O | 9.98 | 2.20 | 0.00 | 5.53 | 17.47 | 31.30 | 44.85 | 57.15 |
| 1 | 3 | O=C1CCC(=O)CC1 | 18.03 | 3.86 | 0.00 | 10.53 | 32.70 | 58.25 | 83.34 | 106.29 |
| 1 | 3 | O=C1CCC(=O)CC1 | 11.88 | 2.47 | 0.00 | 9.07 | 29.29 | 54.83 | 81.99 | 108.33 |
| 1 | 3 | O=C1CCC(=O)CC1 | 11.90 | 2.84 | 0.00 | 6.97 | 24.28 | 46.90 | 71.67 | 96.78 |
| 1 | 3 | CC(=O)O | 9.68 | 1.98 | 0.00 | 6.36 | 19.50 | 34.75 | 49.82 | 63.62 |
| 1 | 3 | CC(=O)N | 9.77 | 2.13 | 0.00 | 5.62 | 17.67 | 31.68 | 45.49 | 58.11 |
| 1 | 3 | CCC(=O)N | 8.73 | 1.74 | 0.00 | 5.91 | 17.88 | 31.63 | 45.12 | 57.41 |
| 1 | 4 | CCC#N | 10.33 | 2.09 | 0.00 | 6.80 | 20.74 | 36.93 | 53.02 | 67.99 |
| 1 | 4 | CC#C | 11.00 | 2.29 | 0.00 | 7.09 | 22.03 | 39.63 | 57.28 | 73.70 |
| 1 | 5 | CN([CH2])C | 7.90 | 0.95 | 0.00 | 8.40 | 23.18 | 39.65 | 55.57 | 69.89 |
| 1 | 5 | CN([CH2])C | 7.90 | 0.95 | 0.00 | 8.40 | 23.18 | 39.65 | 55.57 | 69.89 |
| 1 | 5 | CN([CH2])C | 7.75 | 0.87 | 0.00 | 8.60 | 23.62 | 40.34 | 56.51 | 71.08 |
| 1 | 5 | CN([CH2])C | 7.90 | 0.95 | 0.00 | 8.40 | 23.18 | 39.65 | 55.57 | 69.89 |
| 1 | 5 | CN([CH2])C | 7.90 | 0.95 | 0.00 | 8.40 | 23.18 | 39.65 | 55.57 | 69.89 |
| 1 | 5 | CN([CH2])C | 7.75 | 0.87 | 0.00 | 8.60 | 23.62 | 40.34 | 56.51 | 71.08 |
| 1 | 5 | CC(=O)[O] | 8.21 | 1.15 | 0.00 | 7.75 | 21.59 | 36.90 | 51.53 | 64.56 |
| 1 | 5 | CC(=O)[O] | 8.34 | 1.22 | 0.00 | 7.64 | 21.39 | 36.63 | 51.22 | 64.23 |
| 1 | 5 | C[NH3] | 7.23 | 0.65 | 0.00 | 8.89 | 24.09 | 40.94 | 57.23 | 71.89 |
| 1 | 5 | C[NH3] | 7.22 | 0.65 | 0.00 | 8.89 | 24.09 | 40.95 | 57.23 | 71.90 |
| 1 | 5 | C[NH3] | 7.22 | 0.65 | 0.00 | 8.89 | 24.09 | 40.95 | 57.23 | 71.90 |
| 1 | 5 | C[SH2] | 8.06 | 1.02 | 0.00 | 8.38 | 23.26 | 39.88 | 55.99 | 70.54 |
| 1 | 5 | C[SH2] | 7.78 | 0.89 | 0.00 | 8.57 | 23.51 | 40.08 | 56.04 | 70.34 |
| 1 | 5 | C[SH2] | 8.06 | 1.02 | 0.00 | 8.38 | 23.26 | 39.88 | 55.99 | 70.54 |
| 1 | 5 | C[SH]C | 8.12 | 1.04 | 0.00 | 8.34 | 23.15 | 39.67 | 55.68 | 70.12 |
| 1 | 5 | C[SH]C | 8.12 | 1.05 | 0.00 | 8.30 | 23.09 | 39.64 | 55.70 | 70.20 |
| 1 | 5 | C[SH]C | 7.99 | 0.98 | 0.00 | 8.39 | 23.16 | 39.58 | 55.41 | 69.61 |
| 1 | 5 | C[SH]C | 8.12 | 1.04 | 0.00 | 8.34 | 23.15 | 39.67 | 55.68 | 70.12 |
| 1 | 5 | C[SH]C | 7.98 | 0.98 | 0.00 | 8.39 | 23.16 | 39.58 | 55.42 | 69.62 |
| 1 | 5 | C[SH]C | 8.13 | 1.05 | 0.00 | 8.30 | 23.08 | 39.63 | 55.70 | 70.20 |
| 1 | 5 | CC=N | 7.57 | 0.87 | 0.00 | 8.26 | 22.64 | 38.59 | 53.96 | 67.76 |
| 1 | 5 | CC=N | 7.70 | 0.93 | 0.00 | 8.14 | 22.39 | 38.19 | 53.38 | 66.94 |
| 1 | 5 | CC=N | 7.70 | 0.93 | 0.00 | 8.14 | 22.39 | 38.19 | 53.38 | 66.94 |
| 1 | 5 | C/N=C/C | 6.98 | 0.67 | 0.00 | 8.24 | 22.21 | 37.47 | 51.97 | 64.80 |
| 1 | 5 | C/N=C/C | 7.57 | 0.86 | 0.00 | 8.30 | 22.75 | 38.80 | 54.28 | 68.20 |
| 1 | 5 | C/N=C/C | 7.62 | 0.89 | 0.00 | 8.23 | 22.57 | 38.44 | 53.66 | 67.22 |
| 1 | 5 | C/N=C/C | 7.62 | 0.89 | 0.00 | 8.23 | 22.57 | 38.44 | 53.66 | 67.22 |
| 1 | 5 | C/N=C/C | 6.74 | 0.52 | 0.00 | 8.69 | 23.25 | 39.18 | 54.37 | 67.87 |
| 1 | 5 | C/N=C/C | 6.98 | 0.67 | 0.00 | 8.24 | 22.21 | 37.47 | 51.97 | 64.80 |
| 1 | 5 | COOC | 8.69 | 1.29 | 0.00 | 7.95 | 22.40 | 38.59 | 54.24 | 68.29 |
| 1 | 5 | COOC | 8.86 | 1.41 | 0.00 | 7.46 | 21.17 | 36.45 | 51.12 | 64.24 |
| 1 | 5 | COOC | 8.73 | 1.35 | 0.00 | 7.61 | 21.50 | 36.95 | 51.79 | 65.04 |
| 1 | 5 | COOC | 8.69 | 1.29 | 0.00 | 7.95 | 22.40 | 38.59 | 54.24 | 68.29 |
| 1 | 5 | COOC | 8.86 | 1.41 | 0.00 | 7.46 | 21.17 | 36.45 | 51.12 | 64.24 |
| 1 | 5 | COOC | 8.73 | 1.35 | 0.00 | 7.61 | 21.50 | 36.95 | 51.79 | 65.04 |
| 1 | 5 | C[N]([O])(C)C | 8.82 | 1.29 | 0.00 | 8.15 | 22.98 | 39.62 | 55.75 | 70.27 |
| 1 | 5 | C[N]([O])(C)C | 9.81 | 1.81 | 0.00 | 6.99 | 20.53 | 35.82 | 50.61 | 63.89 |
| 1 | 5 | C[N]([O])(C)C | 8.80 | 1.29 | 0.00 | 8.16 | 22.99 | 39.63 | 55.76 | 70.27 |
| 1 | 5 | C[N]([O])(C)C | 8.80 | 1.29 | 0.00 | 8.16 | 22.99 | 39.63 | 55.76 | 70.27 |
| 1 | 5 | C[N]([O])(C)C | 9.79 | 1.81 | 0.00 | 7.00 | 20.54 | 35.83 | 50.62 | 63.89 |
| 1 | 5 | C[N]([O])(C)C | 8.82 | 1.29 | 0.00 | 8.15 | 22.98 | 39.62 | 55.75 | 70.27 |
| 1 | 5 | C[N]([O])(C)C | 8.80 | 1.29 | 0.00 | 8.16 | 22.99 | 39.63 | 55.76 | 70.27 |
| 1 | 5 | C[N]([O])(C)C | 9.79 | 1.81 | 0.00 | 7.00 | 20.54 | 35.83 | 50.62 | 63.90 |
| 1 | 5 | C[N]([O])(C)C | 8.80 | 1.29 | 0.00 | 8.16 | 22.99 | 39.63 | 55.76 | 70.27 |
| 1 | 5 | C[N]N([O])C | 8.53 | 1.20 | 0.00 | 8.14 | 22.76 | 39.05 | 54.75 | 68.78 |
| 1 | 5 | C[N]N([O])C | 9.03 | 1.39 | 0.00 | 8.02 | 22.76 | 39.32 | 55.35 | 69.75 |
| 1 | 5 | C[N]N([O])C | 9.03 | 1.38 | 0.00 | 8.02 | 22.76 | 39.32 | 55.35 | 69.75 |
| 1 | 5 | C[N]N([O])C | 9.73 | 1.63 | 0.00 | 7.90 | 22.80 | 39.63 | 55.97 | 70.64 |
| 1 | 5 | C[N]N([O])C | 9.04 | 1.49 | 0.00 | 7.37 | 21.02 | 36.20 | 50.72 | 63.60 |
| 1 | 5 | C[N]N([O])C | 9.04 | 1.49 | 0.00 | 7.37 | 21.02 | 36.20 | 50.72 | 63.60 |
| 1 | 5 | CNN | 7.99 | 1.03 | 0.00 | 8.09 | 22.39 | 38.26 | 53.52 | 67.18 |
| 1 | 5 | CNN | 8.85 | 1.52 | 0.00 | 6.80 | 19.48 | 33.55 | 47.01 | 59.03 |
| 1 | 5 | CNN | 8.02 | 1.06 | 0.00 | 7.96 | 22.08 | 37.74 | 52.77 | 66.23 |
| 1 | 5 | C[SiH3] | 8.24 | 1.13 | 0.00 | 8.02 | 22.38 | 38.41 | 53.88 | 67.77 |
| 1 | 5 | C[SiH3] | 8.23 | 1.13 | 0.00 | 8.02 | 22.38 | 38.41 | 53.88 | 67.78 |
| 1 | 5 | C[SiH3] | 8.24 | 1.13 | 0.00 | 8.02 | 22.38 | 38.41 | 53.88 | 67.77 |
| 1 | 5 | C/C=C/C | 8.22 | 1.16 | 0.00 | 7.78 | 21.70 | 37.17 | 51.99 | 65.21 |
| 1 | 5 | C/C=C/C | 8.22 | 1.16 | 0.00 | 7.78 | 21.71 | 37.17 | 51.99 | 65.21 |
| 1 | 5 | C/C=C/C | 8.11 | 1.09 | 0.00 | 8.00 | 22.25 | 38.13 | 53.45 | 67.24 |
| 1 | 5 | C/C=C/C | 8.22 | 1.16 | 0.00 | 7.78 | 21.70 | 37.17 | 51.99 | 65.21 |
| 1 | 5 | C/C=C/C | 8.22 | 1.16 | 0.00 | 7.78 | 21.70 | 37.17 | 51.99 | 65.21 |
| 1 | 5 | C/C=C/C | 8.11 | 1.09 | 0.00 | 8.00 | 22.25 | 38.12 | 53.45 | 67.24 |
| 1 | 5 | CCC=C | 8.04 | 1.08 | 0.00 | 7.93 | 22.04 | 37.73 | 52.87 | 66.47 |
| 1 | 5 | CCC=C | 8.12 | 1.11 | 0.00 | 7.87 | 21.93 | 37.59 | 52.72 | 66.33 |
| 1 | 5 | CCC=C | 8.03 | 1.09 | 0.00 | 7.79 | 21.67 | 37.08 | 51.92 | 65.25 |
| 1 | 5 | CCC=C | 8.14 | 1.16 | 0.00 | 7.62 | 21.26 | 36.39 | 50.90 | 63.87 |
| 1 | 5 | CCC=C | 8.01 | 1.05 | 0.00 | 7.99 | 22.18 | 37.96 | 53.20 | 66.89 |
| 1 | 5 | [CH2]/C=C(\C)/[CH2] | 7.94 | 1.00 | 0.00 | 8.19 | 22.63 | 38.69 | 54.18 | 68.10 |
| 1 | 5 | [CH2]/C=C(\C)/[CH2] | 8.22 | 1.16 | 0.00 | 7.80 | 21.80 | 37.37 | 52.36 | 65.78 |
| 1 | 5 | [CH2]/C=C(\C)/[CH2] | 8.22 | 1.16 | 0.00 | 7.80 | 21.80 | 37.37 | 52.36 | 65.77 |
| 1 | 5 | CC(I)(C)C | 8.35 | 1.18 | 0.00 | 7.86 | 22.11 | 38.01 | 53.39 | 67.23 |
| 1 | 5 | CC(I)(C)C | 8.75 | 1.39 | 0.00 | 7.42 | 21.18 | 36.58 | 51.47 | 64.87 |
| 1 | 5 | CC(I)(C)C | 8.35 | 1.18 | 0.00 | 7.86 | 22.11 | 38.01 | 53.39 | 67.23 |
| 1 | 5 | CC(I)(C)C | 8.35 | 1.18 | 0.00 | 7.86 | 22.11 | 38.01 | 53.39 | 67.23 |
| 1 | 5 | CC(I)(C)C | 8.75 | 1.39 | 0.00 | 7.41 | 21.18 | 36.58 | 51.47 | 64.87 |
| 1 | 5 | CC(I)(C)C | 8.35 | 1.18 | 0.00 | 7.86 | 22.11 | 38.01 | 53.39 | 67.23 |
| 1 | 5 | CC(I)(C)C | 8.35 | 1.18 | 0.00 | 7.86 | 22.11 | 38.01 | 53.39 | 67.23 |
| 1 | 5 | CC(I)(C)C | 8.75 | 1.39 | 0.00 | 7.42 | 21.18 | 36.58 | 51.47 | 64.87 |
| 1 | 5 | CC(I)(C)C | 8.35 | 1.18 | 0.00 | 7.86 | 22.11 | 38.01 | 53.39 | 67.23 |
| 1 | 5 | CC(=O)C | 8.05 | 1.04 | 0.00 | 8.18 | 22.67 | 38.81 | 54.38 | 68.35 |
| 1 | 5 | CC(=O)C | 8.14 | 1.09 | 0.00 | 8.04 | 22.40 | 38.42 | 53.88 | 67.75 |
| 1 | 5 | CC(=O)C | 8.14 | 1.09 | 0.00 | 8.04 | 22.40 | 38.42 | 53.88 | 67.75 |
| 1 | 5 | CC(=O)C | 8.05 | 1.04 | 0.00 | 8.18 | 22.67 | 38.81 | 54.38 | 68.35 |
| 1 | 5 | CC(=O)C | 8.14 | 1.09 | 0.00 | 8.04 | 22.40 | 38.42 | 53.88 | 67.75 |
| 1 | 5 | CC(=O)C | 8.14 | 1.09 | 0.00 | 8.04 | 22.40 | 38.42 | 53.88 | 67.75 |
| 1 | 5 | CC(F)C | 8.48 | 1.24 | 0.00 | 7.83 | 22.03 | 37.93 | 53.32 | 67.19 |
| 1 | 5 | CC(F)C | 8.57 | 1.27 | 0.00 | 7.84 | 22.13 | 38.16 | 53.70 | 67.73 |
| 1 | 5 | CC(F)C | 8.61 | 1.30 | 0.00 | 7.74 | 21.89 | 37.76 | 53.15 | 67.02 |
| 1 | 5 | CC(F)C | 8.48 | 1.24 | 0.00 | 7.83 | 22.03 | 37.93 | 53.32 | 67.19 |
| 1 | 5 | CC(F)C | 9.08 | 1.51 | 0.00 | 7.29 | 20.85 | 35.99 | 50.55 | 63.60 |
| 1 | 5 | CC(F)C | 8.61 | 1.30 | 0.00 | 7.74 | 21.89 | 37.76 | 53.15 | 67.02 |
| 1 | 5 | CC(F)C | 8.57 | 1.27 | 0.00 | 7.85 | 22.13 | 38.16 | 53.70 | 67.73 |
| 1 | 5 | CC(I)C | 8.21 | 1.12 | 0.00 | 8.02 | 22.34 | 38.32 | 53.74 | 67.61 |
| 1 | 5 | CC(I)C | 8.66 | 1.35 | 0.00 | 7.54 | 21.36 | 36.82 | 51.75 | 65.18 |
| 1 | 5 | CC(I)C | 8.35 | 1.18 | 0.00 | 7.89 | 22.10 | 37.98 | 53.34 | 67.17 |
| 1 | 5 | CC(I)C | 8.22 | 1.12 | 0.00 | 8.02 | 22.34 | 38.31 | 53.74 | 67.61 |
| 1 | 5 | CC(I)C | 8.27 | 1.12 | 0.00 | 8.00 | 22.16 | 37.78 | 52.65 | 65.82 |
| 1 | 5 | CC(I)C | 8.35 | 1.18 | 0.00 | 7.89 | 22.10 | 37.98 | 53.34 | 67.17 |
| 1 | 5 | CC(I)C | 8.66 | 1.35 | 0.00 | 7.54 | 21.36 | 36.82 | 51.75 | 65.18 |
| 1 | 5 | CCC#N | 7.88 | 0.98 | 0.00 | 8.17 | 22.57 | 38.60 | 54.08 | 68.01 |
| 1 | 5 | CCC#N | 7.86 | 0.97 | 0.00 | 8.18 | 22.59 | 38.63 | 54.13 | 68.07 |
| 1 | 5 | CCC#N | 7.86 | 0.98 | 0.00 | 8.18 | 22.59 | 38.63 | 54.13 | 68.07 |
| 1 | 5 | CCC#N | 8.26 | 1.15 | 0.00 | 7.92 | 22.09 | 37.85 | 52.99 | 66.51 |
| 1 | 5 | CCC#N | 8.24 | 1.14 | 0.00 | 7.93 | 22.11 | 37.88 | 53.03 | 66.55 |
| 1 | 5 | FC(F)F | 11.97 | 2.51 | 0.00 | 7.01 | 21.56 | 38.24 | 54.46 | 69.05 |
| 1 | 5 | CC(=O)N | 7.57 | 0.80 | 0.00 | 8.66 | 23.66 | 40.33 | 56.42 | 70.87 |
| 1 | 5 | CC(=O)N | 8.20 | 1.14 | 0.00 | 7.91 | 22.10 | 37.94 | 53.24 | 66.98 |
| 1 | 5 | CC(=O)N | 8.20 | 1.14 | 0.00 | 7.91 | 22.10 | 37.95 | 53.24 | 66.99 |
| 1 | 5 | CSSC | 8.16 | 1.08 | 0.00 | 8.16 | 22.67 | 38.82 | 54.36 | 68.28 |
| 1 | 5 | CSSC | 8.00 | 1.02 | 0.00 | 8.09 | 22.32 | 37.98 | 52.87 | 66.01 |
| 1 | 5 | CSSC | 7.90 | 0.98 | 0.00 | 8.19 | 22.53 | 38.32 | 53.35 | 66.64 |
| 1 | 5 | CSSC | 8.16 | 1.08 | 0.00 | 8.16 | 22.67 | 38.82 | 54.36 | 68.28 |
| 1 | 5 | CSSC | 8.00 | 1.02 | 0.00 | 8.09 | 22.32 | 37.98 | 52.87 | 66.01 |
| 1 | 5 | CSSC | 7.91 | 0.98 | 0.00 | 8.19 | 22.53 | 38.32 | 53.35 | 66.64 |
| 1 | 5 | CS(=O)(=O)C | 8.61 | 1.25 | 0.00 | 8.00 | 22.50 | 38.72 | 54.42 | 68.56 |
| 1 | 5 | CS(=O)(=O)C | 8.43 | 1.15 | 0.00 | 8.30 | 23.16 | 39.77 | 55.88 | 70.44 |
| 1 | 5 | CS(=O)(=O)C | 8.61 | 1.25 | 0.00 | 8.00 | 22.50 | 38.72 | 54.42 | 68.56 |
| 1 | 5 | CS(=O)(=O)C | 8.61 | 1.25 | 0.00 | 8.00 | 22.50 | 38.72 | 54.42 | 68.56 |
| 1 | 5 | CS(=O)(=O)C | 8.43 | 1.15 | 0.00 | 8.30 | 23.16 | 39.77 | 55.88 | 70.44 |
| 1 | 5 | CS(=O)(=O)C | 8.61 | 1.25 | 0.00 | 8.00 | 22.50 | 38.72 | 54.42 | 68.56 |
| 1 | 5 | CNC=O | 7.62 | 0.84 | 0.00 | 8.52 | 23.30 | 39.69 | 55.48 | 69.64 |
| 1 | 5 | CNC=O | 8.05 | 1.09 | 0.00 | 7.83 | 21.68 | 36.96 | 51.53 | 64.48 |
| 1 | 5 | CNC=O | 8.05 | 1.09 | 0.00 | 7.83 | 21.68 | 36.96 | 51.54 | 64.48 |
| 1 | 5 | CNNC | 8.34 | 1.26 | 0.00 | 7.37 | 20.69 | 35.42 | 49.51 | 62.07 |
| 1 | 5 | CNNC | 7.84 | 0.96 | 0.00 | 8.22 | 22.67 | 38.72 | 54.18 | 68.06 |
| 1 | 5 | CNNC | 7.85 | 0.96 | 0.00 | 8.22 | 22.67 | 38.71 | 54.18 | 68.05 |
| 1 | 5 | CNNC | 8.34 | 1.26 | 0.00 | 7.37 | 20.69 | 35.43 | 49.51 | 62.07 |
| 1 | 5 | CNNC | 8.14 | 1.13 | 0.00 | 7.80 | 21.74 | 37.23 | 52.12 | 65.46 |
| 1 | 5 | CNNC | 8.15 | 1.13 | 0.00 | 7.80 | 21.74 | 37.22 | 52.11 | 65.45 |
| 1 | 5 | CP(C)C | 8.75 | 1.34 | 0.00 | 7.73 | 21.90 | 37.79 | 53.18 | 67.05 |
| 1 | 5 | CP(C)C | 9.06 | 1.51 | 0.00 | 7.27 | 20.81 | 35.89 | 50.34 | 63.21 |
| 1 | 5 | CP(C)C | 8.75 | 1.34 | 0.00 | 7.73 | 21.90 | 37.79 | 53.18 | 67.05 |
| 1 | 5 | CP(C)C | 8.75 | 1.34 | 0.00 | 7.73 | 21.90 | 37.79 | 53.18 | 67.05 |
| 1 | 5 | CP(C)C | 9.06 | 1.51 | 0.00 | 7.27 | 20.81 | 35.89 | 50.34 | 63.21 |
| 1 | 5 | CP(C)C | 8.75 | 1.34 | 0.00 | 7.73 | 21.90 | 37.79 | 53.18 | 67.05 |
| 1 | 5 | CP(C)C | 8.75 | 1.34 | 0.00 | 7.73 | 21.90 | 37.79 | 53.18 | 67.05 |
| 1 | 5 | CP(C)C | 8.75 | 1.34 | 0.00 | 7.73 | 21.90 | 37.79 | 53.18 | 67.05 |
| 1 | 5 | CP(C)C | 9.06 | 1.51 | 0.00 | 7.27 | 20.81 | 35.89 | 50.34 | 63.21 |
| 1 | 5 | CN(=O)=O | 8.41 | 1.12 | 0.00 | 8.39 | 23.38 | 40.13 | 56.32 | 70.84 |
| 1 | 5 | CN(=O)=O | 8.59 | 1.22 | 0.00 | 8.18 | 22.96 | 39.51 | 55.53 | 69.91 |
| 1 | 5 | CN(=O)=O | 8.59 | 1.22 | 0.00 | 8.18 | 22.96 | 39.51 | 55.53 | 69.91 |
| 1 | 5 | C[SiH](C)C | 8.42 | 1.22 | 0.00 | 7.82 | 21.99 | 37.81 | 53.10 | 66.84 |
| 1 | 5 | C[SiH](C)C | 8.33 | 1.18 | 0.00 | 7.90 | 22.13 | 38.02 | 53.35 | 67.13 |
| 1 | 5 | C[SiH](C)C | 8.42 | 1.22 | 0.00 | 7.82 | 21.98 | 37.81 | 53.10 | 66.84 |
| 1 | 5 | C[SiH](C)C | 8.34 | 1.18 | 0.00 | 7.90 | 22.13 | 38.02 | 53.35 | 67.13 |
| 1 | 5 | C[SiH](C)C | 8.33 | 1.18 | 0.00 | 7.90 | 22.13 | 38.02 | 53.35 | 67.13 |
| 1 | 5 | C[SiH](C)C | 8.42 | 1.22 | 0.00 | 7.82 | 21.98 | 37.81 | 53.10 | 66.84 |
| 1 | 5 | C[SiH](C)C | 8.33 | 1.18 | 0.00 | 7.90 | 22.13 | 38.02 | 53.35 | 67.13 |
| 1 | 5 | C[SiH](C)C | 8.33 | 1.18 | 0.00 | 7.90 | 22.13 | 38.02 | 53.35 | 67.13 |
| 1 | 5 | C[SiH](C)C | 8.34 | 1.18 | 0.00 | 7.90 | 22.13 | 38.02 | 53.35 | 67.13 |
| 1 | 5 | ICC=C | 8.71 | 1.27 | 0.00 | 7.98 | 22.33 | 38.22 | 53.35 | 66.72 |
| 1 | 5 | ICC=C | 8.64 | 1.24 | 0.00 | 8.07 | 22.57 | 38.65 | 54.03 | 67.69 |
| 1 | 5 | CC(=C)C=O | 8.08 | 1.07 | 0.00 | 8.04 | 22.36 | 38.31 | 53.73 | 67.62 |
| 1 | 5 | CC(=C)C=O | 7.96 | 1.02 | 0.00 | 8.13 | 22.50 | 38.47 | 53.85 | 67.61 |
| 1 | 5 | CC(=C)C=O | 7.96 | 1.02 | 0.00 | 8.12 | 22.49 | 38.46 | 53.82 | 67.57 |
| 1 | 5 | C/C=C/O | 8.21 | 1.14 | 0.00 | 7.88 | 22.00 | 37.74 | 52.94 | 66.61 |
| 1 | 5 | C/C=C/O | 8.19 | 1.15 | 0.00 | 7.74 | 21.56 | 36.83 | 51.41 | 64.34 |
| 1 | 5 | C/C=C/O | 8.19 | 1.15 | 0.00 | 7.74 | 21.56 | 36.83 | 51.41 | 64.34 |
| 1 | 5 | C1=CCC=CC1 | 8.35 | 1.27 | 0.00 | 7.35 | 20.63 | 35.30 | 49.25 | 61.58 |
| 1 | 5 | C1=CCC=CC1 | 8.34 | 1.26 | 0.00 | 7.35 | 20.64 | 35.30 | 49.25 | 61.58 |
| 1 | 5 | C1=CCC=CC1 | 8.36 | 1.27 | 0.00 | 7.34 | 20.63 | 35.29 | 49.25 | 61.57 |
| 1 | 5 | C1=CCC=CC1 | 8.32 | 1.26 | 0.00 | 7.36 | 20.64 | 35.31 | 49.26 | 61.59 |
| 1 | 5 | O=C1CCCC1 | 7.93 | 1.01 | 0.00 | 8.13 | 22.46 | 38.37 | 53.66 | 67.35 |
| 1 | 5 | O=C1CCCC1 | 8.18 | 1.15 | 0.00 | 7.75 | 21.66 | 37.12 | 51.99 | 65.29 |
| 1 | 5 | O=C1CCCC1 | 8.18 | 1.15 | 0.00 | 7.76 | 21.66 | 37.14 | 52.01 | 65.31 |
| 1 | 5 | O=C1CCCC1 | 7.93 | 1.00 | 0.00 | 8.13 | 22.46 | 38.36 | 53.65 | 67.33 |
| 1 | 5 | O=C1CCCC1 | 7.89 | 1.04 | 0.00 | 7.83 | 21.69 | 37.05 | 51.83 | 65.10 |
| 1 | 5 | O=C1CCCC1 | 7.86 | 0.99 | 0.00 | 8.08 | 22.26 | 37.95 | 53.01 | 66.49 |
| 1 | 5 | O=C1CCCC1 | 7.90 | 1.04 | 0.00 | 7.83 | 21.69 | 37.05 | 51.83 | 65.10 |
| 1 | 5 | O=C1CCCC1 | 7.86 | 0.99 | 0.00 | 8.07 | 22.25 | 37.94 | 53.00 | 66.48 |
| 1 | 5 | C1CCCCC1 | 8.25 | 1.22 | 0.00 | 7.47 | 20.98 | 36.04 | 50.56 | 63.64 |
| 1 | 5 | C1CCCCC1 | 7.97 | 1.08 | 0.00 | 7.74 | 21.47 | 36.69 | 51.31 | 64.42 |
| 1 | 5 | C1CCCCC1 | 7.97 | 1.08 | 0.00 | 7.74 | 21.48 | 36.69 | 51.31 | 64.42 |
| 1 | 5 | C1CCCCC1 | 8.25 | 1.22 | 0.00 | 7.47 | 20.98 | 36.04 | 50.56 | 63.63 |
| 1 | 5 | C1CCCCC1 | 8.25 | 1.22 | 0.00 | 7.47 | 20.98 | 36.03 | 50.55 | 63.63 |
| 1 | 5 | C1CCCCC1 | 7.96 | 1.08 | 0.00 | 7.74 | 21.48 | 36.69 | 51.31 | 64.42 |
| 1 | 5 | C1CCCCC1 | 7.96 | 1.08 | 0.00 | 7.74 | 21.48 | 36.70 | 51.31 | 64.42 |
| 1 | 5 | C1CCCCC1 | 8.25 | 1.22 | 0.00 | 7.47 | 20.98 | 36.03 | 50.55 | 63.63 |
| 1 | 5 | C1CCCCC1 | 7.96 | 1.08 | 0.00 | 7.74 | 21.48 | 36.70 | 51.31 | 64.42 |
| 1 | 5 | C1CCCCC1 | 8.25 | 1.22 | 0.00 | 7.47 | 20.98 | 36.03 | 50.55 | 63.63 |
| 1 | 5 | C1CCCCC1 | 7.97 | 1.08 | 0.00 | 7.74 | 21.48 | 36.69 | 51.31 | 64.42 |
| 1 | 5 | C1CCCCC1 | 8.25 | 1.22 | 0.00 | 7.47 | 20.98 | 36.04 | 50.56 | 63.63 |
| 1 | 5 | OCC(O)(C)C | 8.52 | 1.29 | 0.00 | 7.61 | 21.49 | 37.04 | 52.11 | 65.71 |
| 1 | 5 | OCC(O)(C)C | 8.19 | 1.10 | 0.00 | 8.11 | 22.61 | 38.81 | 54.50 | 68.62 |
| 1 | 5 | OCC(O)(C)C | 8.93 | 1.48 | 0.00 | 7.18 | 20.51 | 35.37 | 49.62 | 62.38 |
| 1 | 5 | OCC(O)(C)C | 8.54 | 1.24 | 0.00 | 7.90 | 22.18 | 38.11 | 53.48 | 67.26 |
| 1 | 5 | OCC(O)(C)C | 8.28 | 1.16 | 0.00 | 7.89 | 22.10 | 37.99 | 53.36 | 67.22 |
| 1 | 5 | OCC(O)(C)C | 8.43 | 1.24 | 0.00 | 7.71 | 21.68 | 37.31 | 52.44 | 66.08 |
| 1 | 5 | OCC(O)(C)C | 8.43 | 1.23 | 0.00 | 7.77 | 21.85 | 37.63 | 52.91 | 66.68 |
| 1 | 5 | CC(C(C)(C)C)C | 7.78 | 1.06 | 0.00 | 7.49 | 20.77 | 35.45 | 49.54 | 62.18 |
| 1 | 5 | CC(C(C)(C)C)C | 8.41 | 1.23 | 0.00 | 7.75 | 21.81 | 37.53 | 52.74 | 66.45 |
| 1 | 5 | CC(C(C)(C)C)C | 8.14 | 1.12 | 0.00 | 7.85 | 21.88 | 37.51 | 52.58 | 66.13 |
| 1 | 5 | CC(C(C)(C)C)C | 8.29 | 1.16 | 0.00 | 7.91 | 22.13 | 38.01 | 53.35 | 67.16 |
| 1 | 5 | CC(C(C)(C)C)C | 8.34 | 1.19 | 0.00 | 7.83 | 21.94 | 37.70 | 52.93 | 66.62 |
| 1 | 5 | CC(C(C)(C)C)C | 8.40 | 1.24 | 0.00 | 7.64 | 21.50 | 37.00 | 52.00 | 65.54 |
| 1 | 5 | CC(C(C)(C)C)C | 8.26 | 1.18 | 0.00 | 7.76 | 21.74 | 37.33 | 52.38 | 65.94 |
| 1 | 5 | CC(C(C)(C)C)C | 8.26 | 1.18 | 0.00 | 7.77 | 21.75 | 37.36 | 52.43 | 66.01 |
| 1 | 5 | CC(C(C)(C)C)C | 8.41 | 1.22 | 0.00 | 7.77 | 21.85 | 37.59 | 52.82 | 66.55 |
| 1 | 5 | CC(C(C)(C)C)C | 8.40 | 1.24 | 0.00 | 7.64 | 21.50 | 37.01 | 52.01 | 65.54 |
| 1 | 5 | CC(C(C)(C)C)C | 8.34 | 1.19 | 0.00 | 7.83 | 21.94 | 37.70 | 52.93 | 66.62 |
| 1 | 5 | CC(C(C)(C)C)C | 8.26 | 1.18 | 0.00 | 7.76 | 21.74 | 37.33 | 52.38 | 65.94 |
| 1 | 5 | CC(C(C)(C)C)C | 8.13 | 1.12 | 0.00 | 7.85 | 21.88 | 37.51 | 52.58 | 66.14 |
| 1 | 5 | CC(C(C)(C)C)C | 8.40 | 1.23 | 0.00 | 7.76 | 21.81 | 37.54 | 52.75 | 66.45 |
| 1 | 5 | CC(C(C)(C)C)C | 8.30 | 1.16 | 0.00 | 7.91 | 22.13 | 38.01 | 53.35 | 67.16 |
| 1 | 5 | CC(CN(=O)=O)C | 7.84 | 1.01 | 0.00 | 7.91 | 21.88 | 37.40 | 52.35 | 65.77 |
| 1 | 5 | CC(CN(=O)=O)C | 8.32 | 1.21 | 0.00 | 7.71 | 21.65 | 37.25 | 52.35 | 65.98 |
| 1 | 5 | CC(CN(=O)=O)C | 7.94 | 1.01 | 0.00 | 8.11 | 22.45 | 38.40 | 53.80 | 67.65 |
| 1 | 5 | CC(CN(=O)=O)C | 8.23 | 1.17 | 0.00 | 7.78 | 21.78 | 37.41 | 52.53 | 66.16 |
| 1 | 5 | CC(CN(=O)=O)C | 8.58 | 1.24 | 0.00 | 8.02 | 22.52 | 38.74 | 54.41 | 68.49 |
| 1 | 5 | CC(CN(=O)=O)C | 8.74 | 1.33 | 0.00 | 7.81 | 22.06 | 38.01 | 53.41 | 67.23 |
| 1 | 5 | CC(CN(=O)=O)C | 8.04 | 1.06 | 0.00 | 8.03 | 22.29 | 38.18 | 53.53 | 67.34 |
| 1 | 5 | CC(CN(=O)=O)C | 8.33 | 1.21 | 0.00 | 7.73 | 21.71 | 37.35 | 52.50 | 66.18 |
| 1 | 5 | CC(CN(=O)=O)C | 8.30 | 1.18 | 0.00 | 7.81 | 21.91 | 37.70 | 52.99 | 66.80 |
| 1 | 5 | CCS(=O)(=O)C | 8.72 | 1.34 | 0.00 | 7.71 | 21.91 | 37.80 | 53.14 | 66.91 |
| 1 | 5 | CCS(=O)(=O)C | 8.63 | 1.27 | 0.00 | 7.97 | 22.44 | 38.63 | 54.31 | 68.43 |
| 1 | 5 | CCS(=O)(=O)C | 8.65 | 1.26 | 0.00 | 8.05 | 22.64 | 38.97 | 54.79 | 69.06 |
| 1 | 5 | CCS(=O)(=O)C | 8.41 | 1.13 | 0.00 | 8.33 | 23.22 | 39.85 | 55.97 | 70.54 |
| 1 | 5 | CCS(=O)(=O)C | 8.23 | 1.15 | 0.00 | 7.90 | 22.10 | 37.93 | 53.21 | 66.93 |
| 1 | 5 | CCS(=O)(=O)C | 8.04 | 1.03 | 0.00 | 8.22 | 22.81 | 39.10 | 54.88 | 69.10 |
| 1 | 5 | CCS(=O)(=O)C | 8.29 | 1.15 | 0.00 | 7.99 | 22.32 | 38.35 | 53.86 | 67.83 |
| 1 | 5 | CCS(=O)(=O)C | 8.56 | 1.25 | 0.00 | 7.93 | 22.36 | 38.53 | 54.18 | 68.28 |
| 1 | 5 | COP(=O)OC | 8.04 | 0.99 | 0.00 | 8.41 | 23.17 | 39.54 | 55.27 | 69.32 |
| 1 | 5 | COP(=O)OC | 8.37 | 1.18 | 0.00 | 7.91 | 22.10 | 37.91 | 53.14 | 66.79 |
| 1 | 5 | COP(=O)OC | 8.33 | 1.16 | 0.00 | 8.00 | 22.34 | 38.33 | 53.76 | 67.61 |
| 1 | 5 | COP(=O)OC | 8.21 | 1.10 | 0.00 | 8.10 | 22.54 | 38.62 | 54.13 | 68.05 |
| 1 | 5 | COP(=O)OC | 8.05 | 0.99 | 0.00 | 8.39 | 23.14 | 39.51 | 55.23 | 69.27 |
| 1 | 5 | COP(=O)OC | 8.41 | 1.21 | 0.00 | 7.83 | 21.92 | 37.62 | 52.75 | 66.33 |
| 1 | 5 | C=CS(=O)(=O)C | 8.52 | 1.20 | 0.00 | 8.18 | 22.92 | 39.40 | 55.38 | 69.80 |
| 1 | 5 | C=CS(=O)(=O)C | 8.58 | 1.23 | 0.00 | 8.10 | 22.70 | 39.00 | 54.74 | 68.92 |
| 1 | 5 | C=CS(=O)(=O)C | 8.54 | 1.22 | 0.00 | 8.04 | 22.54 | 38.76 | 54.48 | 68.66 |
| 1 | 5 | O=C1CCC(=O)CC1 | 7.70 | 0.91 | 0.00 | 8.24 | 22.64 | 38.60 | 53.97 | 67.75 |
| 1 | 5 | O=C1CCC(=O)CC1 | 8.29 | 1.22 | 0.00 | 7.55 | 21.24 | 36.51 | 51.24 | 64.47 |
| 1 | 5 | O=C1CCC(=O)CC1 | 7.69 | 0.91 | 0.00 | 8.24 | 22.64 | 38.60 | 53.97 | 67.75 |
| 1 | 5 | O=C1CCC(=O)CC1 | 8.28 | 1.22 | 0.00 | 7.55 | 21.24 | 36.51 | 51.24 | 64.47 |
| 1 | 5 | O=C1CCC(=O)CC1 | 7.69 | 0.91 | 0.00 | 8.24 | 22.63 | 38.59 | 53.95 | 67.73 |
| 1 | 5 | O=C1CCC(=O)CC1 | 7.69 | 0.91 | 0.00 | 8.24 | 22.63 | 38.59 | 53.95 | 67.72 |
| 1 | 5 | O=C1CCC(=O)CC1 | 8.29 | 1.23 | 0.00 | 7.55 | 21.23 | 36.51 | 51.24 | 64.47 |
| 1 | 5 | C[N]N([O])C | 9.03 | 1.39 | 0.00 | 8.02 | 22.76 | 39.32 | 55.35 | 69.75 |
| 1 | 5 | C[N]N([O])C | 8.53 | 1.20 | 0.00 | 8.14 | 22.76 | 39.05 | 54.75 | 68.78 |
| 1 | 5 | C[N]N([O])C | 9.04 | 1.49 | 0.00 | 7.37 | 21.02 | 36.20 | 50.72 | 63.60 |
| 1 | 5 | C[N]N([O])C | 9.04 | 1.49 | 0.00 | 7.37 | 21.02 | 36.20 | 50.72 | 63.60 |
| 1 | 5 | C[N]N([O])C | 9.73 | 1.63 | 0.00 | 7.90 | 22.79 | 39.63 | 55.97 | 70.64 |
| 1 | 5 | C[N]N([O])C | 9.03 | 1.39 | 0.00 | 8.02 | 22.76 | 39.32 | 55.35 | 69.75 |
| 1 | 5 | CC1=CC1 | 8.03 | 1.04 | 0.00 | 8.10 | 22.41 | 38.28 | 53.49 | 67.04 |
| 1 | 5 | CC1=CC1 | 8.23 | 1.15 | 0.00 | 7.86 | 21.93 | 37.56 | 52.57 | 65.96 |
| 1 | 5 | CC1=CC1 | 8.35 | 1.21 | 0.00 | 7.70 | 21.52 | 36.78 | 51.25 | 63.92 |
| 1 | 5 | CC1CC1 | 8.25 | 1.17 | 0.00 | 7.76 | 21.69 | 37.19 | 52.12 | 65.49 |
| 1 | 5 | CC1CC1 | 8.20 | 1.13 | 0.00 | 7.90 | 22.03 | 37.79 | 53.01 | 66.71 |
| 1 | 5 | CC1CC1 | 8.25 | 1.17 | 0.00 | 7.76 | 21.70 | 37.21 | 52.13 | 65.51 |
| 1 | 5 | FCCCF | 9.40 | 1.60 | 0.00 | 7.43 | 21.44 | 37.19 | 52.44 | 66.14 |
| 1 | 5 | FCCCF | 8.89 | 1.43 | 0.00 | 7.49 | 21.41 | 37.10 | 52.34 | 66.12 |
| 1 | 5 | FCCCF | 8.90 | 1.44 | 0.00 | 7.48 | 21.39 | 37.07 | 52.31 | 66.08 |
| 1 | 5 | FCCCF | 9.21 | 1.54 | 0.00 | 7.42 | 21.27 | 36.78 | 51.74 | 65.14 |
| 1 | 5 | FCCCF | 9.39 | 1.60 | 0.00 | 7.43 | 21.43 | 37.18 | 52.41 | 66.10 |
| 1 | 5 | FCCCF | 9.23 | 1.54 | 0.00 | 7.42 | 21.26 | 36.77 | 51.73 | 65.14 |
| 1 | 5 | CC(=O)C=C | 7.77 | 0.91 | 0.00 | 8.42 | 23.15 | 39.57 | 55.44 | 69.71 |
| 1 | 5 | CC(=O)C=C | 8.19 | 1.13 | 0.00 | 7.93 | 22.14 | 38.00 | 53.29 | 67.01 |
| 1 | 5 | CC(=O)C=C | 8.20 | 1.13 | 0.00 | 7.93 | 22.15 | 38.00 | 53.30 | 67.02 |
| 1 | 5 | CNC(=O)C | 8.18 | 1.16 | 0.00 | 7.68 | 21.37 | 36.48 | 50.89 | 63.70 |
| 1 | 5 | CNC(=O)C | 8.18 | 1.15 | 0.00 | 7.68 | 21.37 | 36.48 | 50.90 | 63.71 |
| 1 | 5 | CNC(=O)C | 7.52 | 0.78 | 0.00 | 8.67 | 23.67 | 40.31 | 56.36 | 70.76 |
| 1 | 5 | CNC(=O)C | 8.20 | 1.14 | 0.00 | 7.91 | 22.07 | 37.87 | 53.10 | 66.76 |
| 1 | 5 | CNC(=O)C | 8.20 | 1.13 | 0.00 | 7.91 | 22.08 | 37.88 | 53.11 | 66.77 |
| 1 | 5 | CNC(=O)C | 7.91 | 0.91 | 0.00 | 8.65 | 23.83 | 40.77 | 57.19 | 71.96 |
| 1 | 5 | CC(=O)O | 7.96 | 0.97 | 0.00 | 8.39 | 23.18 | 39.65 | 55.57 | 69.86 |
| 1 | 5 | CC(=O)O | 8.16 | 1.08 | 0.00 | 8.16 | 22.72 | 38.98 | 54.71 | 68.84 |
| 1 | 5 | CC(=O)O | 8.16 | 1.08 | 0.00 | 8.16 | 22.72 | 38.99 | 54.72 | 68.85 |
| 1 | 5 | CCC=O | 7.84 | 0.97 | 0.00 | 8.17 | 22.54 | 38.49 | 53.86 | 67.65 |
| 1 | 5 | CCC=O | 8.16 | 1.13 | 0.00 | 7.85 | 21.91 | 37.61 | 52.79 | 66.46 |
| 1 | 5 | CCC=O | 8.03 | 1.06 | 0.00 | 7.98 | 22.16 | 37.97 | 53.24 | 66.99 |
| 1 | 5 | CCC=O | 8.23 | 1.19 | 0.00 | 7.61 | 21.35 | 36.66 | 51.40 | 64.63 |
| 1 | 5 | CCC=O | 7.75 | 0.95 | 0.00 | 8.15 | 22.47 | 38.38 | 53.72 | 67.51 |
| 1 | 5 | CCSSC | 8.17 | 1.09 | 0.00 | 8.15 | 22.66 | 38.80 | 54.33 | 68.24 |
| 1 | 5 | CCSSC | 7.92 | 0.99 | 0.00 | 8.17 | 22.49 | 38.27 | 53.28 | 66.55 |
| 1 | 5 | CCSSC | 8.00 | 1.02 | 0.00 | 8.10 | 22.33 | 38.01 | 52.92 | 66.09 |
| 1 | 5 | CCSSC | 8.21 | 1.17 | 0.00 | 7.76 | 21.74 | 37.33 | 52.35 | 65.85 |
| 1 | 5 | CCSSC | 7.98 | 1.04 | 0.00 | 8.04 | 22.30 | 38.17 | 53.49 | 67.26 |
| 1 | 5 | CCSSC | 8.04 | 1.05 | 0.00 | 8.07 | 22.42 | 38.42 | 53.89 | 67.84 |
| 1 | 5 | CCSSC | 8.05 | 1.08 | 0.00 | 7.93 | 22.06 | 37.73 | 52.75 | 66.15 |
| 1 | 5 | CCSSC | 8.12 | 1.12 | 0.00 | 7.80 | 21.68 | 36.96 | 51.48 | 64.32 |
| 1 | 5 | CCN(=O)=O | 8.26 | 1.13 | 0.00 | 8.06 | 22.49 | 38.63 | 54.24 | 68.30 |
| 1 | 5 | CCN(=O)=O | 8.32 | 1.16 | 0.00 | 7.98 | 22.34 | 38.42 | 54.00 | 68.05 |
| 1 | 5 | CCN(=O)=O | 8.32 | 1.16 | 0.00 | 7.98 | 22.34 | 38.41 | 54.00 | 68.04 |
| 1 | 5 | CCN(=O)=O | 8.37 | 1.14 | 0.00 | 8.17 | 22.77 | 39.04 | 54.70 | 68.72 |
| 1 | 5 | CCN(=O)=O | 8.37 | 1.14 | 0.00 | 8.17 | 22.78 | 39.05 | 54.71 | 68.73 |
| 1 | 5 | O=CN(C)C | 8.09 | 1.10 | 0.00 | 7.85 | 21.78 | 37.18 | 51.88 | 64.95 |
| 1 | 5 | O=CN(C)C | 7.72 | 0.87 | 0.00 | 8.50 | 23.31 | 39.73 | 55.53 | 69.68 |
| 1 | 5 | O=CN(C)C | 8.15 | 1.14 | 0.00 | 7.70 | 21.41 | 36.53 | 50.94 | 63.71 |
| 1 | 5 | O=CN(C)C | 8.15 | 1.14 | 0.00 | 7.70 | 21.42 | 36.54 | 50.95 | 63.73 |
| 1 | 5 | O=CN(C)C | 7.85 | 0.94 | 0.00 | 8.34 | 22.95 | 39.18 | 54.81 | 68.82 |
| 1 | 5 | O=CN(C)C | 8.09 | 1.10 | 0.00 | 7.85 | 21.79 | 37.18 | 51.89 | 64.97 |
| 1 | 5 | CNC=C | 8.08 | 1.06 | 0.00 | 8.08 | 22.41 | 38.35 | 53.70 | 67.46 |
| 1 | 5 | CNC=C | 8.61 | 1.38 | 0.00 | 7.16 | 20.24 | 34.69 | 48.46 | 60.70 |
| 1 | 5 | CNC=C | 8.00 | 1.03 | 0.00 | 8.11 | 22.41 | 38.26 | 53.47 | 67.03 |
| 1 | 5 | ClC(Cl)Cl | 9.83 | 1.86 | 0.00 | 4.85 | 13.46 | 24.39 | 38.81 | 55.15 |
| 1 | 5 | COC(=O)C | 8.49 | 1.25 | 0.00 | 7.79 | 21.94 | 37.78 | 53.10 | 66.86 |
| 1 | 5 | COC(=O)C | 8.20 | 1.05 | 0.00 | 8.35 | 23.13 | 39.57 | 55.41 | 69.60 |
| 1 | 5 | COC(=O)C | 8.27 | 1.12 | 0.00 | 8.08 | 22.48 | 38.48 | 53.88 | 67.66 |
| 1 | 5 | COC(=O)C | 8.28 | 1.12 | 0.00 | 8.07 | 22.47 | 38.47 | 53.86 | 67.63 |
| 1 | 5 | COC(=O)C | 7.52 | 0.76 | 0.00 | 8.81 | 23.98 | 40.79 | 56.98 | 71.46 |
| 1 | 5 | COC(=O)C | 8.47 | 1.25 | 0.00 | 7.79 | 21.94 | 37.78 | 53.10 | 66.86 |
| 1 | 5 | COC=O | 8.17 | 1.04 | 0.00 | 8.39 | 23.22 | 39.71 | 55.60 | 69.83 |
| 1 | 5 | COC=O | 8.25 | 1.11 | 0.00 | 8.10 | 22.52 | 38.56 | 53.99 | 67.80 |
| 1 | 5 | COC=O | 8.26 | 1.11 | 0.00 | 8.10 | 22.51 | 38.54 | 53.96 | 67.75 |
| 1 | 5 | CS(=O)(=O)C | 8.44 | 1.15 | 0.00 | 8.30 | 23.15 | 39.76 | 55.86 | 70.43 |
| 1 | 5 | CS(=O)(=O)C | 8.44 | 1.15 | 0.00 | 8.30 | 23.15 | 39.76 | 55.86 | 70.43 |
| 1 | 5 | CS(=O)(=O)C | 8.58 | 1.24 | 0.00 | 8.02 | 22.52 | 38.74 | 54.44 | 68.59 |
| 1 | 5 | CS(=O)(=O)C | 8.59 | 1.25 | 0.00 | 8.01 | 22.51 | 38.74 | 54.44 | 68.58 |
| 1 | 5 | CS(=O)(=O)C | 8.59 | 1.25 | 0.00 | 8.01 | 22.51 | 38.74 | 54.44 | 68.58 |
| 1 | 5 | CS(=O)(=O)C | 8.58 | 1.24 | 0.00 | 8.02 | 22.52 | 38.74 | 54.44 | 68.59 |
| 1 | 5 | CS(=O)(=O)N | 10.55 | 2.03 | 0.00 | 7.17 | 21.41 | 37.69 | 53.63 | 68.09 |
| 1 | 5 | CS(=O)(=O)N | 10.40 | 1.96 | 0.00 | 7.27 | 21.53 | 37.76 | 53.59 | 67.90 |
| 1 | 5 | CS(=O)(=O)N | 10.40 | 1.96 | 0.00 | 7.28 | 21.54 | 37.77 | 53.60 | 67.91 |
| 1 | 5 | CNS(=O)(=O)C | 9.67 | 1.64 | 0.00 | 7.74 | 22.33 | 38.81 | 54.78 | 69.13 |
| 1 | 5 | CNS(=O)(=O)C | 10.00 | 1.85 | 0.00 | 7.11 | 20.89 | 36.43 | 51.46 | 64.95 |
| 1 | 5 | CNS(=O)(=O)C | 10.54 | 2.14 | 0.00 | 6.39 | 19.27 | 33.82 | 47.81 | 60.29 |
| 1 | 5 | CNS(=O)(=O)C | 10.51 | 2.01 | 0.00 | 7.19 | 21.44 | 37.74 | 53.68 | 68.12 |
| 1 | 5 | CNS(=O)(=O)C | 10.50 | 2.01 | 0.00 | 7.15 | 21.33 | 37.51 | 53.30 | 67.57 |
| 1 | 5 | CNS(=O)(=O)C | 10.40 | 1.96 | 0.00 | 7.27 | 21.58 | 37.90 | 53.81 | 68.19 |
| 1 | 5 | CS(=O)C | 8.38 | 1.14 | 0.00 | 8.22 | 22.93 | 39.33 | 55.20 | 69.49 |
| 1 | 5 | CS(=O)C | 8.62 | 1.28 | 0.00 | 7.83 | 22.01 | 37.81 | 53.01 | 66.61 |
| 1 | 5 | CS(=O)C | 8.79 | 1.36 | 0.00 | 7.76 | 22.01 | 37.99 | 53.44 | 67.33 |
| 1 | 5 | CS(=O)C | 8.79 | 1.36 | 0.00 | 7.76 | 22.01 | 37.99 | 53.44 | 67.33 |
| 1 | 5 | CS(=O)C | 8.62 | 1.28 | 0.00 | 7.83 | 22.01 | 37.81 | 53.01 | 66.61 |
| 1 | 5 | CS(=O)C | 8.38 | 1.14 | 0.00 | 8.22 | 22.93 | 39.33 | 55.20 | 69.49 |
| 1 | 5 | CS(=O)=O | 8.64 | 1.26 | 0.00 | 8.03 | 22.60 | 38.90 | 54.68 | 68.90 |
| 1 | 5 | CS(=O)=O | 8.47 | 1.17 | 0.00 | 8.28 | 23.17 | 39.87 | 56.09 | 70.77 |
| 1 | 5 | CS(=O)=O | 8.64 | 1.26 | 0.00 | 8.03 | 22.60 | 38.90 | 54.68 | 68.90 |
| 1 | 5 | CS=O | 8.77 | 1.35 | 0.00 | 7.78 | 22.07 | 38.11 | 53.63 | 67.62 |
| 1 | 5 | CS=O | 8.37 | 1.16 | 0.00 | 8.09 | 22.54 | 38.58 | 53.97 | 67.71 |
| 1 | 5 | CS=O | 8.38 | 1.15 | 0.00 | 8.20 | 22.89 | 39.30 | 55.15 | 69.41 |
| 1 | 5 | BrCC=O | 7.82 | 0.93 | 0.00 | 8.34 | 22.89 | 38.93 | 54.24 | 67.78 |
| 1 | 5 | BrCC=O | 7.83 | 0.94 | 0.00 | 8.32 | 22.84 | 38.85 | 54.12 | 67.63 |
| 1 | 5 | ClCC=O | 9.10 | 1.49 | 0.00 | 7.52 | 21.72 | 37.85 | 53.26 | 66.77 |
| 1 | 5 | ClCC=O | 9.10 | 1.49 | 0.00 | 7.52 | 21.73 | 37.86 | 53.27 | 66.77 |
| 1 | 5 | FCC=O | 9.40 | 1.60 | 0.00 | 7.45 | 21.46 | 37.20 | 52.38 | 65.96 |
| 1 | 5 | FCC=O | 9.40 | 1.60 | 0.00 | 7.45 | 21.46 | 37.20 | 52.38 | 65.96 |
| 1 | 5 | ICC=O | 8.70 | 1.26 | 0.00 | 8.05 | 22.55 | 38.67 | 54.08 | 67.72 |
| 1 | 5 | ICC=O | 8.68 | 1.26 | 0.00 | 8.06 | 22.57 | 38.70 | 54.12 | 67.78 |
| 1 | 5 | CCN | 8.05 | 1.07 | 0.00 | 8.01 | 22.27 | 38.16 | 53.52 | 67.34 |
| 1 | 5 | CCN | 8.26 | 1.17 | 0.00 | 7.77 | 21.75 | 37.36 | 52.44 | 66.03 |
| 1 | 5 | CCN | 8.07 | 1.07 | 0.00 | 8.01 | 22.27 | 38.16 | 53.51 | 67.30 |
| 1 | 5 | CCN | 8.52 | 1.38 | 0.00 | 6.96 | 19.67 | 33.65 | 46.95 | 58.78 |
| 1 | 5 | CCN | 7.90 | 1.02 | 0.00 | 7.92 | 21.87 | 37.27 | 52.03 | 65.20 |
| 1 | 5 | CCO | 8.43 | 1.24 | 0.00 | 7.71 | 21.72 | 37.41 | 52.61 | 66.32 |
| 1 | 5 | CCO | 8.30 | 1.19 | 0.00 | 7.78 | 21.81 | 37.47 | 52.61 | 66.25 |
| 1 | 5 | CCO | 8.09 | 1.07 | 0.00 | 8.06 | 22.41 | 38.41 | 53.87 | 67.77 |
| 1 | 5 | CCO | 8.57 | 1.34 | 0.00 | 7.35 | 20.74 | 35.60 | 49.84 | 62.58 |
| 1 | 5 | CCO | 8.11 | 1.07 | 0.00 | 8.08 | 22.38 | 38.23 | 53.44 | 67.03 |
| 1 | 5 | CC[S@@](=O)C | 8.93 | 1.40 | 0.00 | 7.76 | 22.12 | 38.26 | 53.89 | 67.95 |
| 1 | 5 | CC[S@@](=O)C | 8.72 | 1.33 | 0.00 | 7.73 | 21.82 | 37.51 | 52.60 | 66.10 |
| 1 | 5 | CC[S@@](=O)C | 8.38 | 1.14 | 0.00 | 8.25 | 23.00 | 39.45 | 55.34 | 69.65 |
| 1 | 5 | CC[S@@](=O)C | 8.11 | 1.09 | 0.00 | 8.02 | 22.29 | 38.16 | 53.44 | 67.12 |
| 1 | 5 | CC[S@@](=O)C | 8.28 | 1.15 | 0.00 | 7.98 | 22.29 | 38.26 | 53.70 | 67.59 |
| 1 | 5 | CC[S@@](=O)C | 8.25 | 1.15 | 0.00 | 7.88 | 22.04 | 37.87 | 53.19 | 67.00 |
| 1 | 5 | CC[S@@](=O)C | 8.80 | 1.40 | 0.00 | 7.49 | 21.31 | 36.71 | 51.48 | 64.68 |
| 1 | 5 | CC[S@@](=O)C | 8.49 | 1.25 | 0.00 | 7.82 | 22.03 | 37.92 | 53.26 | 67.05 |
| 1 | 5 | CN | 8.57 | 1.36 | 0.00 | 7.17 | 20.21 | 34.59 | 48.27 | 60.41 |
| 1 | 5 | CN | 7.97 | 1.01 | 0.00 | 8.15 | 22.51 | 38.42 | 53.69 | 67.33 |
| 1 | 5 | CN | 7.97 | 1.01 | 0.00 | 8.15 | 22.51 | 38.42 | 53.69 | 67.33 |
| 1 | 5 | CNC | 8.10 | 1.07 | 0.00 | 8.05 | 22.32 | 38.16 | 53.39 | 67.01 |
| 1 | 5 | CNC | 9.06 | 1.61 | 0.00 | 6.68 | 19.26 | 33.23 | 46.58 | 58.49 |
| 1 | 5 | CNC | 8.21 | 1.12 | 0.00 | 7.94 | 22.10 | 37.84 | 52.97 | 66.51 |
| 1 | 5 | CNC | 8.21 | 1.12 | 0.00 | 7.94 | 22.11 | 37.84 | 52.97 | 66.51 |
| 1 | 5 | CNC | 9.06 | 1.61 | 0.00 | 6.67 | 19.26 | 33.23 | 46.58 | 58.49 |
| 1 | 5 | CNC | 8.10 | 1.07 | 0.00 | 8.05 | 22.32 | 38.16 | 53.39 | 67.01 |
| 1 | 5 | CO | 8.62 | 1.33 | 0.00 | 7.49 | 21.11 | 36.24 | 50.75 | 63.70 |
| 1 | 5 | CO | 8.62 | 1.33 | 0.00 | 7.49 | 21.11 | 36.24 | 50.75 | 63.70 |
| 1 | 5 | CO | 8.12 | 1.03 | 0.00 | 8.33 | 23.03 | 39.35 | 55.05 | 69.09 |
| 1 | 5 | COC | 8.31 | 1.12 | 0.00 | 8.15 | 22.68 | 38.84 | 54.41 | 68.34 |
| 1 | 5 | COC | 9.08 | 1.55 | 0.00 | 7.08 | 20.35 | 35.17 | 49.42 | 62.21 |
| 1 | 5 | COC | 9.08 | 1.55 | 0.00 | 7.08 | 20.35 | 35.17 | 49.42 | 62.21 |
| 1 | 5 | COC | 8.31 | 1.12 | 0.00 | 8.15 | 22.68 | 38.84 | 54.41 | 68.34 |
| 1 | 5 | COC | 9.08 | 1.55 | 0.00 | 7.08 | 20.35 | 35.17 | 49.42 | 62.21 |
| 1 | 5 | COC | 9.08 | 1.55 | 0.00 | 7.08 | 20.35 | 35.17 | 49.42 | 62.21 |
| 1 | 5 | CS(=O)C | 8.38 | 1.14 | 0.00 | 8.22 | 22.93 | 39.33 | 55.20 | 69.49 |
| 1 | 5 | CS(=O)C | 8.62 | 1.28 | 0.00 | 7.83 | 22.01 | 37.81 | 53.01 | 66.61 |
| 1 | 5 | CS(=O)C | 8.79 | 1.36 | 0.00 | 7.76 | 22.01 | 37.99 | 53.44 | 67.33 |
| 1 | 5 | CS(=O)C | 8.79 | 1.36 | 0.00 | 7.76 | 22.01 | 37.99 | 53.44 | 67.33 |
| 1 | 5 | CS(=O)C | 8.62 | 1.28 | 0.00 | 7.83 | 22.01 | 37.81 | 53.01 | 66.61 |
| 1 | 5 | CS(=O)C | 8.38 | 1.14 | 0.00 | 8.22 | 22.93 | 39.33 | 55.20 | 69.49 |
| 1 | 5 | CBr | 7.64 | 0.82 | 0.00 | 8.60 | 23.41 | 39.73 | 55.30 | 69.11 |
| 1 | 5 | CBr | 7.63 | 0.82 | 0.00 | 8.60 | 23.41 | 39.73 | 55.30 | 69.11 |
| 1 | 5 | CBr | 7.63 | 0.82 | 0.00 | 8.60 | 23.42 | 39.73 | 55.30 | 69.11 |
| 1 | 5 | CC(=O)OC(=O)C | 7.79 | 0.89 | 0.00 | 8.56 | 23.53 | 40.21 | 56.34 | 70.84 |
| 1 | 5 | CC(=O)OC(=O)C | 8.22 | 1.12 | 0.00 | 8.05 | 22.47 | 38.58 | 54.14 | 68.11 |
| 1 | 5 | CC(=O)OC(=O)C | 8.30 | 1.16 | 0.00 | 7.98 | 22.32 | 38.36 | 53.87 | 67.80 |
| 1 | 5 | CC(=O)OC(=O)C | 8.28 | 1.15 | 0.00 | 7.99 | 22.35 | 38.41 | 53.93 | 67.86 |
| 1 | 5 | CC(=O)OC(=O)C | 7.77 | 0.88 | 0.00 | 8.57 | 23.55 | 40.23 | 56.36 | 70.87 |
| 1 | 5 | CC(=O)OC(=O)C | 8.24 | 1.13 | 0.00 | 8.03 | 22.43 | 38.51 | 54.05 | 68.01 |
| 1 | 5 | CC=C | 8.13 | 1.09 | 0.00 | 8.00 | 22.27 | 38.17 | 53.52 | 67.34 |
| 1 | 5 | CC=C | 8.21 | 1.15 | 0.00 | 7.82 | 21.81 | 37.36 | 52.29 | 65.62 |
| 1 | 5 | CC=C | 8.21 | 1.15 | 0.00 | 7.82 | 21.81 | 37.36 | 52.29 | 65.62 |
| 1 | 5 | CC#C | 8.05 | 1.05 | 0.00 | 8.10 | 22.43 | 38.30 | 53.49 | 66.97 |
| 1 | 5 | CC#C | 8.06 | 1.05 | 0.00 | 8.10 | 22.42 | 38.29 | 53.48 | 66.96 |
| 1 | 5 | CC#C | 8.05 | 1.05 | 0.00 | 8.11 | 22.44 | 38.31 | 53.49 | 66.98 |
| 1 | 5 | CC | 7.98 | 1.04 | 0.00 | 8.00 | 22.17 | 37.92 | 53.09 | 66.72 |
| 1 | 5 | CC | 7.98 | 1.04 | 0.00 | 8.00 | 22.17 | 37.92 | 53.09 | 66.72 |
| 1 | 5 | CC | 7.98 | 1.04 | 0.00 | 8.00 | 22.17 | 37.92 | 53.09 | 66.72 |
| 1 | 5 | CC | 7.98 | 1.04 | 0.00 | 8.00 | 22.17 | 37.92 | 53.09 | 66.72 |
| 1 | 5 | CC | 7.98 | 1.04 | 0.00 | 8.00 | 22.17 | 37.92 | 53.09 | 66.72 |
| 1 | 5 | CC | 7.98 | 1.04 | 0.00 | 8.00 | 22.17 | 37.92 | 53.09 | 66.72 |

| | | | | | | | | | | | | |
|---|---|---|---|---|---|---|---|---|---|---|---|---|
| 1 | 5 | CC=O | 8.06 | 1.06 | 0.00 | 8.10 | 22.50 | 38.58 | 54.14 | 68.17 |
| 1 | 5 | CC=O | 8.10 | 1.07 | 0.00 | 8.09 | 22.50 | 38.57 | 54.08 | 67.99 |
| 1 | 5 | CC=O | 8.10 | 1.07 | 0.00 | 8.09 | 22.50 | 38.57 | 54.08 | 67.99 |
| 1 | 5 | CCC | 8.17 | 1.14 | 0.00 | 7.83 | 21.85 | 37.49 | 52.58 | 66.16 |
| 1 | 5 | CCC | 8.05 | 1.08 | 0.00 | 7.93 | 22.04 | 37.74 | 52.89 | 66.49 |
| 1 | 5 | CCC | 8.17 | 1.13 | 0.00 | 7.83 | 21.86 | 37.49 | 52.58 | 66.16 |
| 1 | 5 | CCC | 8.17 | 1.13 | 0.00 | 7.83 | 21.85 | 37.49 | 52.58 | 66.16 |
| 1 | 5 | CCC | 7.89 | 1.04 | 0.00 | 7.80 | 21.60 | 36.89 | 51.58 | 64.76 |
| 1 | 5 | CCC | 7.89 | 1.04 | 0.00 | 7.80 | 21.60 | 36.89 | 51.58 | 64.76 |
| 1 | 5 | CCC | 8.17 | 1.13 | 0.00 | 7.83 | 21.85 | 37.49 | 52.58 | 66.16 |
| 1 | 5 | CCC | 8.07 | 1.08 | 0.00 | 7.93 | 22.04 | 37.74 | 52.88 | 66.49 |
| 1 | 5 | CCI | 8.54 | 1.28 | 0.00 | 7.65 | 21.56 | 37.09 | 52.08 | 65.55 |
| 1 | 5 | CCI | 8.18 | 1.10 | 0.00 | 8.05 | 22.40 | 38.38 | 53.82 | 67.70 |
| 1 | 5 | CCI | 8.17 | 1.10 | 0.00 | 8.05 | 22.40 | 38.39 | 53.82 | 67.70 |
| 1 | 5 | CCI | 8.52 | 1.19 | 0.00 | 8.10 | 22.56 | 38.56 | 53.85 | 67.40 |
| 1 | 5 | CCI | 8.52 | 1.19 | 0.00 | 8.10 | 22.56 | 38.56 | 53.85 | 67.41 |
| 1 | 5 | CCS | 8.25 | 1.18 | 0.00 | 7.75 | 21.72 | 37.28 | 52.30 | 65.78 |
| 1 | 5 | CCS | 8.02 | 1.06 | 0.00 | 7.99 | 22.17 | 37.94 | 53.16 | 66.84 |
| 1 | 5 | CCS | 7.92 | 1.01 | 0.00 | 8.11 | 22.44 | 38.39 | 53.77 | 67.58 |
| 1 | 5 | CCS | 7.79 | 0.97 | 0.00 | 8.07 | 22.17 | 37.62 | 52.25 | 65.14 |
| 1 | 5 | CCS | 7.82 | 0.96 | 0.00 | 8.22 | 22.65 | 38.62 | 53.92 | 67.57 |
| 1 | 5 | CI | 8.40 | 1.12 | 0.00 | 8.32 | 23.08 | 39.45 | 55.12 | 69.05 |
| 1 | 5 | CI | 8.40 | 1.12 | 0.00 | 8.32 | 23.08 | 39.45 | 55.12 | 69.04 |
| 1 | 5 | CI | 8.40 | 1.12 | 0.00 | 8.32 | 23.08 | 39.45 | 55.12 | 69.05 |
| 1 | 5 | CPC | 8.65 | 1.29 | 0.00 | 7.84 | 22.12 | 38.12 | 53.60 | 67.55 |
| 1 | 5 | CPC | 8.71 | 1.34 | 0.00 | 7.62 | 21.50 | 36.92 | 51.67 | 64.77 |
| 1 | 5 | CPC | 8.68 | 1.31 | 0.00 | 7.80 | 22.03 | 38.00 | 53.45 | 67.35 |
| 1 | 5 | CPC | 8.68 | 1.31 | 0.00 | 7.79 | 22.03 | 38.00 | 53.45 | 67.35 |
| 1 | 5 | CPC | 8.71 | 1.34 | 0.00 | 7.62 | 21.50 | 36.92 | 51.67 | 64.77 |
| 1 | 5 | CPC | 8.64 | 1.29 | 0.00 | 7.84 | 22.12 | 38.12 | 53.61 | 67.55 |
| 1 | 5 | CCl | 8.89 | 1.32 | 0.00 | 8.03 | 22.40 | 38.35 | 53.81 | 67.70 |
| 1 | 5 | CCl | 8.89 | 1.32 | 0.00 | 8.03 | 22.40 | 38.35 | 53.81 | 67.69 |
| 1 | 5 | CCl | 8.89 | 1.32 | 0.00 | 8.03 | 22.40 | 38.35 | 53.81 | 67.69 |
| 1 | 5 | ClCCl | 9.89 | 1.82 | 0.00 | 6.20 | 17.67 | 31.11 | 45.65 | 59.89 |
| 1 | 5 | ClCCl | 9.87 | 1.81 | 0.00 | 6.21 | 17.68 | 31.12 | 45.66 | 59.90 |
| 1 | 5 | CF | 9.28 | 1.52 | 0.00 | 7.65 | 21.87 | 37.84 | 53.25 | 67.06 |
| 1 | 5 | CF | 9.28 | 1.52 | 0.00 | 7.65 | 21.87 | 37.84 | 53.25 | 67.06 |
| 1 | 5 | CF | 9.28 | 1.52 | 0.00 | 7.65 | 21.87 | 37.84 | 53.25 | 67.06 |
| 1 | 5 | FCF | 10.64 | 2.06 | 0.00 | 7.08 | 21.07 | 36.90 | 52.21 | 65.92 |
| 1 | 5 | FCF | 10.64 | 2.06 | 0.00 | 7.08 | 21.07 | 36.90 | 52.21 | 65.92 |
| 1 | 5 | CS | 7.84 | 0.92 | 0.00 | 8.51 | 23.40 | 39.91 | 55.80 | 70.02 |
| 1 | 5 | CS | 7.78 | 0.91 | 0.00 | 8.35 | 22.83 | 38.71 | 53.76 | 66.99 |
| 1 | 5 | CS | 7.78 | 0.91 | 0.00 | 8.35 | 22.83 | 38.71 | 53.76 | 66.99 |
| 1 | 5 | CNO | 9.19 | 1.51 | 0.00 | 7.61 | 21.83 | 37.84 | 53.37 | 67.36 |
| 1 | 5 | CNO | 9.42 | 1.67 | 0.00 | 7.03 | 20.36 | 35.27 | 49.59 | 62.38 |
| 1 | 5 | CNO | 8.83 | 1.33 | 0.00 | 7.94 | 22.42 | 38.59 | 54.19 | 68.15 |
| 1 | 5 | CONC | 8.85 | 1.44 | 0.00 | 7.27 | 20.68 | 35.60 | 49.90 | 62.66 |
| 1 | 5 | CONC | 9.25 | 1.54 | 0.00 | 7.55 | 21.70 | 37.64 | 53.10 | 67.03 |
| 1 | 5 | CONC | 9.31 | 1.61 | 0.00 | 7.16 | 20.62 | 35.65 | 50.09 | 62.99 |
| 1 | 5 | CONC | 8.90 | 1.36 | 0.00 | 7.86 | 22.25 | 38.34 | 53.85 | 67.76 |
| 1 | 5 | CONC | 8.53 | 1.23 | 0.00 | 7.97 | 22.36 | 38.44 | 53.97 | 67.90 |
| 1 | 5 | CONC | 8.80 | 1.41 | 0.00 | 7.34 | 20.82 | 35.79 | 50.15 | 62.95 |
| 1 | 5 | CS(=O)(=O)N | 10.34 | 1.92 | 0.00 | 7.39 | 21.82 | 38.27 | 54.37 | 68.94 |
| 1 | 5 | CS(=O)(=O)N | 10.51 | 2.02 | 0.00 | 7.16 | 21.31 | 37.45 | 53.19 | 67.43 |
| 1 | 5 | CS(=O)(=O)N | 10.51 | 2.01 | 0.00 | 7.16 | 21.32 | 37.46 | 53.20 | 67.44 |
| 1 | 5 | CNS(=O)(=O)C | 10.57 | 2.16 | 0.00 | 6.31 | 19.09 | 33.51 | 47.37 | 59.73 |
| 1 | 5 | CNS(=O)(=O)C | 9.65 | 1.63 | 0.00 | 7.76 | 22.39 | 38.87 | 54.83 | 69.12 |
| 1 | 5 | CNS(=O)(=O)C | 9.99 | 1.84 | 0.00 | 7.17 | 21.04 | 36.69 | 51.82 | 65.40 |
| 1 | 5 | CNS(=O)(=O)C | 10.35 | 1.93 | 0.00 | 7.37 | 21.76 | 38.15 | 54.18 | 68.69 |
| 1 | 5 | CNS(=O)(=O)C | 10.60 | 2.06 | 0.00 | 7.04 | 21.07 | 37.07 | 52.71 | 66.86 |
| 1 | 5 | CNS(=O)(=O)C | 10.47 | 2.00 | 0.00 | 7.17 | 21.33 | 37.45 | 53.19 | 67.43 |
| 1 | 5 | C[N]N([O])C | 9.07 | 1.50 | 0.00 | 7.36 | 21.00 | 36.19 | 50.70 | 63.59 |
| 1 | 5 | C[N]N([O])C | 9.75 | 1.64 | 0.00 | 7.89 | 22.77 | 39.60 | 55.93 | 70.60 |
| 1 | 5 | C[N]N([O])C | 9.05 | 1.39 | 0.00 | 8.02 | 22.75 | 39.30 | 55.33 | 69.74 |
| 1 | 5 | C[N]N([O])C | 9.05 | 1.39 | 0.00 | 8.02 | 22.75 | 39.31 | 55.34 | 69.74 |
| 1 | 5 | C[N]N([O])C | 8.56 | 1.21 | 0.00 | 8.13 | 22.74 | 39.04 | 54.73 | 68.77 |
| 1 | 5 | C[N]N([O])C | 9.07 | 1.50 | 0.00 | 7.36 | 21.00 | 36.18 | 50.70 | 63.58 |
| 1 | 5 | CCOC=C | 8.13 | 1.09 | 0.00 | 8.03 | 22.34 | 38.31 | 53.74 | 67.62 |
| 1 | 5 | CCOC=C | 8.16 | 1.10 | 0.00 | 8.01 | 22.34 | 38.36 | 53.87 | 67.85 |
| 1 | 5 | CCOC=C | 8.10 | 1.08 | 0.00 | 8.03 | 22.35 | 38.31 | 53.74 | 67.61 |
| 1 | 5 | CCOC=C | 8.62 | 1.35 | 0.00 | 7.37 | 20.82 | 35.78 | 50.13 | 62.97 |
| 1 | 5 | CCOC=C | 8.61 | 1.35 | 0.00 | 7.37 | 20.83 | 35.79 | 50.14 | 62.98 |
| 1 | 5 | C1CCCN1 | 8.29 | 1.22 | 0.00 | 7.45 | 20.83 | 35.58 | 49.61 | 62.08 |
| 1 | 5 | C1CCCN1 | 8.11 | 1.14 | 0.00 | 7.58 | 21.04 | 35.84 | 49.93 | 62.44 |
| 1 | 5 | C1CCCN1 | 8.15 | 1.13 | 0.00 | 7.73 | 21.41 | 36.42 | 50.66 | 63.25 |
| 1 | 5 | C1CCCN1 | 7.93 | 1.05 | 0.00 | 7.83 | 21.70 | 37.07 | 51.83 | 65.06 |
| 1 | 5 | C1CCCN1 | 8.01 | 1.05 | 0.00 | 8.02 | 22.19 | 37.90 | 52.99 | 66.51 |
| 1 | 5 | C1CCCN1 | 8.08 | 1.12 | 0.00 | 7.70 | 21.44 | 36.70 | 51.39 | 64.58 |
| 1 | 5 | C1CCCN1 | 8.19 | 1.13 | 0.00 | 7.85 | 21.88 | 37.48 | 52.51 | 66.01 |
| 1 | 5 | C1CCCN1 | 8.07 | 1.07 | 0.00 | 7.97 | 22.08 | 37.71 | 52.70 | 66.11 |
| 1 | 5 | C1CCCNC1 | 9.17 | 1.70 | 0.00 | 6.32 | 18.45 | 31.94 | 44.85 | 56.41 |
| 1 | 5 | C1CCCNC1 | 7.97 | 1.06 | 0.00 | 7.85 | 21.72 | 37.05 | 51.73 | 64.83 |
| 1 | 5 | C1CCCNC1 | 9.17 | 1.70 | 0.00 | 6.32 | 18.44 | 31.94 | 44.85 | 56.41 |
| 1 | 5 | C1CCCNC1 | 7.97 | 1.06 | 0.00 | 7.85 | 21.72 | 37.05 | 51.73 | 64.83 |
| 1 | 5 | C1CCCNC1 | 7.99 | 1.08 | 0.00 | 7.81 | 21.68 | 37.09 | 51.93 | 65.26 |
| 1 | 5 | C1CCCNC1 | 8.13 | 1.15 | 0.00 | 7.67 | 21.42 | 36.73 | 51.49 | 64.77 |
| 1 | 5 | C1CCCNC1 | 7.95 | 1.06 | 0.00 | 7.79 | 21.60 | 36.89 | 51.58 | 64.75 |
| 1 | 5 | C1CCCNC1 | 8.27 | 1.23 | 0.00 | 7.45 | 20.93 | 35.96 | 50.46 | 63.52 |
| 1 | 5 | C1CCCNC1 | 8.00 | 1.08 | 0.00 | 7.80 | 21.68 | 37.09 | 51.93 | 65.26 |
| 1 | 5 | C1CCCNC1 | 8.13 | 1.15 | 0.00 | 7.67 | 21.42 | 36.73 | 51.50 | 64.78 |
| 1 | 5 | CC1=CCC1 | 8.18 | 1.12 | 0.00 | 7.92 | 22.10 | 37.89 | 53.14 | 66.84 |
| 1 | 5 | CC1=CCC1 | 8.27 | 1.17 | 0.00 | 7.80 | 21.81 | 37.37 | 52.30 | 65.63 |
| 1 | 5 | CC1=CCC1 | 8.28 | 1.18 | 0.00 | 7.79 | 21.78 | 37.33 | 52.25 | 65.56 |
| 1 | 5 | CC1CCC1 | 8.29 | 1.20 | 0.00 | 7.71 | 21.61 | 37.12 | 52.08 | 65.53 |
| 1 | 5 | CC1CCC1 | 8.21 | 1.15 | 0.00 | 7.83 | 21.86 | 37.51 | 52.61 | 66.20 |
| 1 | 5 | CC1CCC1 | 8.30 | 1.20 | 0.00 | 7.71 | 21.61 | 37.11 | 52.08 | 65.53 |
| 1 | 5 | O=C1CCC(=O)CC1 | 7.70 | 0.91 | 0.00 | 8.24 | 22.64 | 38.60 | 53.97 | 67.75 |
| 1 | 5 | O=C1CCC(=O)CC1 | 8.29 | 1.22 | 0.00 | 7.55 | 21.24 | 36.51 | 51.24 | 64.47 |
| 1 | 5 | O=C1CCC(=O)CC1 | 7.69 | 0.91 | 0.00 | 8.24 | 22.64 | 38.60 | 53.97 | 67.75 |
| 1 | 5 | O=C1CCC(=O)CC1 | 8.28 | 1.22 | 0.00 | 7.55 | 21.24 | 36.51 | 51.24 | 64.47 |
| 1 | 5 | O=C1CCC(=O)CC1 | 8.30 | 1.23 | 0.00 | 7.55 | 21.23 | 36.51 | 51.24 | 64.47 |
| 1 | 5 | O=C1CCC(=O)CC1 | 7.69 | 0.91 | 0.00 | 8.24 | 22.63 | 38.59 | 53.95 | 67.73 |
| 1 | 5 | O=C1CCC(=O)CC1 | 7.69 | 0.91 | 0.00 | 8.24 | 22.63 | 38.59 | 53.95 | 67.72 |
| 1 | 5 | O=C1CCC(=O)CC1 | 8.29 | 1.23 | 0.00 | 7.55 | 21.23 | 36.51 | 51.24 | 64.47 |
| 1 | 5 | OCC(O)(C)C | 8.52 | 1.29 | 0.00 | 7.61 | 21.49 | 37.04 | 52.11 | 65.71 |
| 1 | 5 | OCC(O)(C)C | 8.19 | 1.10 | 0.00 | 8.11 | 22.61 | 38.81 | 54.50 | 68.62 |
| 1 | 5 | OCC(O)(C)C | 8.93 | 1.48 | 0.00 | 7.18 | 20.51 | 35.37 | 49.62 | 62.38 |
| 1 | 5 | OCC(O)(C)C | 8.54 | 1.24 | 0.00 | 7.90 | 22.18 | 38.11 | 53.48 | 67.26 |
| 1 | 5 | OCC(O)(C)C | 8.49 | 1.27 | 0.00 | 7.69 | 21.71 | 37.43 | 52.66 | 66.40 |
| 1 | 5 | OCC(O)(C)C | 8.28 | 1.16 | 0.00 | 7.89 | 22.10 | 37.99 | 53.36 | 67.22 |
| 1 | 5 | OCC(O)(C)C | 8.43 | 1.24 | 0.00 | 7.71 | 21.68 | 37.31 | 52.44 | 66.08 |
| 1 | 5 | OCC(O)(C)C | 8.43 | 1.23 | 0.00 | 7.77 | 21.85 | 37.63 | 52.91 | 66.68 |
| 1 | 5 | CC=N | 7.57 | 0.87 | 0.00 | 8.26 | 22.64 | 38.59 | 53.97 | 67.77 |
| 1 | 5 | CC=N | 7.70 | 0.93 | 0.00 | 8.14 | 22.39 | 38.19 | 53.37 | 66.94 |
| 1 | 5 | CC=N | 7.70 | 0.93 | 0.00 | 8.14 | 22.39 | 38.19 | 53.37 | 66.94 |
| 1 | 5 | CN=C(C)C | 7.11 | 0.73 | 0.00 | 8.11 | 21.92 | 37.00 | 51.31 | 63.95 |
| 1 | 5 | CN=C(C)C | 7.67 | 0.91 | 0.00 | 8.21 | 22.59 | 38.56 | 53.97 | 67.84 |
| 1 | 5 | CN=C(C)C | 7.63 | 0.90 | 0.00 | 8.21 | 22.53 | 38.39 | 53.61 | 67.19 |
| 1 | 5 | CN=C(C)C | 7.63 | 0.90 | 0.00 | 8.21 | 22.53 | 38.39 | 53.61 | 67.19 |
| 1 | 5 | CN=C(C)C | 7.87 | 0.96 | 0.00 | 8.30 | 22.93 | 39.24 | 55.03 | 69.25 |
| 1 | 5 | CN=C(C)C | 7.85 | 1.01 | 0.00 | 7.97 | 22.05 | 37.67 | 52.70 | 66.15 |
| 1 | 5 | CN=C(C)C | 7.85 | 1.01 | 0.00 | 7.97 | 22.05 | 37.67 | 52.70 | 66.15 |
| 1 | 5 | CN=C(C)C | 6.73 | 0.51 | 0.00 | 8.68 | 23.18 | 39.00 | 54.05 | 67.38 |
| 1 | 5 | CN=C(C)C | 7.11 | 0.73 | 0.00 | 8.11 | 21.92 | 37.00 | 51.31 | 63.95 |
| 1 | 5 | C[GeH3] | 8.13 | 1.08 | 0.00 | 8.07 | 22.44 | 38.45 | 53.90 | 67.78 |
| 1 | 5 | C[GeH3] | 8.13 | 1.08 | 0.00 | 8.07 | 22.44 | 38.45 | 53.90 | 67.78 |
| 1 | 5 | C[GeH3] | 8.13 | 1.08 | 0.00 | 8.07 | 22.44 | 38.45 | 53.90 | 67.78 |
| 1 | 5 | C[Pb]=[Pb] | 8.34 | 1.19 | 0.00 | 7.77 | 21.69 | 37.09 | 51.74 | 64.62 |
| 1 | 5 | C[Pb]=[Pb] | 8.08 | 1.05 | 0.00 | 8.10 | 22.39 | 38.20 | 53.30 | 66.68 |
| 1 | 5 | C[Pb]=[Pb] | 8.12 | 1.07 | 0.00 | 8.08 | 22.37 | 38.18 | 53.28 | 66.67 |
| 1 | 5 | C[Te] | 7.82 | 1.09 | 0.00 | 8.24 | 22.57 | 38.27 | 53.14 | 66.17 |
| 1 | 5 | C[Te] | 7.82 | 0.94 | 0.00 | 8.03 | 22.25 | 37.98 | 52.95 | 66.06 |
| 1 | 5 | C[Te] | 7.82 | 0.94 | 0.00 | 8.03 | 22.25 | 37.98 | 52.95 | 66.06 |
| 1 | 5 | Cc1ccccc1 | 8.09 | 1.07 | 0.00 | 8.05 | 22.37 | 38.33 | 53.74 | 67.61 |
| 1 | 5 | Cc1ccccc1 | 8.16 | 1.12 | 0.00 | 7.88 | 21.93 | 37.54 | 52.53 | 65.90 |
| 1 | 5 | Cc1ccccc1 | 8.18 | 1.14 | 0.00 | 7.83 | 21.82 | 37.35 | 52.24 | 65.50 |
| 1 | 5 | BOC | 8.65 | 1.32 | 0.00 | 7.65 | 21.61 | 37.19 | 52.20 | 65.66 |
| 1 | 5 | BOC | 7.70 | 0.93 | 0.00 | 8.04 | 22.06 | 37.48 | 52.21 | 65.33 |
| 1 | 5 | BOC | 7.70 | 0.93 | 0.00 | 8.04 | 22.06 | 37.48 | 52.21 | 65.33 |
| 1 | 5 | COBC=C | 8.18 | 1.19 | 0.00 | 7.46 | 20.84 | 35.63 | 49.79 | 62.43 |
| 1 | 5 | COBC=C | 8.13 | 1.17 | 0.00 | 7.52 | 20.96 | 35.82 | 50.04 | 62.74 |
| 1 | 5 | COBC=C | 8.58 | 1.28 | 0.00 | 7.72 | 21.69 | 37.23 | 52.15 | 65.46 |
| 1 | 5 | COBC | 8.18 | 1.19 | 0.00 | 7.48 | 20.90 | 35.73 | 49.91 | 62.57 |
| 1 | 5 | COBC | 8.45 | 1.21 | 0.00 | 7.93 | 22.26 | 38.28 | 53.77 | 67.70 |
| 1 | 5 | COBC | 8.88 | 1.41 | 0.00 | 7.60 | 21.68 | 37.51 | 52.86 | 66.67 |
| 1 | 5 | COBC | 8.95 | 1.45 | 0.00 | 7.50 | 21.46 | 37.18 | 52.41 | 66.13 |
| 1 | 5 | COBC | 8.53 | 1.26 | 0.00 | 7.74 | 21.73 | 37.28 | 52.20 | 65.52 |
| 1 | 5 | COBC | 8.23 | 1.21 | 0.00 | 7.43 | 20.80 | 35.58 | 49.72 | 62.35 |
| 1 | 5 | CN([CH2])C | 7.67 | 0.85 | 0.00 | 8.57 | 23.50 | 40.11 | 56.15 | 70.58 |
| 1 | 5 | CN([CH2])C | 7.67 | 0.85 | 0.00 | 8.57 | 23.50 | 40.11 | 56.16 | 70.59 |
| 1 | 5 | CN([CH2])C | 7.58 | 0.79 | 0.00 | 8.73 | 23.84 | 40.65 | 56.89 | 71.51 |
| 1 | 5 | CN([CH2])C | 7.68 | 0.86 | 0.00 | 8.56 | 23.46 | 40.05 | 56.07 | 70.47 |
| 1 | 5 | CN([CH2])C | 7.68 | 0.86 | 0.00 | 8.56 | 23.46 | 40.05 | 56.07 | 70.47 |
| 1 | 5 | CN([CH2])C | 7.57 | 0.79 | 0.00 | 8.74 | 23.87 | 40.69 | 56.94 | 71.58 |
| 1 | 5 | ON=C(C)C | 8.04 | 1.06 | 0.00 | 8.03 | 22.27 | 38.08 | 53.25 | 66.78 |
| 1 | 5 | ON=C(C)C | 8.04 | 1.06 | 0.00 | 8.03 | 22.27 | 38.08 | 53.25 | 66.78 |
| 1 | 5 | ON=C(C)C | 7.90 | 0.92 | 0.00 | 8.59 | 23.67 | 40.49 | 56.76 | 71.37 |
| 1 | 5 | ON=C(C)C | 8.33 | 1.21 | 0.00 | 7.73 | 21.68 | 37.24 | 52.23 | 65.66 |
| 1 | 5 | ON=C(C)C | 8.33 | 1.21 | 0.00 | 7.73 | 21.68 | 37.24 | 52.23 | 65.66 |
| 1 | 5 | ON=C(C)C | 7.97 | 1.01 | 0.00 | 8.21 | 22.74 | 38.92 | 54.57 | 68.64 |
| 1 | 5 | CC(=O)O | 7.92 | 0.96 | 0.00 | 8.41 | 23.20 | 39.68 | 55.59 | 69.87 |
| 1 | 5 | CC(=O)O | 8.11 | 1.06 | 0.00 | 8.19 | 22.77 | 39.06 | 54.80 | 68.95 |
| 1 | 5 | CC(=O)O | 8.11 | 1.06 | 0.00 | 8.19 | 22.78 | 39.07 | 54.81 | 68.96 |
| 1 | 5 | CCOC=C | 7.98 | 1.03 | 0.00 | 8.10 | 22.47 | 38.52 | 54.05 | 68.05 |
| 1 | 5 | CCOC=C | 7.91 | 1.00 | 0.00 | 8.13 | 22.49 | 38.48 | 53.92 | 67.80 |
| 1 | 5 | CCOC=C | 7.91 | 1.00 | 0.00 | 8.13 | 22.49 | 38.49 | 53.93 | 67.82 |
| 1 | 5 | CCOC=C | 7.97 | 1.08 | 0.00 | 7.74 | 21.47 | 36.65 | 51.20 | 64.21 |
| 1 | 5 | CCOC=C | 7.97 | 1.08 | 0.00 | 7.74 | 21.47 | 36.65 | 51.19 | 64.20 |
| 1 | 5 | CC[PbH3] | 8.01 | 1.08 | 0.00 | 7.81 | 21.63 | 36.90 | 51.52 | 64.55 |
| 1 | 5 | CC[PbH3] | 8.15 | 1.14 | 0.00 | 7.69 | 21.40 | 36.57 | 51.07 | 63.96 |
| 1 | 5 | CC[PbH3] | 8.09 | 1.11 | 0.00 | 7.82 | 21.77 | 37.25 | 52.15 | 65.50 |
| 1 | 5 | CC[PbH3] | 8.10 | 1.11 | 0.00 | 7.82 | 21.77 | 37.25 | 52.15 | 65.50 |
| 1 | 5 | CC[PbH3] | 8.00 | 1.08 | 0.00 | 7.81 | 21.63 | 36.90 | 51.52 | 64.55 |
| 1 | 5 | CC[SeH] | 7.99 | 1.05 | 0.00 | 8.01 | 22.20 | 37.97 | 53.17 | 66.80 |
| 1 | 5 | CC[SeH] | 8.02 | 1.06 | 0.00 | 7.99 | 22.19 | 38.00 | 53.27 | 67.02 |
| 1 | 5 | CC[SeH] | 8.02 | 1.06 | 0.00 | 7.99 | 22.18 | 38.00 | 53.27 | 67.02 |
| 1 | 5 | CC[SeH] | 7.64 | 0.89 | 0.00 | 8.18 | 22.31 | 37.76 | 52.40 | 65.33 |
| 1 | 5 | CC[SeH] | 7.63 | 0.88 | 0.00 | 8.18 | 22.31 | 37.76 | 52.41 | 65.33 |
| 1 | 5 | CC[SnH3] | 8.01 | 1.09 | 0.00 | 7.77 | 21.57 | 36.83 | 51.46 | 64.51 |
| 1 | 5 | CC[SnH3] | 8.15 | 1.14 | 0.00 | 7.69 | 21.40 | 36.58 | 51.10 | 64.02 |
| 1 | 5 | CC[SnH3] | 8.12 | 1.12 | 0.00 | 7.79 | 21.70 | 37.16 | 52.04 | 65.38 |
| 1 | 5 | CC[SnH3] | 8.12 | 1.12 | 0.00 | 7.79 | 21.70 | 37.16 | 52.04 | 65.38 |
| 1 | 5 | CC[SnH3] | 8.01 | 1.08 | 0.00 | 7.78 | 21.57 | 36.84 | 51.46 | 64.51 |
| 1 | 5 | CC[TeH] | 7.96 | 1.04 | 0.00 | 7.99 | 22.10 | 37.76 | 52.81 | 66.28 |
| 1 | 5 | CC[TeH] | 8.03 | 1.06 | 0.00 | 7.96 | 22.10 | 37.83 | 53.01 | 66.67 |
| 1 | 5 | CC[TeH] | 8.02 | 1.06 | 0.00 | 7.96 | 22.10 | 37.84 | 53.02 | 66.68 |
| 1 | 5 | CC[TeH] | 7.67 | 0.90 | 0.00 | 8.18 | 22.34 | 37.87 | 52.63 | 65.67 |
| 1 | 5 | CC[TeH] | 7.65 | 0.89 | 0.00 | 8.18 | 22.35 | 37.88 | 52.63 | 65.67 |
| 1 | 5 | C/N=C/C=C | 8.12 | 1.10 | 0.00 | 7.94 | 22.05 | 37.66 | 52.60 | 65.87 |
| 1 | 5 | C/N=C/C=C | 8.13 | 1.10 | 0.00 | 7.93 | 22.04 | 37.65 | 52.58 | 65.84 |
| 1 | 5 | C/N=C/C=C | 8.93 | 1.53 | 0.00 | 6.95 | 19.95 | 34.43 | 48.35 | 60.84 |
| 1 | 5 | COC=C | 7.64 | 0.84 | 0.00 | 8.57 | 23.44 | 39.90 | 55.75 | 69.93 |
| 1 | 5 | COC=C | 8.06 | 1.09 | 0.00 | 7.82 | 21.72 | 37.10 | 51.85 | 65.01 |
| 1 | 5 | COC=C | 8.09 | 1.11 | 0.00 | 7.81 | 21.70 | 37.08 | 51.82 | 64.98 |
| 1 | 5 | C[PbH2]C | 8.26 | 1.14 | 0.00 | 7.97 | 22.20 | 38.03 | 53.24 | 66.85 |
| 1 | 5 | C[PbH2]C | 8.26 | 1.14 | 0.00 | 7.97 | 22.20 | 38.03 | 53.24 | 66.85 |
| 1 | 5 | C[PbH2]C | 8.23 | 1.13 | 0.00 | 7.97 | 22.20 | 38.03 | 53.25 | 66.86 |
| 1 | 5 | C[PbH2]C | 8.23 | 1.13 | 0.00 | 7.97 | 22.20 | 38.03 | 53.25 | 66.86 |
| 1 | 5 | C[PbH2]C | 8.23 | 1.13 | 0.00 | 7.97 | 22.20 | 38.03 | 53.25 | 66.86 |
| 1 | 5 | C[PbH2]C | 8.24 | 1.13 | 0.00 | 7.97 | 22.20 | 38.02 | 53.25 | 66.86 |
| 1 | 5 | [PbH2].[PbH3].[CH3] | 8.70 | 1.37 | 0.00 | 7.45 | 21.12 | 36.38 | 51.09 | 64.27 |
| 1 | 5 | [PbH2].[PbH3].[CH3] | 8.34 | 1.20 | 0.00 | 7.77 | 21.74 | 37.27 | 52.19 | 65.49 |
| 1 | 5 | [PbH2].[PbH3].[CH3] | 8.70 | 1.37 | 0.00 | 7.45 | 21.12 | 36.38 | 51.09 | 64.27 |
| 1 | 5 | C[Se]C | 7.98 | 1.00 | 0.00 | 8.27 | 22.85 | 39.07 | 54.70 | 68.72 |
| 1 | 5 | C[Se]C | 8.05 | 1.06 | 0.00 | 8.00 | 22.07 | 37.55 | 52.25 | 65.21 |
| 1 | 5 | C[Se]C | 8.05 | 1.06 | 0.00 | 8.00 | 22.07 | 37.55 | 52.25 | 65.21 |
| 1 | 5 | C[Se]C | 7.98 | 1.00 | 0.00 | 8.27 | 22.85 | 39.07 | 54.70 | 68.72 |
| 1 | 5 | C[Se]C | 8.05 | 1.06 | 0.00 | 8.00 | 22.07 | 37.55 | 52.25 | 65.21 |
| 1 | 5 | C[Se]C | 8.05 | 1.06 | 0.00 | 8.00 | 22.07 | 37.55 | 52.25 | 65.21 |
| 1 | 5 | C[SeH] | 7.87 | 0.94 | 0.00 | 8.39 | 23.10 | 39.43 | 55.16 | 69.27 |
| 1 | 5 | C[SeH] | 7.77 | 0.91 | 0.00 | 8.30 | 22.71 | 38.53 | 53.57 | 66.85 |
| 1 | 5 | C[SeH] | 7.77 | 0.91 | 0.00 | 8.30 | 22.71 | 38.53 | 53.57 | 66.84 |
| 1 | 5 | C[SnH2]C | 8.24 | 1.14 | 0.00 | 7.90 | 22.03 | 37.76 | 52.89 | 66.43 |
| 1 | 5 | C[SnH2]C | 8.22 | 1.14 | 0.00 | 7.91 | 22.04 | 37.76 | 52.89 | 66.43 |
| 1 | 5 | C[SnH2]C | 8.22 | 1.13 | 0.00 | 7.92 | 22.09 | 37.85 | 53.03 | 66.63 |
| 1 | 5 | C[SnH2]C | 8.22 | 1.13 | 0.00 | 7.92 | 22.09 | 37.85 | 53.03 | 66.63 |
| 1 | 5 | C[SnH2]C | 8.21 | 1.13 | 0.00 | 7.93 | 22.10 | 37.86 | 53.04 | 66.63 |
| 1 | 5 | C[SnH2]C | 8.22 | 1.13 | 0.00 | 7.92 | 22.09 | 37.85 | 53.03 | 66.63 |
| 1 | 5 | C[SnH3] | 8.17 | 1.11 | 0.00 | 7.98 | 22.21 | 38.03 | 53.27 | 66.90 |
| 1 | 5 | C[SnH3] | 8.15 | 1.10 | 0.00 | 7.99 | 22.21 | 38.04 | 53.27 | 66.91 |
| 1 | 5 | C[SnH3] | 8.19 | 1.12 | 0.00 | 7.97 | 22.20 | 38.02 | 53.26 | 66.90 |
| 1 | 5 | C[Te]C | 7.98 | 1.00 | 0.00 | 8.27 | 22.83 | 39.01 | 54.61 | 68.60 |
| 1 | 5 | C[Te]C | 7.96 | 1.01 | 0.00 | 8.12 | 22.36 | 38.03 | 52.93 | 66.08 |
| 1 | 5 | C[Te]C | 7.96 | 1.01 | 0.00 | 8.12 | 22.36 | 38.03 | 52.93 | 66.08 |
| 1 | 5 | C[Te]C | 7.98 | 1.00 | 0.00 | 8.27 | 22.83 | 39.01 | 54.61 | 68.60 |
| 1 | 5 | C[Te]C | 7.96 | 1.01 | 0.00 | 8.12 | 22.36 | 38.02 | 52.93 | 66.08 |
| 1 | 5 | C[Te]C | 7.96 | 1.01 | 0.00 | 8.12 | 22.36 | 38.02 | 52.93 | 66.08 |
| 1 | 5 | C[TeH] | 7.78 | 0.91 | 0.00 | 8.34 | 22.81 | 38.74 | 53.90 | 67.30 |
| 1 | 5 | C[TeH] | 7.78 | 0.91 | 0.00 | 8.34 | 22.81 | 38.74 | 53.90 | 67.31 |
| 1 | 5 | Cn1cccc1 | 7.78 | 0.90 | 0.00 | 8.49 | 23.32 | 39.81 | 55.73 | 70.03 |
| 1 | 5 | Cn1cccc1 | 7.93 | 1.00 | 0.00 | 8.13 | 22.42 | 38.24 | 53.41 | 66.92 |
| 1 | 5 | Cn1cccc1 | 7.93 | 1.00 | 0.00 | 8.13 | 22.42 | 38.25 | 53.42 | 66.94 |
| 1 | 5 | CC(=O)N | 7.57 | 0.80 | 0.00 | 8.66 | 23.66 | 40.33 | 56.42 | 70.87 |
| 1 | 5 | CC(=O)N | 8.21 | 1.14 | 0.00 | 7.91 | 22.10 | 37.95 | 53.24 | 66.98 |
| 1 | 5 | CC(=O)N | 8.21 | 1.14 | 0.00 | 7.91 | 22.10 | 37.95 | 53.24 | 66.98 |
| 1 | 5 | CCC(=O)N | 7.92 | 1.01 | 0.00 | 8.06 | 22.30 | 38.12 | 53.37 | 67.05 |
| 1 | 5 | CCC(=O)N | 8.15 | 1.13 | 0.00 | 7.84 | 21.87 | 37.50 | 52.59 | 66.15 |
| 1 | 5 | CCC(=O)N | 7.45 | 0.80 | 0.00 | 8.42 | 22.97 | 39.07 | 54.55 | 68.42 |
| 1 | 5 | CCC(=O)N | 8.16 | 1.16 | 0.00 | 7.64 | 21.36 | 36.62 | 51.32 | 64.50 |
| 1 | 5 | CCC(=O)N | 7.98 | 1.04 | 0.00 | 8.03 | 22.29 | 38.17 | 53.52 | 67.33 |
| 1 | 5 | CS(=O)(=O)C | 8.59 | 1.25 | 0.00 | 8.01 | 22.51 | 38.74 | 54.44 | 68.58 |
| 1 | 5 | CS(=O)(=O)C | 8.44 | 1.15 | 0.00 | 8.30 | 23.15 | 39.76 | 55.86 | 70.43 |
| 1 | 5 | CS(=O)(=O)C | 8.59 | 1.25 | 0.00 | 8.01 | 22.51 | 38.74 | 54.44 | 68.58 |
| 1 | 5 | CS(=O)(=O)C | 8.58 | 1.24 | 0.00 | 8.02 | 22.52 | 38.74 | 54.44 | 68.58 |
| 1 | 5 | CS(=O)(=O)C | 8.44 | 1.15 | 0.00 | 8.30 | 23.15 | 39.76 | 55.86 | 70.43 |
| 1 | 5 | CS(=O)(=O)C | 8.58 | 1.24 | 0.00 | 8.02 | 22.52 | 38.74 | 54.44 | 68.59 |
| 1 | 5 | CO[SiH2]OC | 8.33 | 1.15 | 0.00 | 8.01 | 22.29 | 38.14 | 53.36 | 66.95 |
| 1 | 5 | CO[SiH2]OC | 8.72 | 1.37 | 0.00 | 7.44 | 21.06 | 36.23 | 50.82 | 63.88 |
| 1 | 5 | CO[SiH2]OC | 8.66 | 1.34 | 0.00 | 7.50 | 21.19 | 36.42 | 51.06 | 64.16 |
| 1 | 5 | CO[SiH2]OC | 8.31 | 1.14 | 0.00 | 8.02 | 22.31 | 38.17 | 53.40 | 66.99 |
| 1 | 5 | CO[SiH2]OC | 8.72 | 1.37 | 0.00 | 7.45 | 21.08 | 36.27 | 50.88 | 63.97 |
| 1 | 5 | CO[SiH2]OC | 8.66 | 1.34 | 0.00 | 7.51 | 21.22 | 36.48 | 51.14 | 64.26 |
| 1 | 5 | CN=C(C)C | 7.95 | 1.06 | 0.00 | 7.88 | 21.88 | 37.42 | 52.35 | 65.69 |
| 1 | 5 | CN=C(C)C | 6.71 | 0.50 | 0.00 | 8.69 | 23.20 | 39.04 | 54.11 | 67.46 |
| 1 | 5 | CN=C(C)C | 7.27 | 0.81 | 0.00 | 7.97 | 21.66 | 36.64 | 50.86 | 63.44 |
| 1 | 5 | CN=C(C)C | 7.26 | 0.81 | 0.00 | 7.97 | 21.67 | 36.65 | 50.87 | 63.46 |
| 1 | 5 | CN=C(C)C | 7.60 | 0.86 | 0.00 | 8.34 | 22.86 | 38.94 | 54.43 | 68.32 |
| 1 | 5 | CN=C(C)C | 7.67 | 0.92 | 0.00 | 8.16 | 22.44 | 38.24 | 53.39 | 66.91 |
| 1 | 5 | CN=C(C)C | 7.67 | 0.92 | 0.00 | 8.16 | 22.43 | 38.23 | 53.38 | 66.90 |
| 1 | 5 | CN=C(C)C | 7.51 | 0.82 | 0.00 | 8.43 | 23.00 | 39.13 | 54.63 | 68.50 |
| 1 | 5 | CN=C(C)C | 7.95 | 1.06 | 0.00 | 7.88 | 21.88 | 37.41 | 52.34 | 65.68 |
| 1 | 6 | COOC | 8.91 | 1.33 | 0.00 | 7.98 | 22.29 | 37.95 | 52.58 | 65.00 |
| 1 | 6 | COOC | 8.91 | 1.33 | 0.00 | 7.98 | 22.29 | 37.95 | 52.58 | 65.00 |
| 1 | 6 | OCC(O)(C)C | 9.15 | 1.39 | 0.00 | 8.18 | 23.04 | 39.57 | 55.40 | 69.43 |
| 1 | 6 | OCC(O)(C)C | 9.73 | 1.80 | 0.00 | 6.92 | 20.32 | 35.51 | 50.33 | 63.81 |
| 1 | 6 | CCO | 9.49 | 1.56 | 0.00 | 7.84 | 22.39 | 38.66 | 54.30 | 68.20 |
| 1 | 6 | CO | 9.71 | 1.60 | 0.00 | 8.05 | 23.02 | 39.75 | 55.75 | 69.83 |
| 1 | 6 | COC | 9.86 | 1.59 | 0.00 | 8.22 | 23.26 | 39.81 | 55.37 | 68.78 |
| 1 | 6 | COC | 9.86 | 1.59 | 0.00 | 8.22 | 23.25 | 39.81 | 55.37 | 68.78 |
| 1 | 6 | CCOC=C | 8.95 | 1.36 | 0.00 | 7.85 | 22.00 | 37.63 | 52.53 | 65.73 |
| 1 | 6 | OCC(O)(C)C | 9.15 | 1.39 | 0.00 | 8.18 | 23.04 | 39.57 | 55.40 | 69.43 |
| 1 | 6 | OCC(O)(C)C | 9.73 | 1.80 | 0.00 | 6.92 | 20.32 | 35.51 | 50.33 | 63.81 |
| 1 | 6 | BOC | 16.20 | 4.34 | 0.00 | 4.53 | 17.57 | 33.26 | 48.72 | 62.54 |
| 1 | 6 | COBC=C | 14.11 | 3.46 | 0.00 | 5.66 | 19.38 | 35.60 | 51.68 | 66.37 |
| 1 | 6 | COBC | 14.15 | 3.47 | 0.00 | 5.67 | 19.40 | 35.65 | 51.76 | 66.46 |
| 1 | 6 | CO[SiH2]OC | 11.66 | 2.41 | 0.00 | 7.07 | 21.64 | 38.36 | 54.64 | 69.27 |
| 1 | 6 | CO[SiH2]OC | 11.68 | 2.43 | 0.00 | 7.05 | 21.61 | 38.32 | 54.58 | 69.18 |
| 1 | 8 | CNC=C | 11.01 | 2.29 | 0.00 | 6.87 | 21.07 | 37.25 | 52.77 | 66.21 |
| 1 | 8 | CCN | 9.83 | 1.82 | 0.00 | 7.32 | 21.62 | 37.81 | 53.43 | 67.25 |
| 1 | 8 | CN | 10.49 | 2.07 | 0.00 | 7.18 | 21.63 | 38.11 | 54.01 | 67.99 |
| 1 | 8 | CNC | 10.01 | 1.86 | 0.00 | 7.41 | 21.84 | 38.07 | 53.55 | 67.03 |
| 1 | 8 | CNC | 10.01 | 1.86 | 0.00 | 7.41 | 21.84 | 38.07 | 53.55 | 67.03 |
| 1 | 8 | C1CCCN1 | 12.19 | 2.67 | 0.00 | 7.95 | 25.77 | 47.67 | 70.38 | 92.27 |
| 1 | 8 | C1CCCNC1 | 12.79 | 2.65 | 0.00 | 9.38 | 30.14 | 56.27 | 84.07 | 111.08 |
| 1 | 8 | C1CCCNC1 | 12.35 | 2.76 | 0.00 | 7.50 | 24.46 | 45.12 | 66.30 | 86.43 |
| 1 | 9 | CNC=O | 10.96 | 2.13 | 0.00 | 7.60 | 22.82 | 40.29 | 57.44 | 73.04 |
| 1 | 9 | CNC(=O)C | 9.85 | 1.71 | 0.00 | 7.99 | 23.25 | 40.62 | 57.59 | 72.94 |
| 1 | 9 | O=CN(C)C | 10.67 | 2.05 | 0.00 | 7.54 | 22.53 | 39.73 | 56.58 | 71.87 |
| 1 | 9 | O=CN(C)C | 10.33 | 1.90 | 0.00 | 7.73 | 22.84 | 40.13 | 57.09 | 72.53 |
| 1 | 11 | CC(F)C | 11.70 | 2.42 | 0.00 | 6.82 | 20.82 | 36.93 | 52.82 | 67.49 |
| 1 | 11 | FC(F)F | 13.74 | 2.53 | 0.00 | 9.14 | 26.53 | 45.92 | 64.64 | 81.52 |
| 1 | 11 | FC(F)F | 7.82 | 0.28 | 0.00 | 11.19 | 29.03 | 48.22 | 66.51 | 82.90 |
| 1 | 11 | FC(F)F | 2.91 | -1.57 | 0.00 | 12.85 | 30.99 | 49.95 | 67.81 | 83.72 |
| 1 | 11 | FC(C(F)(F)F)(F)F | 23.16 | 6.20 | 0.00 | 5.47 | 21.67 | 40.95 | 59.99 | 77.30 |
| 1 | 11 | FC(C(F)(F)F)(F)F | 15.81 | 3.39 | 0.00 | 8.12 | 25.03 | 44.22 | 62.87 | 79.70 |
| 1 | 11 | FC(C(F)(F)F)(F)F | 9.67 | 1.05 | 0.00 | 10.30 | 27.74 | 46.80 | 65.06 | 81.41 |
| 1 | 11 | FC(C(F)(F)F)(F)F | 12.10 | 1.82 | 0.00 | 10.14 | 28.18 | 47.96 | 66.86 | 83.68 |
| 1 | 11 | FC(C(F)(F)F)(F)F | 6.40 | -0.34 | 0.00 | 12.10 | 30.54 | 50.12 | 68.56 | 84.87 |
| 1 | 11 | FC(C(F)(F)F)(F)F | 1.65 | -2.13 | 0.00 | 13.68 | 32.40 | 51.73 | 69.73 | 85.53 |
| 1 | 11 | FCCCF | 11.41 | 2.22 | 0.00 | 7.41 | 22.12 | 38.90 | 55.37 | 70.47 |
| 1 | 11 | FCCCF | 11.43 | 2.23 | 0.00 | 7.40 | 22.10 | 38.88 | 55.35 | 70.45 |
| 1 | 11 | FCC=O | 10.34 | 1.67 | 0.00 | 8.56 | 24.39 | 42.16 | 59.40 | 74.99 |
| 1 | 11 | CF | 11.26 | 2.07 | 0.00 | 8.03 | 23.58 | 41.21 | 58.41 | 73.98 |
| 1 | 11 | FCF | 11.30 | 1.80 | 0.00 | 9.20 | 25.97 | 44.61 | 62.61 | 78.86 |
| 1 | 11 | FCF | 5.98 | -0.23 | 0.00 | 11.09 | 28.27 | 46.72 | 64.29 | 80.05 |
| 1 | 12 | ClC(Cl)Cl | 29.20 | 11.15 | 0.00 | -0.95 | 16.03 | 30.02 | 36.51 | 37.09 |
| 1 | 12 | ClC(Cl)Cl | 26.62 | 9.80 | 0.00 | 2.03 | 19.64 | 32.86 | 38.81 | 38.58 |
| 1 | 12 | ClC(Cl)Cl | 23.86 | 8.29 | 0.00 | 5.05 | 23.05 | 35.54 | 40.90 | 39.82 |
| 1 | 12 | ClCC=O | 15.80 | 6.54 | 0.00 | -4.08 | 2.20 | 16.74 | 34.43 | 47.68 |
| 1 | 12 | CCl | 16.05 | 6.47 | 0.00 | -1.27 | 9.44 | 25.93 | 42.56 | 53.92 |
| 1 | 12 | ClCCl | 17.19 | 6.36 | 0.00 | 1.86 | 15.59 | 28.73 | 38.64 | 43.91 |
| 1 | 12 | ClCCl | 15.88 | 5.58 | 0.00 | 3.74 | 17.93 | 30.97 | 40.66 | 45.18 |
| 1 | 13 | BrCC=O | 5.25 | 0.95 | 0.00 | 4.54 | 13.70 | 24.62 | 35.72 | 46.14 |
| 1 | 13 | CBr | 5.51 | 1.03 | 0.00 | 4.53 | 13.79 | 24.84 | 36.05 | 46.54 |
| 1 | 14 | CC(I)(C)C | 2.18 | 0.02 | 0.00 | 4.01 | 11.04 | 19.39 | 28.05 | 36.41 |
| 1 | 14 | CC(I)C | 2.27 | 0.01 | 0.00 | 4.28 | 11.60 | 20.26 | 29.20 | 37.80 |
| 1 | 14 | ICC=C | 2.96 | 0.28 | 0.00 | 4.02 | 11.23 | 19.79 | 28.61 | 36.98 |
| 1 | 14 | ICC=O | 2.27 | -0.08 | 0.00 | 4.82 | 12.93 | 22.42 | 32.12 | 41.24 |
| 1 | 14 | CCI | 2.43 | 0.03 | 0.00 | 4.48 | 12.16 | 21.22 | 30.55 | 39.42 |
| 1 | 14 | CI | 1.25 | -0.56 | 0.00 | 5.58 | 14.26 | 24.23 | 34.27 | 43.57 |
| 1 | 15 | CSSC | 6.69 | 1.35 | 0.00 | 4.31 | 13.38 | 24.24 | 34.89 | 44.42 |
| 1 | 15 | CSSC | 6.69 | 1.35 | 0.00 | 4.31 | 13.38 | 24.24 | 34.89 | 44.42 |
| 1 | 15 | CCSSC | 5.58 | 0.99 | 0.00 | 4.50 | 13.63 | 24.48 | 35.00 | 44.41 |
| 1 | 15 | CCSSC | 6.78 | 1.39 | 0.00 | 4.28 | 13.36 | 24.24 | 34.90 | 44.41 |
| 1 | 15 | CCS | 5.04 | 0.71 | 0.00 | 5.16 | 15.10 | 26.84 | 38.42 | 49.08 |
| 1 | 15 | CS | 6.27 | 1.14 | 0.00 | 4.89 | 14.75 | 26.55 | 38.36 | 49.28 |
| 1 | 16 | C[SH2] | 6.14 | 1.16 | 0.00 | 4.42 | 13.29 | 23.83 | 34.47 | 44.55 |
| 1 | 16 | C[SH]C | 5.32 | 0.75 | 0.00 | 5.29 | 14.94 | 26.08 | 37.18 | 47.59 |
| 1 | 16 | C[SH]C | 5.32 | 0.75 | 0.00 | 5.29 | 14.94 | 26.08 | 37.18 | 47.59 |
| 1 | 17 | CS(=O)C | 6.42 | 1.27 | 0.00 | 4.34 | 12.93 | 22.85 | 32.49 | 41.16 |
| 1 | 17 | CS(=O)C | 6.42 | 1.27 | 0.00 | 4.34 | 12.93 | 22.85 | 32.49 | 41.16 |
| 1 | 17 | CS=O | 6.20 | 1.20 | 0.00 | 4.28 | 12.81 | 22.72 | 32.35 | 40.99 |
| 1 | 17 | CC[S@@](=O)C | 6.32 | 1.34 | 0.00 | 3.70 | 11.74 | 21.43 | 30.92 | 39.37 |
| 1 | 17 | CC[S@@](=O)C | 6.26 | 1.22 | 0.00 | 4.33 | 12.87 | 22.72 | 32.28 | 40.88 |
| 1 | 17 | CS(=O)C | 6.42 | 1.27 | 0.00 | 4.34 | 12.93 | 22.85 | 32.49 | 41.16 |
| 1 | 17 | CS(=O)C | 6.42 | 1.27 | 0.00 | 4.34 | 12.93 | 22.85 | 32.49 | 41.16 |
| 1 | 18 | CS(=O)(=O)C | 7.82 | 1.79 | 0.00 | 4.27 | 13.74 | 25.13 | 36.57 | 47.17 |
| 1 | 18 | CS(=O)(=O)C | 7.82 | 1.79 | 0.00 | 4.27 | 13.74 | 25.13 | 36.57 | 47.17 |
| 1 | 18 | CCS(=O)(=O)C | 7.57 | 1.75 | 0.00 | 3.95 | 13.07 | 24.20 | 35.31 | 45.58 |
| 1 | 18 | CCS(=O)(=O)C | 7.77 | 1.79 | 0.00 | 4.19 | 13.55 | 24.78 | 36.04 | 46.47 |
| 1 | 18 | C=CS(=O)(=O)C | 7.82 | 1.81 | 0.00 | 4.16 | 13.53 | 24.83 | 36.19 | 46.72 |
| 1 | 18 | CS(=O)(=O)C | 7.80 | 1.78 | 0.00 | 4.28 | 13.76 | 25.15 | 36.58 | 47.18 |
| 1 | 18 | CS(=O)(=O)C | 7.80 | 1.78 | 0.00 | 4.28 | 13.76 | 25.15 | 36.58 | 47.18 |
| 1 | 18 | CS(=O)=O | 6.86 | 1.36 | 0.00 | 4.84 | 14.66 | 26.24 | 37.74 | 48.32 |
| 1 | 18 | CS(=O)(=O)C | 7.80 | 1.78 | 0.00 | 4.28 | 13.76 | 25.15 | 36.58 | 47.18 |
| 1 | 18 | CS(=O)(=O)C | 7.80 | 1.78 | 0.00 | 4.28 | 13.76 | 25.15 | 36.58 | 47.18 |
| 1 | 19 | C[SiH3] | 4.86 | 0.72 | 0.00 | 5.07 | 14.64 | 26.17 | 38.20 | 49.83 |
| 1 | 19 | C[SiH](C)C | 4.92 | 0.76 | 0.00 | 4.93 | 14.35 | 25.74 | 37.65 | 49.17 |
| 1 | 19 | C[SiH](C)C | 4.92 | 0.76 | 0.00 | 4.93 | 14.35 | 25.74 | 37.65 | 49.17 |
| 1 | 19 | C[SiH](C)C | 4.92 | 0.76 | 0.00 | 4.93 | 14.35 | 25.74 | 37.65 | 49.17 |
| 1 | 22 | CC1CC1 | 8.66 | 1.58 | 0.00 | 6.87 | 20.44 | 36.16 | 51.73 | 66.03 |
| 1 | 25 | CP(C)C | 6.13 | 1.34 | 0.00 | 4.32 | 13.67 | 24.68 | 35.86 | 46.48 |
| 1 | 25 | CP(C)C | 6.13 | 1.34 | 0.00 | 4.32 | 13.67 | 24.68 | 35.86 | 46.48 |
| 1 | 25 | CP(C)C | 6.13 | 1.34 | 0.00 | 4.32 | 13.67 | 24.68 | 35.86 | 46.48 |
| 1 | 25 | CPC | 6.05 | 1.31 | 0.00 | 4.29 | 13.55 | 24.49 | 35.62 | 46.14 |
| 1 | 25 | CPC | 6.05 | 1.31 | 0.00 | 4.29 | 13.55 | 24.49 | 35.62 | 46.14 |
| 1 | 26 | COBC | 5.94 | 0.88 | 0.00 | 6.01 | 17.41 | 30.79 | 44.36 | 57.18 |
| 1 | 31 | C[GeH3] | 4.65 | 0.83 | 0.00 | 4.21 | 12.81 | 23.28 | 34.16 | 44.65 |
| 1 | 32 | C[Sn] | 5.27 | 1.52 | 0.00 | 1.48 | 6.56 | 13.37 | 20.81 | 28.22 |
| 1 | 32 | CC[SnH3] | 2.76 | 0.29 | 0.00 | 3.81 | 11.00 | 19.70 | 28.78 | 37.59 |
| 1 | 32 | C[SnH2]C | 3.15 | 0.43 | 0.00 | 3.75 | 11.10 | 20.10 | 29.57 | 38.82 |
| 1 | 32 | C[SnH2]C | 3.15 | 0.43 | 0.00 | 3.75 | 11.11 | 20.10 | 29.57 | 38.82 |
| 1 | 32 | C[SnH3] | 2.98 | 0.35 | 0.00 | 3.92 | 11.43 | 20.57 | 30.15 | 39.48 |
| 1 | 33 | C[Pb] | 3.29 | 0.79 | 0.00 | 1.94 | 6.73 | 12.92 | 19.63 | 26.30 |
| 1 | 33 | C[Pb]=[Pb] | 2.85 | 0.59 | 0.00 | 2.16 | 6.84 | 12.51 | 18.19 | 23.22 |
| 1 | 33 | CC[PbH3] | -0.04 | -0.81 | 0.00 | 4.77 | 11.91 | 20.00 | 28.19 | 35.97 |
| 1 | 33 | C[PbH2]C | 0.15 | -0.76 | 0.00 | 4.85 | 12.25 | 20.70 | 29.31 | 37.55 |
| 1 | 33 | C[PbH2]C | 0.15 | -0.76 | 0.00 | 4.85 | 12.25 | 20.70 | 29.31 | 37.55 |
| 1 | 33 | [PbH2].[PbH3].[CH3] | 2.24 | 0.29 | 0.00 | 2.83 | 8.43 | 15.36 | 22.77 | 30.13 |
| 1 | 34 | CC[SeH] | 4.92 | 0.95 | 0.00 | 3.95 | 12.19 | 22.14 | 32.40 | 42.11 |
| 1 | 34 | C[Se]C | 5.27 | 1.05 | 0.00 | 4.08 | 12.74 | 23.27 | 34.10 | 44.27 |
| 1 | 34 | C[SeH] | 4.90 | 0.90 | 0.00 | 4.20 | 12.87 | 23.37 | 34.16 | 44.28 |
| 1 | 35 | C[Te] | 3.49 | 0.50 | 0.00 | 3.92 | 11.48 | 20.49 | 29.67 | 38.23 |
| 1 | 35 | CC[TeH] | 3.13 | 0.37 | 0.00 | 3.96 | 11.42 | 20.30 | 29.40 | 38.01 |
| 1 | 35 | C[Te]C | 3.19 | 0.33 | 0.00 | 4.27 | 12.22 | 21.65 | 31.28 | 40.34 |
| 1 | 35 | C[Te]C | 3.19 | 0.33 | 0.00 | 4.27 | 12.22 | 21.65 | 31.27 | 40.34 |
| 1 | 37 | C/N=C/C=C | 12.05 | 2.74 | 0.00 | 6.05 | 19.30 | 34.35 | 48.65 | 61.01 |
| 1 | 39 | C[NH3] | 9.24 | 1.93 | 0.00 | 5.64 | 17.36 | 30.95 | 44.43 | 56.95 |
| 1 | 39 | CN([CH2])C | 4.17 | -0.23 | 0.00 | 8.43 | 21.68 | 36.13 | 50.09 | 62.83 |
| 1 | 39 | CN([CH2])C | 3.11 | -0.65 | 0.00 | 8.85 | 22.23 | 36.67 | 50.56 | 63.21 |
| 1 | 40 | Cn1cccc1 | 3.43 | -0.90 | 0.00 | 10.84 | 27.20 | 44.82 | 61.65 | 76.75 |
| 1 | 41 | CCOC=C | 11.62 | 2.46 | 0.00 | 6.42 | 19.63 | 34.52 | 48.79 | 61.43 |
| 1 | 41 | COC=C | 10.91 | 2.05 | 0.00 | 7.53 | 22.08 | 38.22 | 53.39 | 66.37 |
| 1 | 43 | C[N]([O])(C)C | 7.49 | 1.19 | 0.00 | 6.14 | 16.84 | 27.83 | 37.41 | 45.18 |
| 1 | 43 | C[N]([O])(C)C | 7.49 | 1.19 | 0.00 | 6.14 | 16.84 | 27.83 | 37.41 | 45.18 |
| 1 | 43 | C[N]([O])(C)C | 7.49 | 1.20 | 0.00 | 6.13 | 16.83 | 27.83 | 37.41 | 45.17 |
| 1 | 43 | C[N]N([O])C | 6.12 | 0.64 | 0.00 | 6.86 | 18.29 | 30.38 | 41.41 | 50.70 |
| 1 | 43 | C[N]N([O])C | 6.12 | 0.64 | 0.00 | 6.86 | 18.29 | 30.38 | 41.41 | 50.70 |
| 1 | 43 | C[N]N([O])C | 6.14 | 0.65 | 0.00 | 6.85 | 18.29 | 30.38 | 41.41 | 50.71 |
| 1 | 46 | CN(=O)=O | 6.43 | 0.74 | 0.00 | 6.93 | 18.74 | 31.44 | 43.23 | 53.34 |
| 1 | 46 | CC(CN(=O)=O)C | 6.68 | 0.94 | 0.00 | 6.29 | 17.42 | 29.61 | 41.15 | 51.28 |
| 1 | 46 | CCN(=O)=O | 6.72 | 0.94 | 0.00 | 6.37 | 17.58 | 29.83 | 41.38 | 51.45 |
| 1 | 50 | Cc1ccccc1 | 9.64 | 1.95 | 0.00 | 6.55 | 20.15 | 36.15 | 52.16 | 66.95 |
| 1 | 56 | C[C@H]1CC(=O)O1 | 6.29 | 0.63 | 0.00 | 7.79 | 21.47 | 36.88 | 51.96 | 65.69 |
| 1 | 56 | CC1CCC1 | 7.36 | 1.14 | 0.00 | 6.94 | 19.96 | 34.86 | 49.56 | 63.04 |
| 1 | 57 | CC1=CCC1 | 7.33 | 0.98 | 0.00 | 7.69 | 21.67 | 37.57 | 53.20 | 67.49 |
| 1 | 72 | C/N=C/C | 13.62 | 3.47 | 0.00 | 4.91 | 17.33 | 31.78 | 45.61 | 57.60 |
| 1 | 72 | CN=C(C)C | 12.71 | 3.08 | 0.00 | 5.38 | 17.94 | 32.31 | 45.93 | 57.65 |
| 1 | 72 | CN=C(C)C | 12.19 | 2.85 | 0.00 | 5.69 | 18.44 | 32.92 | 46.64 | 58.46 |
| 1 | 75 | COC(=O)C | 9.03 | 1.45 | 0.00 | 7.65 | 21.79 | 37.65 | 53.01 | 66.82 |
| 1 | 75 | COC=O | 9.05 | 1.46 | 0.00 | 7.61 | 21.68 | 37.49 | 52.81 | 66.62 |
| 1 | 109 | C[N]N([O])C | 10.62 | 2.28 | 0.00 | 6.01 | 18.55 | 32.68 | 46.07 | 57.68 |
| 1 | 109 | C[N]N([O])C | 10.62 | 2.28 | 0.00 | 6.01 | 18.55 | 32.68 | 46.07 | 57.68 |
| 1 | 109 | C[N]N([O])C | 10.67 | 2.30 | 0.00 | 5.99 | 18.52 | 32.65 | 46.05 | 57.66 |
| 1 | 110 | CN([CH2])C | 8.69 | 1.56 | 0.00 | 6.64 | 19.40 | 33.94 | 48.24 | 61.40 |
| 1 | 110 | CN([CH2])C | 8.69 | 1.56 | 0.00 | 6.64 | 19.40 | 33.94 | 48.24 | 61.40 |
| 1 | 145 | CONC | 18.66 | 5.23 | 0.00 | 3.82 | 16.62 | 31.98 | 46.79 | 59.54 |
| 1 | 146 | CNO | 11.89 | 2.66 | 0.00 | 6.29 | 19.96 | 35.62 | 50.67 | 63.86 |
| 1 | 146 | CONC | 11.29 | 2.47 | 0.00 | 6.26 | 19.58 | 34.75 | 49.25 | 61.92 |
| 1 | 150 | CNN | 11.26 | 2.40 | 0.00 | 6.51 | 20.06 | 35.32 | 49.79 | 62.36 |
| 1 | 150 | CNNC | 12.64 | 3.03 | 0.00 | 5.62 | 18.72 | 33.87 | 48.41 | 61.02 |
| 1 | 150 | CNNC | 12.63 | 3.03 | 0.00 | 5.63 | 18.73 | 33.88 | 48.43 | 61.04 |
| 1 | 154 | CS(=O)(=O)N | 3.73 | -0.02 | 0.00 | 6.85 | 18.01 | 30.52 | 42.80 | 54.09 |
| 1 | 154 | CNS(=O)(=O)C | 2.71 | -0.39 | 0.00 | 7.07 | 18.29 | 30.71 | 42.80 | 53.80 |
| 1 | 154 | CS(=O)(=O)N | 3.71 | -0.03 | 0.00 | 6.86 | 18.03 | 30.54 | 42.80 | 54.07 |
| 1 | 154 | CNS(=O)(=O)C | 4.02 | 0.11 | 0.00 | 6.69 | 17.72 | 30.10 | 42.23 | 53.38 |
| 1 | 155 | CNS(=O)(=O)C | 15.82 | 4.29 | 0.00 | 4.47 | 17.58 | 33.55 | 49.55 | 64.19 |
| 1 | 155 | CNS(=O)(=O)C | 16.32 | 4.53 | 0.00 | 4.05 | 16.82 | 32.54 | 48.40 | 62.98 |
| 1 | 159 | COP(=O)OC | 10.89 | 2.16 | 0.00 | 7.01 | 21.11 | 37.32 | 53.27 | 67.82 |
| 1 | 159 | COP(=O)OC | 11.38 | 2.38 | 0.00 | 6.71 | 20.62 | 36.69 | 52.53 | 67.02 |
| 2 | 2 | [NH3]C=C | 18.37 | 3.47 | 0.00 | 12.97 | 38.30 | 66.77 | 93.89 | 117.01 |
| 2 | 2 | C=CC(=O)C=C | 21.09 | 4.79 | 0.00 | 10.59 | 33.89 | 60.67 | 86.30 | 107.49 |
| 2 | 2 | C=CC(=O)C=C | 21.06 | 4.78 | 0.00 | 10.60 | 33.91 | 60.69 | 86.32 | 107.51 |
| 2 | 2 | C=CC(=O)C=O | 21.27 | 4.88 | 0.00 | 10.40 | 33.53 | 60.15 | 85.64 | 106.77 |
| 2 | 2 | C=CC=C=O | 21.80 | 5.16 | 0.00 | 9.89 | 32.63 | 59.11 | 84.95 | 107.82 |
| 2 | 2 | C=CC=C=O | 40.77 | 14.40 | 0.00 | -6.74 | 2.16 | 17.56 | 34.86 | 51.77 |
| 2 | 2 | C=CN(=O)=O | 18.75 | 3.62 | 0.00 | 12.81 | 38.06 | 66.45 | 93.36 | 115.50 |
| 2 | 2 | Oc1ccccc1 | 28.98 | 8.04 | 0.00 | 9.41 | 38.90 | 76.84 | 114.49 | 146.18 |
| 2 | 2 | Oc1ccccc1 | 61.46 | 18.89 | 0.00 | 4.85 | 38.31 | 81.82 | 126.41 | 167.75 |
| 2 | 2 | Oc1ccccc1 | 39.20 | 12.69 | 0.00 | 1.81 | 24.83 | 57.87 | 94.33 | 130.30 |
| 2 | 2 | Oc1ccccc1 | 41.03 | 13.41 | 0.00 | 1.22 | 24.41 | 58.12 | 95.55 | 132.62 |
| 2 | 2 | Oc1ccccc1 | 27.80 | 7.48 | 0.00 | 10.38 | 40.75 | 79.53 | 117.90 | 150.07 |
| 2 | 2 | C=Cc1ccccc1 | 19.75 | 4.22 | 0.00 | 11.32 | 35.01 | 62.02 | 88.05 | 110.98 |
| 2 | 2 | NC=C | 19.87 | 4.27 | 0.00 | 11.34 | 35.27 | 62.80 | 89.76 | 114.43 |
| 2 | 2 | OC=C | 21.15 | 4.72 | 0.00 | 11.14 | 35.33 | 63.27 | 90.59 | 115.47 |
| 2 | 2 | [SiH3]C=C | 17.20 | 3.20 | 0.00 | 12.47 | 36.65 | 63.80 | 89.75 | 112.17 |
| 2 | 2 | C/C=C/C | 18.51 | 3.61 | 0.00 | 12.48 | 37.21 | 65.12 | 91.91 | 115.66 |
| 2 | 2 | CCC=C | 19.07 | 3.83 | 0.00 | 12.26 | 36.96 | 64.95 | 91.83 | 115.40 |
| 2 | 2 | [CH2]/C=C(\C)/[CH2] | 20.73 | 4.79 | 0.00 | 9.97 | 32.22 | 57.82 | 82.57 | 104.61 |
| 2 | 2 | [CH2]/C=C(\C)/[CH2] | 37.78 | 13.17 | 0.00 | -5.33 | 4.16 | 19.70 | 36.91 | 53.62 |
| 2 | 2 | [CH2]/C=C(\C)/[CH2] | 21.58 | 5.09 | 0.00 | 9.78 | 32.17 | 58.07 | 83.18 | 105.58 |
| 2 | 2 | ICC=C | 19.17 | 3.88 | 0.00 | 12.18 | 36.79 | 64.67 | 91.46 | 115.16 |
| 2 | 2 | CC(=C)C=O | 21.06 | 4.80 | 0.00 | 10.47 | 33.60 | 60.22 | 85.88 | 107.82 |
| 2 | 2 | C=CC=O | 21.61 | 4.99 | 0.00 | 10.43 | 33.76 | 60.66 | 86.55 | 108.47 |
| 2 | 2 | C/C=C/O | 20.57 | 4.49 | 0.00 | 11.33 | 35.45 | 63.13 | 90.01 | 114.14 |
| 2 | 2 | c1nccs1 | 57.01 | 18.74 | 0.00 | -1.55 | 21.54 | 53.72 | 87.39 | 119.19 |
| 2 | 2 | C1=CCC=CC1 | 38.86 | 8.00 | 0.00 | 23.96 | 72.90 | 128.51 | 182.08 | 229.74 |
| 2 | 2 | C=CS(=O)(=O)C=C | 19.88 | 4.12 | 0.00 | 12.05 | 36.69 | 64.56 | 91.20 | 113.92 |
| 2 | 2 | C=CS(=O)(=O)C=C | 19.91 | 4.13 | 0.00 | 12.04 | 36.68 | 64.55 | 91.18 | 113.90 |
| 2 | 2 | C=CS(=O)(=O)C | 19.99 | 4.16 | 0.00 | 12.01 | 36.66 | 64.59 | 91.31 | 114.12 |
| 2 | 2 | C=C[N]N(C=C)[O] | 23.10 | 5.78 | 0.00 | 8.34 | 28.61 | 51.46 | 72.43 | 87.60 |
| 2 | 2 | C=C[N]N(C=C)[O] | 31.50 | 10.04 | 0.00 | -0.10 | 11.88 | 25.33 | 21.59 | -5.12 |
| 2 | 2 | NC=C | 21.39 | 4.86 | 0.00 | 10.75 | 34.54 | 62.13 | 89.22 | 114.02 |
| 2 | 2 | [CH]C=C.[CH2].[C] | 22.04 | 5.23 | 0.00 | 9.80 | 32.34 | 58.36 | 83.46 | 105.35 |
| 2 | 2 | [CH2]/C=C\1/[CH]CC1 | 22.30 | 5.39 | 0.00 | 9.43 | 31.63 | 57.40 | 82.41 | 104.70 |
| 2 | 2 | C=CC=C | 21.82 | 5.18 | 0.00 | 9.70 | 32.06 | 57.93 | 83.00 | 105.32 |
| 2 | 2 | C=CC=C | 38.80 | 13.58 | 0.00 | -5.77 | 3.58 | 19.14 | 36.45 | 53.29 |
| 2 | 2 | C=CC=C | 21.82 | 5.18 | 0.00 | 9.71 | 32.07 | 57.94 | 83.00 | 105.33 |
| 2 | 2 | CC(=O)C=C | 21.58 | 4.99 | 0.00 | 10.37 | 33.59 | 60.34 | 86.08 | 107.74 |
| 2 | 2 | CNC=C | 19.75 | 4.23 | 0.00 | 11.32 | 35.12 | 62.45 | 89.05 | 113.06 |
| 2 | 2 | c1ccco1 | 79.12 | 28.22 | 0.00 | -12.68 | 6.65 | 39.16 | 74.89 | 108.84 |
| 2 | 2 | c1ccco1 | 62.01 | 20.58 | 0.00 | -1.10 | 26.53 | 65.22 | 105.83 | 143.70 |
| 2 | 2 | C=CNS(=O)(=O)N | 20.31 | 4.33 | 0.00 | 11.70 | 36.22 | 64.28 | 91.52 | 115.88 |
| 2 | 2 | IC=C | 19.86 | 4.14 | 0.00 | 11.94 | 36.45 | 64.19 | 90.78 | 114.22 |
| 2 | 2 | BrC=C | 19.60 | 3.99 | 0.00 | 12.28 | 37.12 | 65.19 | 92.06 | 115.57 |
| 2 | 2 | ClC=C | 17.02 | 2.78 | 0.00 | 13.48 | 37.57 | 64.05 | 89.82 | 112.68 |
| 2 | 2 | FC=C | 20.46 | 4.23 | 0.00 | 12.40 | 37.77 | 66.50 | 94.01 | 117.77 |
| 2 | 2 | [GeH3]C=C | 19.56 | 4.12 | 0.00 | 11.57 | 35.53 | 62.77 | 88.96 | 111.88 |
| 2 | 2 | c1ccc[nH]1 | 69.20 | 23.90 | 0.00 | -6.69 | 16.53 | 51.94 | 90.21 | 126.67 |
| 2 | 2 | c1ccc[nH]1 | 61.33 | 20.38 | 0.00 | -0.71 | 27.98 | 68.42 | 111.02 | 150.71 |
| 2 | 2 | c1ncc[nH]1 | 64.54 | 21.20 | 0.00 | -0.79 | 27.20 | 65.15 | 103.53 | 137.82 |
| 2 | 2 | C=CC(=O)C(=O)C=C | 21.13 | 4.82 | 0.00 | 10.51 | 33.74 | 60.45 | 86.03 | 107.18 |
| 2 | 2 | C=CC(=O)C(=O)C=C | 21.13 | 4.82 | 0.00 | 10.51 | 33.74 | 60.45 | 86.03 | 107.18 |
| 2 | 2 | c1cccnc1 | 44.09 | 14.57 | 0.00 | 0.19 | 22.87 | 55.83 | 91.96 | 127.13 |
| 2 | 2 | c1cccnc1 | 63.05 | 19.62 | 0.00 | 3.40 | 35.09 | 76.41 | 118.46 | 157.31 |
| 2 | 2 | [CH]/C(=C\C=C\[CH])/[N]N[O] | 12.87 | 2.03 | 0.00 | 5.09 | 28.57 | 51.44 | 72.70 | 92.16 |
| 2 | 2 | C=CC1CC1 | 19.26 | 3.93 | 0.00 | 12.03 | 36.48 | 64.17 | 90.75 | 114.13 |
| 2 | 2 | C=C | 19.50 | 3.99 | 0.00 | 12.20 | 37.04 | 65.24 | 92.37 | 116.38 |
| 2 | 2 | CC=C | 19.11 | 3.84 | 0.00 | 12.28 | 37.04 | 65.07 | 91.99 | 115.55 |
| 2 | 2 | CCOC=C | 19.49 | 4.08 | 0.00 | 11.66 | 35.76 | 63.31 | 90.03 | 114.02 |
| 2 | 2 | C=CON=N | 26.51 | 6.75 | 0.00 | 9.38 | 33.40 | 61.75 | 89.49 | 114.04 |
| 2 | 2 | COBC=C | 22.76 | 5.49 | 0.00 | 9.78 | 32.76 | 59.53 | 85.54 | 108.23 |
| 2 | 2 | O=COC(=O)C=C | 19.24 | 3.92 | 0.00 | 12.12 | 36.74 | 64.68 | 91.34 | 113.50 |
| 2 | 2 | PC=C | 20.70 | 4.56 | 0.00 | 11.13 | 34.94 | 62.21 | 88.59 | 112.07 |
| 2 | 2 | C=CPC=C | 20.72 | 4.58 | 0.00 | 11.02 | 34.68 | 61.76 | 87.95 | 111.18 |
| 2 | 2 | C=CPC=C | 20.71 | 4.58 | 0.00 | 11.02 | 34.68 | 61.77 | 87.96 | 111.18 |
| 2 | 2 | CCOC=C | 21.75 | 5.12 | 0.00 | 9.99 | 32.94 | 59.78 | 86.29 | 110.73 |
| 2 | 2 | C/N=C/C=C | 40.50 | 14.27 | 0.00 | -6.59 | 2.28 | 17.42 | 34.27 | 50.58 |
| 2 | 2 | C/N=C/C=C | 21.55 | 5.03 | 0.00 | 10.07 | 32.86 | 59.18 | 84.69 | 107.22 |
| 2 | 2 | COC=C | 22.13 | 5.19 | 0.00 | 10.28 | 33.75 | 61.10 | 88.00 | 112.62 |
| 2 | 2 | Cn1cccc1 | 52.35 | 16.38 | 0.00 | 4.23 | 34.56 | 74.68 | 116.18 | 154.96 |
| 2 | 2 | Cn1cccc1 | 33.57 | 10.56 | 0.00 | 2.29 | 21.39 | 47.33 | 75.42 | 104.11 |
| 2 | 2 | c1ccc[nH]1 | 51.55 | 16.01 | 0.00 | 4.84 | 35.66 | 76.15 | 117.93 | 156.90 |
| 2 | 2 | c1ccc[nH]1 | 33.51 | 10.57 | 0.00 | 2.15 | 21.09 | 46.94 | 74.98 | 103.67 |
| 2 | 2 | [NH3]n1cccc1 | 48.15 | 14.36 | 0.00 | 7.56 | 40.21 | 81.82 | 124.23 | 163.59 |
| 2 | 2 | [NH3]n1cccc1 | 33.93 | 10.62 | 0.00 | 2.61 | 22.39 | 49.25 | 78.40 | 108.23 |
| 2 | 2 | C=CN1C=C[CH]C=C1 | 59.51 | 17.76 | 0.00 | 7.35 | 43.02 | 88.30 | 134.21 | 176.61 |
| 2 | 2 | C=CN1C=C[CH]C=C1 | 42.71 | 13.85 | 0.00 | 1.78 | 26.33 | 61.48 | 100.12 | 138.04 |
| 2 | 2 | C=CN1C=C[CH]C=C1 | 31.81 | 9.19 | 0.00 | 8.36 | 37.95 | 76.72 | 115.64 | 148.99 |
| 2 | 2 | C=CN1C=C[CH]C=C1 | 20.03 | 4.26 | 0.00 | 11.59 | 35.76 | 63.25 | 89.70 | 113.07 |
| 2 | 2 | C=CC#C | 22.44 | 5.40 | 0.00 | 9.59 | 31.97 | 57.82 | 82.73 | 104.67 |
| 2 | 2 | C=CC1CCC1 | 20.11 | 4.30 | 0.00 | 11.57 | 35.81 | 63.47 | 90.14 | 113.65 |
| 2 | 3 | C=CC(=O)C=C | 39.11 | 14.01 | 0.00 | -7.65 | -0.97 | 11.51 | 25.62 | 39.40 |
| 2 | 3 | C=CC(=O)C=C | 39.10 | 14.01 | 0.00 | -7.64 | -0.96 | 11.52 | 25.63 | 39.41 |
| 2 | 3 | C=CC(=O)C=O | 43.74 | 15.88 | 0.00 | -9.67 | -3.64 | 8.97 | 23.63 | 38.15 |
| 2 | 3 | CC(=C)C=O | 36.80 | 12.89 | 0.00 | -5.75 | 2.29 | 15.67 | 30.31 | 44.33 |
| 2 | 3 | C=CC=O | 36.73 | 12.84 | 0.00 | -5.60 | 2.63 | 16.20 | 31.01 | 45.15 |
| 2 | 3 | CC(=O)C=C | 37.53 | 13.26 | 0.00 | -6.41 | 1.14 | 14.17 | 28.56 | 42.41 |
| 2 | 3 | C=CC(=O)C(=O)C=C | 42.90 | 15.53 | 0.00 | -9.16 | -2.74 | 10.20 | 25.15 | 39.90 |
| 2 | 3 | C=CC(=O)C(=O)C=C | 42.92 | 15.53 | 0.00 | -9.16 | -2.74 | 10.20 | 25.14 | 39.89 |
| 2 | 3 | O=COC(=O)C=C | 11.10 | 2.45 | 0.00 | 6.20 | 19.64 | 35.32 | 50.81 | 64.97 |
| 2 | 4 | C=CC#C | 43.46 | 15.27 | 0.00 | -6.76 | 3.27 | 20.23 | 39.15 | 57.61 |
| 2 | 5 | [NH3]C=C | 8.73 | 1.16 | 0.00 | 8.77 | 24.48 | 42.09 | 59.17 | 74.58 |
| 2 | 5 | [NH3]C=C | 9.27 | 1.46 | 0.00 | 8.06 | 22.99 | 39.83 | 56.21 | 71.02 |
| 2 | 5 | [NH3]C=C | 7.54 | 0.74 | 0.00 | 8.96 | 24.37 | 41.46 | 57.96 | 72.81 |
| 2 | 5 | CN([CH2])C | 9.08 | 1.33 | 0.00 | 8.44 | 23.83 | 41.14 | 57.96 | 73.15 |
| 2 | 5 | CN([CH2])C | 9.08 | 1.33 | 0.00 | 8.45 | 23.84 | 41.14 | 57.96 | 73.15 |
| 2 | 5 | C=CC(=O)C=C | 9.03 | 1.32 | 0.00 | 8.45 | 23.82 | 41.11 | 57.90 | 73.02 |
| 2 | 5 | C=CC(=O)C=C | 8.92 | 1.29 | 0.00 | 8.37 | 23.53 | 40.48 | 56.85 | 71.52 |
| 2 | 5 | C=CC(=O)C=C | 8.91 | 1.28 | 0.00 | 8.38 | 23.53 | 40.48 | 56.85 | 71.53 |
| 2 | 5 | C=CC(=O)C=C | 9.02 | 1.31 | 0.00 | 8.45 | 23.83 | 41.12 | 57.90 | 73.02 |
| 2 | 5 | C=CC(=O)C=C | 8.80 | 1.27 | 0.00 | 8.22 | 23.09 | 39.70 | 55.70 | 70.04 |
| 2 | 5 | C=CC(=O)C=C | 8.81 | 1.28 | 0.00 | 8.22 | 23.09 | 39.69 | 55.69 | 70.03 |
| 2 | 5 | C=CC(=O)C=O | 8.88 | 1.26 | 0.00 | 8.45 | 23.70 | 40.76 | 57.23 | 72.00 |
| 2 | 5 | C=CC(=O)C=O | 9.05 | 1.33 | 0.00 | 8.41 | 23.74 | 40.97 | 57.68 | 72.73 |
| 2 | 5 | C=CC(=O)C=O | 8.60 | 1.16 | 0.00 | 8.56 | 23.89 | 41.04 | 57.64 | 72.53 |
| 2 | 5 | C=CC=C=O | 8.68 | 1.17 | 0.00 | 8.57 | 23.89 | 40.98 | 57.46 | 72.22 |
| 2 | 5 | C=CC=C=O | 9.00 | 1.33 | 0.00 | 8.23 | 23.21 | 39.98 | 56.18 | 70.74 |
| 2 | 5 | C=CC=C=O | 8.80 | 1.30 | 0.00 | 8.08 | 22.76 | 39.17 | 55.02 | 69.26 |
| 2 | 5 | C=CC=C=O | 9.20 | 1.42 | 0.00 | 8.14 | 23.12 | 39.95 | 56.26 | 70.93 |
| 2 | 5 | C=CN(=O)=O | 8.87 | 1.24 | 0.00 | 8.57 | 24.00 | 41.27 | 57.98 | 72.98 |
| 2 | 5 | C=CN(=O)=O | 8.93 | 1.25 | 0.00 | 8.63 | 24.24 | 41.79 | 58.85 | 74.23 |
| 2 | 5 | C=CN(=O)=O | 8.59 | 1.08 | 0.00 | 8.95 | 24.79 | 42.46 | 59.52 | 74.81 |
| 2 | 5 | CC=N | 9.34 | 1.55 | 0.00 | 7.60 | 21.70 | 37.42 | 52.45 | 65.81 |
| 2 | 5 | C/N=C/C | 10.24 | 2.05 | 0.00 | 6.37 | 19.03 | 33.21 | 46.79 | 58.90 |
| 2 | 5 | Oc1ccccc1 | 8.69 | 1.22 | 0.00 | 8.29 | 23.18 | 39.76 | 55.75 | 70.10 |
| 2 | 5 | Oc1ccccc1 | 8.68 | 1.22 | 0.00 | 8.28 | 23.15 | 39.72 | 55.71 | 70.06 |
| 2 | 5 | Oc1ccccc1 | 8.77 | 1.26 | 0.00 | 8.28 | 23.24 | 39.96 | 56.13 | 70.68 |
| 2 | 5 | Oc1ccccc1 | 9.03 | 1.40 | 0.00 | 7.92 | 22.45 | 38.71 | 54.44 | 68.61 |
| 2 | 5 | Oc1ccccc1 | 8.67 | 1.21 | 0.00 | 8.32 | 23.25 | 39.88 | 55.92 | 70.30 |
| 2 | 5 | C=Cc1ccccc1 | 8.88 | 1.36 | 0.00 | 7.89 | 22.30 | 38.41 | 53.94 | 67.88 |
| 2 | 5 | C=Cc1ccccc1 | 9.38 | 1.49 | 0.00 | 8.06 | 22.97 | 39.72 | 55.91 | 70.43 |
| 2 | 5 | C=Cc1ccccc1 | 8.64 | 1.15 | 0.00 | 8.60 | 23.94 | 41.01 | 57.45 | 72.16 |
| 2 | 5 | NC=C | 8.34 | 1.00 | 0.00 | 8.91 | 24.56 | 41.94 | 58.68 | 73.66 |
| 2 | 5 | NC=C | 9.09 | 1.37 | 0.00 | 8.16 | 23.08 | 39.78 | 55.92 | 70.43 |
| 2 | 5 | NC=C | 8.55 | 1.15 | 0.00 | 8.46 | 23.54 | 40.34 | 56.51 | 70.98 |
| 2 | 5 | OC=C | 10.16 | 1.89 | 0.00 | 7.17 | 21.08 | 36.78 | 51.98 | 65.65 |
| 2 | 5 | OC=C | 9.25 | 1.41 | 0.00 | 8.30 | 23.52 | 40.61 | 57.17 | 72.06 |
| 2 | 5 | OC=C | 8.77 | 1.17 | 0.00 | 8.76 | 24.40 | 41.87 | 58.75 | 73.88 |
| 2 | 5 | [SiH3]C=C | 9.22 | 1.47 | 0.00 | 7.80 | 22.18 | 38.24 | 53.66 | 67.42 |
| 2 | 5 | [SiH3]C=C | 9.15 | 1.41 | 0.00 | 8.07 | 22.85 | 39.38 | 55.31 | 69.59 |
| 2 | 5 | [SiH3]C=C | 9.01 | 1.35 | 0.00 | 8.16 | 22.99 | 39.52 | 55.40 | 69.58 |
| 2 | 5 | C/C=C/C | 9.20 | 1.47 | 0.00 | 7.83 | 22.29 | 38.48 | 54.11 | 68.15 |
| 2 | 5 | C/C=C/C | 9.21 | 1.47 | 0.00 | 7.83 | 22.29 | 38.48 | 54.11 | 68.15 |
| 2 | 5 | CCC=C | 8.85 | 1.26 | 0.00 | 8.38 | 23.47 | 40.31 | 56.54 | 71.06 |
| 2 | 5 | CCC=C | 9.14 | 1.40 | 0.00 | 8.14 | 23.04 | 39.73 | 55.85 | 70.31 |
| 2 | 5 | CCC=C | 9.10 | 1.42 | 0.00 | 7.89 | 22.39 | 38.62 | 54.27 | 68.33 |
| 2 | 5 | [CH2]/C=C(\C)/[CH2] | 8.84 | 1.32 | 0.00 | 7.99 | 22.52 | 38.76 | 54.42 | 68.47 |
| 2 | 5 | [CH2]/C=C(\C)/[CH2] | 9.15 | 1.37 | 0.00 | 8.30 | 23.48 | 40.51 | 57.00 | 71.81 |
| 2 | 5 | [CH2]/C=C(\C)/[CH2] | 8.77 | 1.22 | 0.00 | 8.44 | 23.62 | 40.55 | 56.88 | 71.50 |
| 2 | 5 | [CH2]/C=C(\C)/[CH2] | 9.03 | 1.33 | 0.00 | 8.29 | 23.35 | 40.19 | 56.46 | 71.05 |
| 2 | 5 | [CH2]/C=C(\C)/[CH2] | 9.05 | 1.36 | 0.00 | 8.20 | 23.14 | 39.88 | 56.05 | 70.58 |
| 2 | 5 | ICC=C | 8.86 | 1.29 | 0.00 | 8.20 | 23.05 | 39.65 | 55.66 | 70.02 |
| 2 | 5 | ICC=C | 9.11 | 1.37 | 0.00 | 8.20 | 23.18 | 39.95 | 56.16 | 70.72 |
| 2 | 5 | ICC=C | 8.76 | 1.20 | 0.00 | 8.52 | 23.79 | 40.82 | 57.24 | 71.94 |
| 2 | 5 | CC(=C)C=O | 9.12 | 1.38 | 0.00 | 8.23 | 23.28 | 40.17 | 56.52 | 71.24 |
| 2 | 5 | CC(=C)C=O | 9.12 | 1.37 | 0.00 | 8.27 | 23.37 | 40.30 | 56.68 | 71.38 |
| 2 | 5 | C=CC=O | 9.01 | 1.35 | 0.00 | 8.20 | 23.19 | 40.02 | 56.33 | 71.01 |
| 2 | 5 | C=CC=O | 9.11 | 1.37 | 0.00 | 8.23 | 23.29 | 40.20 | 56.59 | 71.34 |
| 2 | 5 | C=CC=O | 8.89 | 1.26 | 0.00 | 8.45 | 23.72 | 40.80 | 57.30 | 72.10 |
| 2 | 5 | C/C=C/O | 10.35 | 1.99 | 0.00 | 6.99 | 20.73 | 36.29 | 51.37 | 64.95 |
| 2 | 5 | C/C=C/O | 9.30 | 1.48 | 0.00 | 7.98 | 22.74 | 39.31 | 55.34 | 69.75 |
| 2 | 5 | C1=CCC=CC1 | 8.85 | 1.30 | 0.00 | 8.15 | 22.89 | 39.31 | 55.12 | 69.27 |
| 2 | 5 | C1=CCC=CC1 | 8.85 | 1.30 | 0.00 | 8.15 | 22.89 | 39.32 | 55.12 | 69.27 |
| 2 | 5 | C1=CCC=CC1 | 8.85 | 1.30 | 0.00 | 8.15 | 22.89 | 39.31 | 55.12 | 69.27 |
| 2 | 5 | C1=CCC=CC1 | 8.85 | 1.30 | 0.00 | 8.15 | 22.89 | 39.32 | 55.12 | 69.27 |
| 2 | 5 | C=CS(=O)(=O)C=C | 9.56 | 1.56 | 0.00 | 8.01 | 22.94 | 39.76 | 56.05 | 70.69 |
| 2 | 5 | C=CS(=O)(=O)C=C | 9.04 | 1.33 | 0.00 | 8.36 | 23.57 | 40.64 | 57.16 | 72.03 |
| 2 | 5 | C=CS(=O)(=O)C=C | 9.11 | 1.36 | 0.00 | 8.29 | 23.42 | 40.36 | 56.70 | 71.33 |
| 2 | 5 | C=CS(=O)(=O)C=C | 9.07 | 1.35 | 0.00 | 8.31 | 23.44 | 40.38 | 56.72 | 71.35 |
| 2 | 5 | C=CS(=O)(=O)C=C | 9.01 | 1.32 | 0.00 | 8.37 | 23.58 | 40.65 | 57.18 | 72.04 |
| 2 | 5 | C=CS(=O)(=O)C=C | 9.55 | 1.55 | 0.00 | 8.01 | 22.94 | 39.75 | 56.05 | 70.68 |
| 2 | 5 | C=CS(=O)(=O)C | 9.06 | 1.34 | 0.00 | 8.34 | 23.54 | 40.59 | 57.10 | 71.95 |
| 2 | 5 | C=CS(=O)(=O)C | 9.08 | 1.35 | 0.00 | 8.32 | 23.46 | 40.40 | 56.75 | 71.39 |
| 2 | 5 | C=CS(=O)(=O)C | 9.57 | 1.55 | 0.00 | 8.04 | 23.03 | 39.93 | 56.30 | 71.01 |
| 2 | 5 | C=C[N]N(C=C)[O] | 9.55 | 1.98 | 0.00 | 5.66 | 16.98 | 29.56 | 41.35 | 51.44 |
| 2 | 5 | C=C[N]N(C=C)[O] | 10.21 | 1.76 | 0.00 | 8.18 | 23.59 | 41.01 | 57.87 | 72.96 |
| 2 | 5 | C=C[N]N(C=C)[O] | 8.52 | 1.12 | 0.00 | 8.74 | 24.17 | 41.34 | 57.86 | 72.62 |
| 2 | 5 | C=C[N]N(C=C)[O] | 8.88 | 1.26 | 0.00 | 8.59 | 23.99 | 41.23 | 57.89 | 72.84 |
| 2 | 5 | C=C[N]N(C=C)[O] | 8.95 | 1.30 | 0.00 | 8.55 | 23.95 | 41.23 | 57.96 | 73.00 |
| 2 | 5 | C=C[N]N(C=C)[O] | 9.84 | 1.77 | 0.00 | 7.34 | 21.11 | 36.33 | 50.71 | 63.27 |
| 2 | 5 | NC=C | 8.55 | 1.09 | 0.00 | 8.77 | 24.32 | 41.63 | 58.31 | 73.25 |
| 2 | 5 | NC=C | 9.14 | 1.39 | 0.00 | 8.15 | 23.07 | 39.79 | 55.95 | 70.48 |
| 2 | 5 | NC=C | 8.91 | 1.28 | 0.00 | 8.35 | 23.44 | 40.30 | 56.56 | 71.14 |
| 2 | 5 | [CH]C=C.[CH2].[C] | 8.87 | 1.26 | 0.00 | 8.43 | 23.65 | 40.67 | 57.10 | 71.84 |
| 2 | 5 | [CH]C=C.[CH2].[C] | 9.01 | 1.32 | 0.00 | 8.30 | 23.41 | 40.35 | 56.74 | 71.47 |
| 2 | 5 | [CH]C=C.[CH2].[C] | 8.95 | 1.33 | 0.00 | 8.13 | 22.92 | 39.43 | 55.33 | 69.55 |
| 2 | 5 | [CH2]/C=C\1/[CH]CC1 | 8.82 | 1.24 | 0.00 | 8.42 | 23.57 | 40.50 | 56.82 | 71.44 |
| 2 | 5 | [CH2]/C=C\1/[CH]CC1 | 9.10 | 1.37 | 0.00 | 8.22 | 23.24 | 40.08 | 56.38 | 71.03 |
| 2 | 5 | [CH2]/C=C\1/[CH]CC1 | 9.09 | 1.42 | 0.00 | 7.88 | 22.40 | 38.64 | 54.32 | 68.39 |
| 2 | 5 | C=CC=C | 8.82 | 1.24 | 0.00 | 8.43 | 23.61 | 40.56 | 56.92 | 71.57 |
| 2 | 5 | C=CC=C | 9.05 | 1.35 | 0.00 | 8.22 | 23.20 | 39.99 | 56.22 | 70.80 |
| 2 | 5 | C=CC=C | 8.98 | 1.38 | 0.00 | 7.95 | 22.52 | 38.83 | 54.59 | 68.74 |
| 2 | 5 | C=CC=C | 8.98 | 1.38 | 0.00 | 7.95 | 22.52 | 38.83 | 54.59 | 68.74 |
| 2 | 5 | C=CC=C | 9.07 | 1.36 | 0.00 | 8.21 | 23.20 | 39.99 | 56.22 | 70.79 |
| 2 | 5 | C=CC=C | 8.81 | 1.24 | 0.00 | 8.43 | 23.61 | 40.57 | 56.92 | 71.58 |
| 2 | 5 | CC(=O)C=C | 8.88 | 1.26 | 0.00 | 8.41 | 23.61 | 40.59 | 56.97 | 71.65 |
| 2 | 5 | CC(=O)C=C | 8.83 | 1.28 | 0.00 | 8.22 | 23.16 | 39.88 | 56.07 | 70.62 |
| 2 | 5 | CC(=O)C=C | 9.19 | 1.38 | 0.00 | 8.34 | 23.60 | 40.75 | 57.38 | 72.32 |
| 2 | 5 | CNC=C | 8.48 | 1.07 | 0.00 | 8.77 | 24.27 | 41.52 | 58.13 | 72.99 |
| 2 | 5 | CNC=C | 8.76 | 1.25 | 0.00 | 8.25 | 23.14 | 39.76 | 55.79 | 70.17 |
| 2 | 5 | CNC=C | 9.10 | 1.38 | 0.00 | 8.13 | 23.02 | 39.70 | 55.82 | 70.32 |
| 2 | 5 | IC=C | 9.00 | 1.32 | 0.00 | 8.30 | 23.36 | 40.17 | 56.37 | 70.84 |
| 2 | 5 | IC=C | 8.90 | 1.27 | 0.00 | 8.41 | 23.61 | 40.59 | 56.98 | 71.66 |
| 2 | 5 | IC=C | 8.87 | 1.25 | 0.00 | 8.42 | 23.49 | 40.18 | 56.14 | 70.26 |
| 2 | 5 | BrC=C | 8.96 | 1.29 | 0.00 | 8.40 | 23.60 | 40.60 | 57.00 | 71.68 |
| 2 | 5 | BrC=C | 8.90 | 1.27 | 0.00 | 8.44 | 23.70 | 40.77 | 57.27 | 72.08 |
| 2 | 5 | BrC=C | 8.84 | 1.23 | 0.00 | 8.51 | 23.73 | 40.60 | 56.76 | 71.09 |
| 2 | 5 | ClC=C | 9.27 | 1.51 | 0.00 | 7.75 | 22.37 | 39.03 | 55.39 | 70.31 |
| 2 | 5 | ClC=C | 8.60 | 1.07 | 0.00 | 9.01 | 24.65 | 41.61 | 57.84 | 72.42 |
| 2 | 5 | ClC=C | 8.11 | 0.92 | 0.00 | 8.39 | 22.79 | 38.95 | 54.99 | 69.75 |
| 2 | 5 | FC=C | 9.08 | 1.29 | 0.00 | 8.62 | 24.24 | 41.74 | 58.69 | 73.91 |
| 2 | 5 | FC=C | 9.38 | 1.45 | 0.00 | 8.31 | 23.62 | 40.85 | 57.57 | 72.61 |
| 2 | 5 | FC=C | 10.23 | 1.77 | 0.00 | 7.99 | 23.16 | 40.27 | 56.84 | 71.69 |
| 2 | 5 | [GeH3]C=C | 9.04 | 1.37 | 0.00 | 8.09 | 22.82 | 39.24 | 55.01 | 69.08 |
| 2 | 5 | [GeH3]C=C | 9.13 | 1.40 | 0.00 | 8.06 | 22.80 | 39.27 | 55.13 | 69.34 |
| 2 | 5 | [GeH3]C=C | 9.14 | 1.44 | 0.00 | 7.87 | 22.30 | 38.37 | 53.79 | 67.52 |
| 2 | 5 | C=CC(=O)C(=O)C=C | 8.33 | 1.01 | 0.00 | 8.85 | 24.45 | 41.82 | 58.53 | 73.43 |
| 2 | 5 | C=CC(=O)C(=O)C=C | 8.32 | 1.01 | 0.00 | 8.85 | 24.45 | 41.82 | 58.53 | 73.44 |
| 2 | 5 | C=CC(=O)C(=O)C=C | 9.07 | 1.34 | 0.00 | 8.38 | 23.66 | 40.84 | 57.50 | 72.49 |
| 2 | 5 | C=CC(=O)C(=O)C=C | 8.84 | 1.25 | 0.00 | 8.46 | 23.70 | 40.73 | 57.17 | 71.91 |
| 2 | 5 | C=CC(=O)C(=O)C=C | 8.84 | 1.25 | 0.00 | 8.46 | 23.70 | 40.73 | 57.17 | 71.91 |
| 2 | 5 | C=CC(=O)C(=O)C=C | 9.06 | 1.33 | 0.00 | 8.38 | 23.67 | 40.85 | 57.51 | 72.50 |
| 2 | 5 | [CH]/C(=C\C=C\[CH])/[N]N[O] | 24.82 | 7.73 | 0.00 | 0.88 | 12.38 | 27.41 | 42.59 | 56.37 |
| 2 | 5 | [CH]/C(=C\C=C\[CH])/[N]N[O] | 23.03 | 6.87 | 0.00 | 2.37 | 15.14 | 31.32 | 47.64 | 62.60 |
| 2 | 5 | [CH]/C(=C\C=C\[CH])/[N]N[O] | 24.43 | 7.55 | 0.00 | 1.24 | 13.23 | 28.95 | 45.03 | 59.98 |
| 2 | 5 | C=CC1CC1 | 9.08 | 1.45 | 0.00 | 7.71 | 21.93 | 37.80 | 53.07 | 66.74 |
| 2 | 5 | C=CC1CC1 | 9.23 | 1.43 | 0.00 | 8.11 | 23.01 | 39.74 | 55.91 | 70.42 |
| 2 | 5 | C=CC1CC1 | 8.79 | 1.22 | 0.00 | 8.46 | 23.65 | 40.60 | 56.93 | 71.54 |
| 2 | 5 | C=C | 8.86 | 1.26 | 0.00 | 8.38 | 23.50 | 40.37 | 56.63 | 71.19 |
| 2 | 5 | C=C | 8.86 | 1.26 | 0.00 | 8.39 | 23.50 | 40.38 | 56.64 | 71.19 |
| 2 | 5 | C=C | 8.86 | 1.26 | 0.00 | 8.38 | 23.50 | 40.37 | 56.63 | 71.19 |
| 2 | 5 | C=C | 8.86 | 1.26 | 0.00 | 8.39 | 23.50 | 40.38 | 56.64 | 71.19 |
| 2 | 5 | CC=C | 8.82 | 1.24 | 0.00 | 8.41 | 23.55 | 40.43 | 56.70 | 71.27 |
| 2 | 5 | CC=C | 9.13 | 1.39 | 0.00 | 8.16 | 23.08 | 39.79 | 55.93 | 70.40 |
| 2 | 5 | CC=C | 8.95 | 1.35 | 0.00 | 8.03 | 22.67 | 39.01 | 54.76 | 68.87 |
| 2 | 5 | CCOC=C | 8.61 | 1.16 | 0.00 | 8.51 | 23.67 | 40.55 | 56.78 | 71.27 |
| 2 | 5 | CCOC=C | 9.07 | 1.33 | 0.00 | 8.39 | 23.67 | 40.79 | 57.37 | 72.25 |
| 2 | 5 | CCOC=C | 8.48 | 1.06 | 0.00 | 8.82 | 24.40 | 41.76 | 58.48 | 73.47 |
| 2 | 5 | CC=N | 9.37 | 1.56 | 0.00 | 7.58 | 21.68 | 37.39 | 52.42 | 65.78 |
| 2 | 5 | C=CON=N | 23.85 | 7.23 | 0.00 | 1.98 | 14.74 | 31.20 | 48.00 | 63.56 |
| 2 | 5 | C=CON=N | 23.58 | 7.11 | 0.00 | 2.19 | 15.16 | 31.89 | 48.99 | 64.84 |
| 2 | 5 | C=CON=N | 23.16 | 6.89 | 0.00 | 2.66 | 16.11 | 33.27 | 50.74 | 66.90 |
| 2 | 5 | COBC=C | 12.23 | 2.63 | 0.00 | 6.80 | 21.14 | 37.63 | 53.69 | 68.14 |
| 2 | 5 | COBC=C | 11.83 | 2.47 | 0.00 | 7.00 | 21.47 | 38.08 | 54.28 | 68.90 |
| 2 | 5 | COBC=C | 12.37 | 2.73 | 0.00 | 6.50 | 20.49 | 36.62 | 52.34 | 66.51 |
| 2 | 5 | O=COC(=O)C=C | 12.13 | 2.55 | 0.00 | 7.11 | 21.92 | 38.98 | 55.66 | 70.74 |
| 2 | 5 | O=COC(=O)C=C | 11.65 | 2.36 | 0.00 | 7.30 | 22.23 | 39.40 | 56.24 | 71.50 |
| 2 | 5 | O=COC(=O)C=C | 11.89 | 2.45 | 0.00 | 7.21 | 22.07 | 39.14 | 55.82 | 70.89 |
| 2 | 5 | PC=C | 9.02 | 1.34 | 0.00 | 8.23 | 23.20 | 39.93 | 56.04 | 70.44 |
| 2 | 5 | PC=C | 9.17 | 1.41 | 0.00 | 8.07 | 22.87 | 39.44 | 55.45 | 69.82 |
| 2 | 5 | PC=C | 8.96 | 1.34 | 0.00 | 8.07 | 22.72 | 39.01 | 54.62 | 68.48 |
| 2 | 5 | C=CPC=C | 9.15 | 1.40 | 0.00 | 8.09 | 22.91 | 39.50 | 55.51 | 69.86 |
| 2 | 5 | C=CPC=C | 9.02 | 1.34 | 0.00 | 8.22 | 23.16 | 39.84 | 55.89 | 70.23 |
| 2 | 5 | C=CPC=C | 9.01 | 1.34 | 0.00 | 8.22 | 23.16 | 39.84 | 55.90 | 70.24 |
| 2 | 5 | C=CPC=C | 9.15 | 1.41 | 0.00 | 8.09 | 22.91 | 39.50 | 55.51 | 69.86 |
| 2 | 5 | C=CPC=C | 9.14 | 1.42 | 0.00 | 7.93 | 22.49 | 38.74 | 54.36 | 68.29 |
| 2 | 5 | C=CPC=C | 9.14 | 1.43 | 0.00 | 7.93 | 22.49 | 38.74 | 54.36 | 68.29 |
| 2 | 5 | CN([CH2])C | 8.85 | 1.25 | 0.00 | 8.45 | 23.74 | 40.89 | 57.53 | 72.53 |
| 2 | 5 | CN([CH2])C | 8.84 | 1.25 | 0.00 | 8.46 | 23.75 | 40.91 | 57.55 | 72.56 |
| 2 | 5 | CCOC=C | 10.60 | 2.05 | 0.00 | 7.11 | 21.18 | 37.15 | 52.64 | 66.55 |
| 2 | 5 | CCOC=C | 9.30 | 1.41 | 0.00 | 8.39 | 23.75 | 40.99 | 57.66 | 72.61 |
| 2 | 5 | CCOC=C | 8.82 | 1.20 | 0.00 | 8.68 | 24.22 | 41.57 | 58.33 | 73.36 |
| 2 | 5 | C/N=C/C=C | 9.10 | 1.38 | 0.00 | 8.18 | 23.14 | 39.92 | 56.14 | 70.73 |
| 2 | 5 | C/N=C/C=C | 8.81 | 1.23 | 0.00 | 8.49 | 23.74 | 40.79 | 57.24 | 71.98 |
| 2 | 5 | C/N=C/C=C | 10.50 | 2.16 | 0.00 | 6.25 | 18.94 | 33.25 | 47.02 | 59.35 |
| 2 | 5 | C/N=C/C=C | 8.74 | 1.24 | 0.00 | 8.30 | 23.30 | 40.09 | 56.31 | 70.88 |
| 2 | 5 | COC=C | 11.04 | 2.28 | 0.00 | 6.57 | 20.06 | 35.45 | 50.43 | 63.96 |
| 2 | 5 | COC=C | 9.17 | 1.37 | 0.00 | 8.33 | 23.56 | 40.65 | 57.19 | 72.06 |
| 2 | 5 | COC=C | 8.83 | 1.20 | 0.00 | 8.70 | 24.29 | 41.72 | 58.55 | 73.65 |
| 2 | 5 | Cn1cccc1 | 8.85 | 1.21 | 0.00 | 8.67 | 24.22 | 41.58 | 58.36 | 73.44 |
| 2 | 5 | Cn1cccc1 | 8.87 | 1.22 | 0.00 | 8.66 | 24.21 | 41.56 | 58.33 | 73.41 |
| 2 | 5 | Cn1cccc1 | 9.19 | 1.35 | 0.00 | 8.53 | 24.04 | 41.43 | 58.28 | 73.45 |
| 2 | 5 | Cn1cccc1 | 9.14 | 1.32 | 0.00 | 8.58 | 24.13 | 41.55 | 58.41 | 73.58 |
| 2 | 5 | c1ccc[nH]1 | 8.92 | 1.22 | 0.00 | 8.73 | 24.42 | 41.96 | 58.93 | 74.20 |
| 2 | 5 | c1ccc[nH]1 | 8.81 | 1.19 | 0.00 | 8.71 | 24.30 | 41.71 | 58.53 | 73.65 |
| 2 | 5 | c1ccc[nH]1 | 8.81 | 1.19 | 0.00 | 8.71 | 24.30 | 41.71 | 58.53 | 73.65 |
| 2 | 5 | c1ccc[nH]1 | 8.92 | 1.22 | 0.00 | 8.73 | 24.42 | 41.96 | 58.93 | 74.20 |
| 2 | 5 | [NH3]n1cccc1 | 8.81 | 1.16 | 0.00 | 8.89 | 24.78 | 42.57 | 59.83 | 75.41 |
| 2 | 5 | [NH3]n1cccc1 | 8.80 | 1.16 | 0.00 | 8.89 | 24.78 | 42.58 | 59.83 | 75.42 |
| 2 | 5 | [NH3]n1cccc1 | 9.17 | 1.36 | 0.00 | 8.43 | 23.85 | 41.19 | 58.06 | 73.34 |
| 2 | 5 | [NH3]n1cccc1 | 9.17 | 1.36 | 0.00 | 8.43 | 23.85 | 41.19 | 58.06 | 73.34 |
| 2 | 5 | C=CN1C=C[CH]C=C1 | 8.88 | 1.27 | 0.00 | 8.42 | 23.66 | 40.75 | 57.34 | 72.32 |
| 2 | 5 | C=CN1C=C[CH]C=C1 | 8.71 | 1.19 | 0.00 | 8.58 | 23.99 | 41.23 | 57.96 | 73.06 |
| 2 | 5 | C=CN1C=C[CH]C=C1 | 8.95 | 1.23 | 0.00 | 8.82 | 24.72 | 42.61 | 60.02 | 75.77 |
| 2 | 5 | C=CN1C=C[CH]C=C1 | 8.46 | 1.10 | 0.00 | 8.59 | 23.88 | 40.93 | 57.41 | 72.21 |
| 2 | 5 | C=CN1C=C[CH]C=C1 | 9.34 | 1.43 | 0.00 | 8.33 | 23.62 | 40.80 | 57.42 | 72.33 |
| 2 | 5 | C=CN1C=C[CH]C=C1 | 8.71 | 1.15 | 0.00 | 8.76 | 24.42 | 41.95 | 58.93 | 74.22 |
| 2 | 5 | C=CN1C=C[CH]C=C1 | 8.58 | 1.13 | 0.00 | 8.68 | 24.16 | 41.48 | 58.26 | 73.41 |
| 2 | 5 | C=CN1C=C[CH]C=C1 | 8.72 | 1.19 | 0.00 | 8.59 | 24.00 | 41.26 | 58.01 | 73.13 |
| 2 | 5 | C=CC#C | 8.81 | 1.23 | 0.00 | 8.49 | 23.75 | 40.80 | 57.26 | 72.01 |
| 2 | 5 | C=CC#C | 8.90 | 1.28 | 0.00 | 8.40 | 23.59 | 40.59 | 57.03 | 71.79 |
| 2 | 5 | C=CC#C | 9.18 | 1.46 | 0.00 | 7.81 | 22.22 | 38.28 | 53.66 | 67.32 |
| 2 | 5 | C=CC1CCC1 | 8.95 | 1.31 | 0.00 | 8.29 | 23.31 | 40.08 | 56.24 | 70.71 |
| 2 | 5 | C=CC1CCC1 | 9.07 | 1.36 | 0.00 | 8.21 | 23.18 | 39.93 | 56.10 | 70.60 |
| 2 | 5 | C=CC1CCC1 | 9.02 | 1.37 | 0.00 | 8.05 | 22.78 | 39.26 | 55.17 | 69.46 |
| 2 | 6 | Oc1ccccc1 | 15.07 | 3.33 | 0.00 | 7.56 | 23.57 | 41.65 | 59.01 | 74.36 |
| 2 | 6 | CCOC=C | 11.27 | 1.64 | 0.00 | 9.91 | 27.25 | 45.90 | 63.18 | 77.94 |
| 2 | 8 | NC=C | 9.90 | 1.12 | 0.00 | 10.95 | 29.96 | 50.94 | 71.13 | 89.27 |
| 2 | 8 | CNC=C | 14.22 | 3.11 | 0.00 | 7.65 | 23.86 | 42.24 | 59.81 | 75.13 |
| 2 | 9 | NC=C | 7.83 | 0.44 | 0.00 | 11.07 | 29.37 | 49.36 | 68.50 | 85.64 |
| 2 | 9 | C=CNS(=O)(=O)N | 10.71 | 1.82 | 0.00 | 8.68 | 25.02 | 43.42 | 61.37 | 77.74 |
| 2 | 11 | FC=C | 11.66 | 1.90 | 0.00 | 9.25 | 26.15 | 44.89 | 62.88 | 79.05 |
| 2 | 12 | ClC=C | 17.75 | 7.13 | 0.00 | -2.34 | 6.16 | 18.11 | 32.85 | 48.75 |
| 2 | 13 | BrC=C | 6.77 | 1.45 | 0.00 | 4.43 | 14.05 | 25.61 | 37.35 | 48.33 |
| 2 | 14 | IC=C | 5.01 | 0.97 | 0.00 | 4.08 | 12.48 | 22.60 | 33.06 | 42.96 |
| 2 | 18 | C=CS(=O)(=O)C=C | 9.00 | 2.26 | 0.00 | 3.69 | 12.37 | 23.17 | 34.40 | 44.85 |
| 2 | 18 | C=CS(=O)(=O)C=C | 9.07 | 2.29 | 0.00 | 3.66 | 12.33 | 23.12 | 34.35 | 44.80 |
| 2 | 18 | C=CS(=O)(=O)C | 8.69 | 2.10 | 0.00 | 4.04 | 13.12 | 24.24 | 35.66 | 46.22 |
| 2 | 19 | [SiH3]C=C | 5.51 | 0.97 | 0.00 | 4.88 | 14.44 | 26.16 | 38.49 | 50.36 |
| 2 | 22 | C=CC1CC1 | 12.81 | 3.12 | 0.00 | 5.73 | 19.53 | 36.03 | 52.57 | 67.82 |
| 2 | 25 | PC=C | 6.77 | 1.56 | 0.00 | 4.16 | 13.56 | 24.75 | 36.29 | 47.25 |
| 2 | 25 | C=CPC=C | 6.02 | 1.24 | 0.00 | 4.56 | 14.19 | 25.45 | 36.94 | 47.78 |
| 2 | 25 | C=CPC=C | 6.01 | 1.24 | 0.00 | 4.56 | 14.20 | 25.46 | 36.95 | 47.80 |
| 2 | 26 | COBC=C | 6.54 | 1.01 | 0.00 | 6.35 | 18.48 | 32.68 | 47.06 | 60.62 |
| 2 | 31 | [GeH3]C=C | 7.38 | 1.96 | 0.00 | 2.90 | 10.97 | 21.34 | 32.36 | 43.10 |
| 2 | 36 | C=C | 9.72 | 1.60 | 0.00 | 8.02 | 23.01 | 39.88 | 56.19 | 70.81 |
| 2 | 37 | c1nccs1 | 67.48 | 23.84 | 0.00 | -10.55 | 5.66 | 33.00 | 63.25 | 92.09 |
| 2 | 37 | c1ncc[nH]1 | 83.39 | 29.22 | 0.00 | -11.45 | 10.45 | 45.24 | 82.30 | 116.48 |
| 2 | 37 | c1ncc[nH]1 | 35.81 | 10.04 | 0.00 | 8.51 | 36.11 | 69.25 | 101.18 | 130.01 |
| 2 | 37 | c1cccnc1 | 45.88 | 14.74 | 0.00 | 1.26 | 25.06 | 58.54 | 94.53 | 128.85 |
| 2 | 37 | c1cccnc1 | 31.99 | 8.66 | 0.00 | 10.78 | 43.28 | 84.96 | 125.82 | 158.37 |
| 2 | 37 | [CH]/C(=C\C=C\[CH])/[N]N[O] | 20.91 | 6.46 | 0.00 | 0.55 | 8.69 | 17.85 | 26.04 | 33.51 |
| 2 | 37 | C/N=C/C=C | 24.94 | 5.61 | 0.00 | 11.58 | 36.10 | 62.86 | 86.43 | 99.81 |
| 2 | 38 | [CH]C=C.[CH2].[C] | 36.50 | 12.55 | 0.00 | -4.35 | 5.69 | 21.27 | 38.07 | 54.01 |
| 2 | 39 | [NH3]C=C | 65.17 | 24.28 | 0.00 | -18.31 | -15.44 | -3.84 | 10.52 | 24.95 |
| 2 | 39 | CN([CH2])C | 23.08 | 4.52 | 0.00 | 14.44 | 42.47 | 73.01 | 100.73 | 121.90 |
| 2 | 40 | c1ccc[nH]1 | 40.97 | 12.41 | 0.00 | 5.25 | 31.51 | 65.11 | 99.20 | 131.37 |
| 2 | 40 | Cn1cccc1 | 66.00 | 22.04 | 0.00 | -2.47 | 25.04 | 64.54 | 106.56 | 146.16 |
| 2 | 40 | Cn1cccc1 | 43.61 | 14.09 | 0.00 | -0.22 | 18.34 | 43.01 | 67.96 | 91.51 |
| 2 | 40 | c1ccc[nH]1 | 66.48 | 22.20 | 0.00 | -2.49 | 25.24 | 65.07 | 107.43 | 147.32 |
| 2 | 40 | c1ccc[nH]1 | 44.88 | 14.62 | 0.00 | -0.85 | 17.38 | 41.97 | 67.07 | 91.03 |
| 2 | 40 | [NH3]n1cccc1 | 65.75 | 22.06 | 0.00 | -3.07 | 23.35 | 61.52 | 102.29 | 140.90 |
| 2 | 40 | [NH3]n1cccc1 | 45.27 | 15.17 | 0.00 | -3.32 | 11.22 | 31.09 | 50.44 | 67.68 |
| 2 | 41 | OC=C | 53.79 | 18.65 | 0.00 | -8.60 | 1.59 | 18.53 | 36.77 | 53.85 |
| 2 | 41 | C/C=C/O | 54.78 | 19.12 | 0.00 | -9.35 | 0.28 | 16.82 | 34.76 | 51.63 |
| 2 | 41 | c1ccco1 | 39.73 | 11.83 | 0.00 | 5.06 | 29.01 | 59.25 | 90.59 | 121.78 |
| 2 | 41 | CCOC=C | 61.12 | 21.31 | 0.00 | -10.74 | -0.92 | 16.11 | 34.40 | 51.35 |
| 2 | 41 | COC=C | 61.91 | 21.74 | 0.00 | -11.57 | -2.38 | 14.20 | 32.16 | 48.86 |
| 2 | 42 | c1nccs1 | 29.15 | 9.99 | 0.00 | -2.52 | 8.01 | 24.97 | 45.02 | 66.34 |
| 2 | 46 | C=CN(=O)=O | 8.20 | 1.36 | 0.00 | 6.57 | 18.56 | 31.65 | 43.93 | 54.57 |
| 2 | 50 | C=Cc1ccccc1 | 14.04 | 3.52 | 0.00 | 5.65 | 19.80 | 36.89 | 54.13 | 70.10 |
| 2 | 57 | [CH2]/C=C\1/[CH]CC1 | 38.41 | 13.36 | 0.00 | -5.33 | 4.28 | 19.90 | 37.11 | 53.75 |
| 2 | 69 | C=CON=N | 17.58 | 4.36 | 0.00 | 6.16 | 21.11 | 38.05 | 53.85 | 66.94 |
| 2 | 72 | CC=N | 22.25 | 4.26 | 0.00 | 14.26 | 41.48 | 70.92 | 97.47 | 118.87 |
| 2 | 72 | C/N=C/C | 20.90 | 3.73 | 0.00 | 14.63 | 41.47 | 69.93 | 94.56 | 111.41 |
| 2 | 72 | CC=N | 22.24 | 4.26 | 0.00 | 14.27 | 41.50 | 70.94 | 97.50 | 118.91 |
| 2 | 72 | CN=C(C)C | 20.68 | 3.66 | 0.00 | 14.70 | 41.62 | 70.19 | 95.51 | 116.31 |
| 2 | 72 | CN=C(C)C | 20.60 | 3.62 | 0.00 | 14.75 | 41.66 | 70.21 | 95.50 | 116.33 |
| 2 | 106 | C=CC=C=O | 25.49 | 6.58 | 0.00 | 8.13 | 29.24 | 53.47 | 76.10 | 94.71 |
| 2 | 108 | ON=C(C)C | 19.62 | 3.41 | 0.00 | 14.16 | 39.84 | 67.01 | 90.05 | 104.93 |
| 2 | 111 | C=CN1C=C[CH]C=C1 | 46.80 | 15.35 | 0.00 | 0.11 | 23.35 | 56.90 | 93.65 | 129.65 |
| 2 | 111 | C=CN1C=C[CH]C=C1 | 35.45 | 10.54 | 0.00 | 6.44 | 33.46 | 67.66 | 98.58 | 118.87 |
| 2 | 111 | C=CN1C=C[CH]C=C1 | 49.52 | 17.72 | 0.00 | -10.09 | -2.88 | 11.05 | 26.79 | 42.07 |
| 2 | 143 | C=C[N]N(C=C)[O] | 664.37 | 263.11 | 0.00 | -283.06 | -397.85 | -437.26 | -443.28 | -435.67 |
| 2 | 144 | C=C[N]N(C=C)[O] | 2.52 | -1.62 | 0.00 | 12.29 | 28.78 | 44.56 | 57.29 | 65.95 |
| 3 | 3 | C=CC(=O)C=O | 99.33 | 38.79 | 0.00 | -37.25 | -45.29 | -40.24 | -30.06 | -18.53 |
| 3 | 3 | O=CC=O | 100.33 | 39.12 | 0.00 | -37.36 | -45.26 | -40.07 | -29.82 | -18.32 |
| 3 | 3 | C1=CC(=O)C(=O)C=C1 | 110.86 | 43.18 | 0.00 | -37.89 | -37.02 | -16.19 | 13.65 | 46.08 |
| 3 | 3 | C=CC(=O)C(=O)C=C | 100.19 | 39.26 | 0.00 | -38.04 | -46.49 | -41.52 | -31.18 | -19.35 |
| 3 | 5 | [O]C=O | 5.06 | 0.30 | 0.00 | 6.64 | 17.12 | 28.02 | 38.00 | 46.72 |
| 3 | 5 | C=CC(=O)C=O | 8.64 | 1.49 | 0.00 | 6.71 | 19.31 | 33.39 | 46.92 | 59.03 |
| 3 | 5 | BrC=O | 7.99 | 1.06 | 0.00 | 7.77 | 21.34 | 36.04 | 49.72 | 61.52 |
| 3 | 5 | ClC=O | 8.75 | 1.26 | 0.00 | 7.80 | 21.04 | 34.94 | 48.00 | 59.95 |
| 3 | 5 | FC=O | 8.65 | 1.26 | 0.00 | 7.97 | 22.35 | 38.30 | 53.55 | 67.06 |
| 3 | 5 | IC=O | 7.33 | 0.86 | 0.00 | 7.69 | 20.76 | 34.68 | 47.40 | 58.10 |
| 3 | 5 | NC=O | 8.36 | 1.32 | 0.00 | 7.09 | 19.97 | 34.18 | 47.70 | 59.68 |
| 3 | 5 | OC=O | 11.83 | 2.65 | 0.00 | 5.88 | 18.61 | 33.12 | 47.11 | 59.60 |
| 3 | 5 | O=CC=O | 8.70 | 1.48 | 0.00 | 6.85 | 19.68 | 34.03 | 47.84 | 60.17 |
| 3 | 5 | O=CC=O | 8.70 | 1.48 | 0.00 | 6.85 | 19.68 | 34.03 | 47.84 | 60.17 |
| 3 | 5 | CNC=O | 8.68 | 1.46 | 0.00 | 6.87 | 19.60 | 33.68 | 47.11 | 59.02 |
| 3 | 5 | CC(=C)C=O | 9.09 | 1.70 | 0.00 | 6.22 | 18.20 | 31.52 | 44.25 | 55.60 |
| 3 | 5 | C=CC=O | 9.16 | 1.73 | 0.00 | 6.19 | 18.15 | 31.48 | 44.22 | 55.60 |
| 3 | 5 | C1CC=C1C=O | 8.81 | 1.58 | 0.00 | 6.43 | 18.56 | 31.98 | 44.76 | 56.11 |
| 3 | 5 | O=CC1CC1 | 8.98 | 1.67 | 0.00 | 6.22 | 18.13 | 31.35 | 43.96 | 55.20 |
| 3 | 5 | CCC=O | 8.83 | 1.58 | 0.00 | 6.41 | 18.49 | 31.86 | 44.58 | 55.89 |
| 3 | 5 | O=CN(C)C | 8.61 | 1.43 | 0.00 | 6.89 | 19.59 | 33.63 | 46.99 | 58.86 |
| 3 | 5 | COC=O | 11.69 | 2.50 | 0.00 | 6.50 | 20.08 | 35.53 | 50.40 | 63.60 |
| 3 | 5 | BrCC=O | 8.69 | 1.47 | 0.00 | 6.82 | 19.48 | 33.48 | 46.81 | 58.64 |
| 3 | 5 | ClCC=O | 9.43 | 1.76 | 0.00 | 6.56 | 19.06 | 32.79 | 45.81 | 57.32 |
| 3 | 5 | FCC=O | 9.07 | 1.63 | 0.00 | 6.63 | 19.20 | 33.16 | 46.45 | 58.23 |
| 3 | 5 | ICC=O | 8.49 | 1.40 | 0.00 | 6.87 | 19.49 | 33.42 | 46.68 | 58.43 |
| 3 | 5 | O=COC=O | 8.43 | 1.20 | 0.00 | 7.96 | 22.25 | 38.11 | 53.31 | 66.78 |
| 3 | 5 | O=COC=O | 8.27 | 1.12 | 0.00 | 8.07 | 22.38 | 38.17 | 53.20 | 66.48 |
| 3 | 5 | C=O | 8.43 | 1.35 | 0.00 | 6.98 | 19.69 | 33.63 | 46.80 | 58.39 |
| 3 | 5 | C=O | 8.43 | 1.35 | 0.00 | 6.98 | 19.69 | 33.63 | 46.80 | 58.39 |
| 3 | 5 | CC=O | 8.59 | 1.47 | 0.00 | 6.59 | 18.82 | 32.28 | 45.06 | 56.38 |
| 3 | 5 | O=CC1CCC1 | 8.70 | 1.53 | 0.00 | 6.45 | 18.53 | 31.83 | 44.45 | 55.64 |
| 3 | 5 | O=COC(=O)C=C | 10.34 | 1.94 | 0.00 | 7.27 | 21.47 | 37.53 | 53.04 | 66.87 |
| 3 | 5 | OC=O | 8.01 | 1.03 | 0.00 | 8.11 | 22.38 | 38.11 | 53.10 | 66.35 |
| 3 | 6 | CC(=O)OC(=O)C | 17.08 | 4.60 | 0.00 | 3.58 | 14.42 | 27.13 | 39.71 | 51.47 |
| 3 | 6 | CC(=O)OC(=O)C | 17.09 | 4.60 | 0.00 | 3.57 | 14.41 | 27.13 | 39.71 | 51.47 |
| 3 | 6 | OC=O | 12.59 | 2.16 | 0.00 | 8.96 | 25.12 | 42.56 | 58.96 | 73.47 |
| 3 | 6 | CC(=O)O | 13.93 | 2.86 | 0.00 | 7.52 | 22.37 | 38.72 | 54.31 | 68.34 |
| 3 | 7 | NC=O | 31.31 | 6.81 | 0.00 | 14.65 | 44.64 | 77.13 | 106.76 | 131.74 |
| 3 | 7 | CC(=O)C | 29.65 | 6.21 | 0.00 | 15.15 | 45.33 | 78.18 | 108.82 | 135.49 |
| 3 | 7 | CC(=O)N | 31.75 | 7.12 | 0.00 | 13.76 | 42.82 | 74.49 | 103.47 | 127.98 |
| 3 | 7 | CNC=O | 31.96 | 7.14 | 0.00 | 13.95 | 43.28 | 75.17 | 104.30 | 128.88 |
| 3 | 7 | CC(=C)C=O | 33.86 | 8.07 | 0.00 | 12.40 | 40.59 | 71.78 | 100.93 | 126.78 |
| 3 | 7 | C=CC=O | 33.84 | 8.05 | 0.00 | 12.46 | 40.69 | 71.84 | 100.82 | 126.60 |
| 3 | 7 | O=C1CCCC1 | 29.54 | 6.10 | 0.00 | 15.47 | 45.91 | 78.91 | 109.65 | 136.39 |
| 3 | 7 | O=C1CCC(=O)CC1 | 29.86 | 6.31 | 0.00 | 14.96 | 44.94 | 77.57 | 107.93 | 134.28 |
| 3 | 7 | O=C1CCC(=O)CC1 | 29.83 | 6.30 | 0.00 | 14.97 | 44.96 | 77.58 | 107.94 | 134.29 |
| 3 | 7 | C1CC=C1C=O | 35.15 | 8.67 | 0.00 | 11.38 | 38.75 | 69.24 | 97.80 | 123.34 |
| 3 | 7 | O=CC1CC1 | 30.54 | 6.54 | 0.00 | 14.88 | 45.00 | 77.83 | 108.44 | 135.11 |
| 3 | 7 | CC(=O)C=C | 34.66 | 8.51 | 0.00 | 11.50 | 38.94 | 69.60 | 98.42 | 124.03 |
| 3 | 7 | CNC(=O)C | 32.54 | 7.56 | 0.00 | 12.80 | 40.95 | 71.82 | 100.15 | 124.15 |
| 3 | 7 | CCC=O | 29.55 | 6.09 | 0.00 | 15.57 | 46.18 | 79.25 | 109.67 | 135.77 |
| 3 | 7 | O=CN(C)C | 32.09 | 7.26 | 0.00 | 13.53 | 42.34 | 73.74 | 102.45 | 126.74 |
| 3 | 7 | BrCC=O | 28.02 | 5.40 | 0.00 | 16.54 | 47.68 | 80.92 | 111.18 | 137.06 |
| 3 | 7 | ClCC=O | 28.11 | 5.59 | 0.00 | 15.66 | 45.87 | 78.29 | 108.06 | 133.91 |
| 3 | 7 | FCC=O | 28.54 | 5.61 | 0.00 | 16.37 | 47.57 | 81.02 | 111.66 | 138.10 |
| 3 | 7 | ICC=O | 28.09 | 5.44 | 0.00 | 16.41 | 47.34 | 80.37 | 110.41 | 136.07 |
| 3 | 7 | CC(=O)OC(=O)C | 24.25 | 3.68 | 0.00 | 18.99 | 51.50 | 85.55 | 116.56 | 143.06 |
| 3 | 7 | CC(=O)OC(=O)C | 24.27 | 3.69 | 0.00 | 18.98 | 51.49 | 85.54 | 116.55 | 143.05 |
| 3 | 7 | C=O | 29.14 | 5.83 | 0.00 | 16.20 | 47.38 | 80.64 | 110.16 | 134.30 |
| 3 | 7 | CC=O | 29.39 | 6.00 | 0.00 | 15.74 | 46.50 | 79.64 | 110.07 | 136.15 |
| 3 | 7 | O=CC1CCC1 | 30.14 | 6.38 | 0.00 | 15.00 | 45.06 | 77.60 | 107.56 | 133.22 |
| 3 | 7 | O=C1CCC(=O)CC1 | 29.86 | 6.31 | 0.00 | 14.96 | 44.94 | 77.57 | 107.93 | 134.28 |
| 3 | 7 | O=C1CCC(=O)CC1 | 29.83 | 6.30 | 0.00 | 14.97 | 44.96 | 77.58 | 107.94 | 134.29 |
| 3 | 7 | OC=O | 28.22 | 5.32 | 0.00 | 17.25 | 49.38 | 83.81 | 115.48 | 142.50 |
| 3 | 7 | CC(=O)O | 29.02 | 5.76 | 0.00 | 16.28 | 47.46 | 80.97 | 111.66 | 137.68 |
| 3 | 7 | CC(=O)N | 31.73 | 7.12 | 0.00 | 13.77 | 42.83 | 74.50 | 103.48 | 127.99 |
| 3 | 7 | CCC(=O)N | 31.95 | 7.23 | 0.00 | 13.52 | 42.36 | 73.84 | 102.68 | 127.09 |
| 3 | 9 | NC=O | 12.20 | 1.89 | 0.00 | 10.16 | 28.27 | 47.92 | 66.56 | 83.28 |
| 3 | 9 | CC(=O)N | 12.40 | 2.10 | 0.00 | 9.45 | 26.70 | 45.54 | 63.46 | 79.58 |
| 3 | 9 | CNC=O | 11.27 | 1.51 | 0.00 | 10.60 | 28.86 | 48.50 | 67.04 | 83.57 |
| 3 | 9 | CNC(=O)C | 11.27 | 1.63 | 0.00 | 10.06 | 27.69 | 46.78 | 64.85 | 81.02 |
| 3 | 9 | O=CN(C)C | 9.55 | 0.77 | 0.00 | 11.59 | 30.39 | 50.32 | 68.99 | 85.56 |
| 3 | 9 | CC(=O)N | 12.39 | 2.10 | 0.00 | 9.45 | 26.71 | 45.55 | 63.46 | 79.58 |
| 3 | 9 | CCC(=O)N | 12.26 | 2.05 | 0.00 | 9.42 | 26.56 | 45.22 | 62.92 | 78.84 |
| 3 | 11 | FC=O | 11.67 | 2.09 | 0.00 | 7.71 | 22.05 | 37.99 | 53.59 | 68.09 |
| 3 | 12 | ClC=O | 15.31 | 6.24 | 0.00 | -2.05 | 4.79 | 14.66 | 26.95 | 38.14 |
| 3 | 13 | BrC=O | 2.71 | 0.08 | 0.00 | 4.53 | 12.33 | 21.39 | 30.63 | 39.47 |
| 3 | 14 | IC=O | -0.78 | -1.19 | 0.00 | 5.25 | 12.45 | 20.43 | 28.38 | 35.86 |
| 3 | 22 | O=CC1CC1 | 12.98 | 3.20 | 0.00 | 5.24 | 17.93 | 32.82 | 47.44 | 60.71 |
| 3 | 47 | [O]C=O | 13.50 | 1.06 | 0.00 | 15.58 | 40.04 | 65.08 | 87.39 | 105.83 |
| 3 | 47 | [O]C=O | 13.53 | 1.07 | 0.00 | 15.57 | 40.04 | 65.08 | 87.39 | 105.83 |
| 3 | 47 | CC(=O)[O] | 13.82 | 1.26 | 0.00 | 15.07 | 38.99 | 63.51 | 85.39 | 103.54 |
| 3 | 47 | CC(=O)[O] | 13.76 | 1.23 | 0.00 | 15.15 | 39.11 | 63.61 | 85.39 | 103.45 |
| 3 | 56 | O=CC1CCC1 | 8.01 | 1.43 | 0.00 | 6.18 | 18.04 | 31.43 | 44.39 | 56.07 |
| 3 | 57 | C1CC=C1C=O | 35.56 | 12.29 | 0.00 | -4.77 | 3.91 | 17.61 | 32.32 | 46.26 |
| 3 | 75 | OC=O | 10.31 | 1.34 | 0.00 | 9.48 | 25.47 | 42.49 | 58.43 | 72.51 |
| 3 | 75 | CC(=O)O | 10.71 | 1.62 | 0.00 | 8.73 | 23.96 | 40.36 | 55.87 | 69.77 |
| 3 | 75 | COC(=O)C | 10.16 | 1.14 | 0.00 | 10.10 | 26.56 | 43.72 | 59.45 | 73.06 |
| 3 | 75 | COC=O | 9.58 | 0.82 | 0.00 | 10.81 | 27.98 | 45.76 | 61.98 | 75.85 |
| 3 | 76 | O=CC=O | 35.29 | 8.72 | 0.00 | 11.30 | 38.36 | 67.93 | 90.62 | 114.89 |
| 3 | 76 | O=CC=O | 35.27 | 8.72 | 0.00 | 11.30 | 38.37 | 67.94 | 90.63 | 114.90 |
| 3 | 77 | OC=O | 21.63 | 2.52 | 0.00 | 20.90 | 55.04 | 90.51 | 122.60 | 149.72 |
| 3 | 77 | CC(=O)O | 26.16 | 4.65 | 0.00 | 17.38 | 48.91 | 82.46 | 113.05 | 138.96 |
| 3 | 78 | COC(=O)C | 27.48 | 5.37 | 0.00 | 15.81 | 45.84 | 78.14 | 107.88 | 133.27 |
| 3 | 78 | COC=O | 26.15 | 4.67 | 0.00 | 17.24 | 48.59 | 82.07 | 112.89 | 139.23 |
| 3 | 80 | BrC=O | 23.35 | 3.08 | 0.00 | 20.49 | 54.34 | 89.26 | 120.72 | 147.82 |
| 3 | 80 | ClC=O | 24.27 | 3.55 | 0.00 | 19.17 | 51.67 | 86.18 | 118.24 | 145.80 |
| 3 | 80 | FC=O | 22.75 | 2.56 | 0.00 | 22.23 | 58.12 | 95.10 | 128.56 | 157.15 |
| 3 | 80 | IC=O | 24.42 | 3.69 | 0.00 | 19.06 | 51.51 | 85.16 | 115.34 | 141.08 |
| 3 | 81 | C=CC(=O)C=C | 37.23 | 9.76 | 0.00 | 9.15 | 34.50 | 63.46 | 91.19 | 115.90 |
| 3 | 81 | C=CC(=O)C=O | 40.51 | 11.20 | 0.00 | 7.09 | 30.91 | 58.07 | 81.81 | 101.85 |
| 3 | 82 | O=COC=O | 27.85 | 4.76 | 0.00 | 19.22 | 53.34 | 89.30 | 122.15 | 150.32 |
| 3 | 82 | O=COC=O | 29.12 | 5.43 | 0.00 | 17.84 | 50.69 | 85.62 | 117.65 | 145.05 |
| 3 | 102 | O=COC(=O)C=C | 21.65 | 2.90 | 0.00 | 18.92 | 50.45 | 83.23 | 112.78 | 137.73 |
| 3 | 102 | O=COC(=O)C=C | 17.85 | 1.16 | 0.00 | 21.83 | 55.80 | 90.80 | 122.51 | 149.66 |
| 3 | 120 | C=CC(=O)C=O | 33.71 | 8.08 | 0.00 | 12.03 | 39.43 | 69.19 | 93.46 | 107.36 |
| 3 | 120 | C=CC(=O)C(=O)C=C | 40.78 | 11.40 | 0.00 | 6.50 | 29.69 | 56.45 | 80.76 | 101.76 |
| 3 | 120 | C=CC(=O)C(=O)C=C | 40.77 | 11.40 | 0.00 | 6.50 | 29.70 | 56.46 | 80.77 | 101.77 |
| 3 | 148 | O=COC=O | 6.06 | 0.19 | 0.00 | 8.49 | 21.61 | 35.51 | 48.80 | 60.98 |
| 3 | 148 | O=COC=O | 5.38 | -0.21 | 0.00 | 9.45 | 23.52 | 38.18 | 51.96 | 64.32 |
| 3 | 149 | O=COC(=O)C=C | 18.84 | 5.15 | 0.00 | 3.63 | 15.24 | 28.86 | 42.26 | 54.64 |
| 3 | 149 | O=COC(=O)C=C | 17.64 | 4.61 | 0.00 | 4.43 | 16.56 | 30.55 | 44.24 | 56.86 |
| 4 | 4 | C#C | 33.56 | 6.73 | 0.00 | 21.17 | 63.25 | 109.88 | 152.99 | 188.22 |
| 4 | 4 | CC#C | 33.93 | 6.94 | 0.00 | 20.67 | 62.33 | 108.75 | 152.06 | 188.30 |
| 4 | 4 | C1CC=C1C#C | 41.13 | 9.89 | 0.00 | 17.13 | 56.93 | 102.29 | 144.95 | 180.98 |
| 4 | 4 | C=CC#C | 36.00 | 7.96 | 0.00 | 18.71 | 58.49 | 103.07 | 144.96 | 180.00 |
| 4 | 10 | CCC#N | 42.17 | 8.79 | 0.00 | 21.96 | 65.05 | 110.54 | 150.79 | 182.90 |
| 4 | 57 | C1CC=C1C#C | 0.26 | -2.30 | 0.00 | 12.97 | 30.87 | 49.62 | 67.35 | 83.27 |
| 4 | 124 | C#C | 10.55 | 1.64 | 0.00 | 9.33 | 26.63 | 46.16 | 65.16 | 82.29 |
| 4 | 124 | C#C | 10.55 | 1.64 | 0.00 | 9.33 | 26.63 | 46.16 | 65.16 | 82.29 |
| 4 | 124 | CC#C | 10.39 | 1.57 | 0.00 | 9.44 | 26.80 | 46.36 | 65.37 | 82.49 |
| 4 | 124 | C1CC=C1C#C | 13.78 | 2.92 | 0.00 | 7.98 | 24.80 | 44.27 | 63.41 | 80.77 |
| 4 | 124 | C=CC#C | 10.58 | 1.65 | 0.00 | 9.36 | 26.70 | 46.28 | 65.32 | 82.48 |
| 5 | 16 | C[SH2] | 8.55 | 1.90 | 0.00 | 4.47 | 15.53 | 28.76 | 41.71 | 53.63 |
| 5 | 16 | C[SH2] | 8.55 | 1.90 | 0.00 | 4.47 | 15.53 | 28.76 | 41.71 | 53.63 |
| 5 | 16 | C[SH]C | 8.38 | 1.84 | 0.00 | 4.51 | 15.63 | 28.87 | 41.74 | 53.48 |
| 5 | 17 | CS=O | 6.25 | 1.28 | 0.00 | 4.14 | 13.23 | 23.21 | 32.56 | 40.85 |
| 5 | 18 | CS(=O)=O | 8.72 | 2.02 | 0.00 | 3.84 | 13.66 | 25.29 | 36.56 | 46.84 |
| 5 | 19 | C[SiH3] | 4.52 | 0.69 | 0.00 | 4.74 | 14.05 | 25.19 | 36.72 | 47.81 |
| 5 | 19 | C[SiH3] | 4.52 | 0.69 | 0.00 | 4.74 | 14.05 | 25.19 | 36.72 | 47.81 |
| 5 | 19 | C[SiH3] | 4.52 | 0.69 | 0.00 | 4.74 | 14.05 | 25.19 | 36.72 | 47.81 |
| 5 | 19 | [SiH3]C=C | 4.53 | 0.70 | 0.00 | 4.73 | 14.02 | 25.10 | 36.55 | 47.52 |
| 5 | 19 | [SiH3]C=C | 4.47 | 0.66 | 0.00 | 4.82 | 14.24 | 25.50 | 37.15 | 48.33 |
| 5 | 19 | [SiH3][SiH2][SiH3] | 4.23 | 0.58 | 0.00 | 4.86 | 14.18 | 25.26 | 36.65 | 47.53 |
| 5 | 19 | [SiH3][SiH2][SiH3] | 4.24 | 0.58 | 0.00 | 4.86 | 14.18 | 25.26 | 36.65 | 47.53 |
| 5 | 19 | [SiH3][SiH2][SiH3] | 4.23 | 0.57 | 0.00 | 4.88 | 14.21 | 25.30 | 36.72 | 47.65 |
| 5 | 19 | [SiH3][SiH2][SiH3] | 4.23 | 0.57 | 0.00 | 4.88 | 14.21 | 25.30 | 36.72 | 47.65 |
| 5 | 19 | [SiH3][SiH2][SiH3] | 4.31 | 0.63 | 0.00 | 4.69 | 13.78 | 24.61 | 35.76 | 46.41 |
| 5 | 19 | [SiH3][SiH2][SiH3] | 4.31 | 0.63 | 0.00 | 4.69 | 13.78 | 24.61 | 35.76 | 46.41 |
| 5 | 19 | [SiH3][SiH2][SiH3] | 4.23 | 0.57 | 0.00 | 4.87 | 14.20 | 25.30 | 36.72 | 47.64 |
| 5 | 19 | [SiH3][SiH2][SiH3] | 4.23 | 0.57 | 0.00 | 4.87 | 14.20 | 25.30 | 36.72 | 47.64 |
| 5 | 19 | [SiH4] | 4.34 | 0.60 | 0.00 | 4.96 | 14.51 | 25.88 | 37.62 | 48.87 |
| 5 | 19 | [SiH4] | 4.34 | 0.60 | 0.00 | 4.96 | 14.51 | 25.88 | 37.62 | 48.87 |
| 5 | 19 | [SiH4] | 4.34 | 0.60 | 0.00 | 4.96 | 14.51 | 25.88 | 37.62 | 48.87 |
| 5 | 19 | [SiH4] | 4.34 | 0.60 | 0.00 | 4.96 | 14.51 | 25.88 | 37.62 | 48.87 |
| 5 | 19 | C[SiH](C)C | 4.87 | 0.88 | 0.00 | 4.31 | 13.17 | 23.88 | 35.03 | 45.82 |
| 5 | 19 | [SiH3][SiH3] | 4.34 | 0.62 | 0.00 | 4.83 | 14.15 | 25.26 | 36.70 | 47.64 |
| 5 | 19 | [SiH3][SiH3] | 4.34 | 0.62 | 0.00 | 4.83 | 14.15 | 25.26 | 36.70 | 47.64 |
| 5 | 19 | [SiH3][SiH3] | 4.34 | 0.62 | 0.00 | 4.83 | 14.15 | 25.26 | 36.70 | 47.64 |
| 5 | 19 | [SiH3][SiH3] | 4.31 | 0.60 | 0.00 | 4.84 | 14.18 | 25.29 | 36.73 | 47.68 |
| 5 | 19 | [SiH3][SiH3] | 4.31 | 0.60 | 0.00 | 4.84 | 14.18 | 25.29 | 36.73 | 47.68 |
| 5 | 19 | [SiH3][SiH3] | 4.31 | 0.60 | 0.00 | 4.84 | 14.18 | 25.29 | 36.73 | 47.68 |
| 5 | 19 | [SiH3]C1CC1 | 4.60 | 0.73 | 0.00 | 4.68 | 13.92 | 25.00 | 36.48 | 47.55 |
| 5 | 19 | [SiH3]C1CC1 | 4.63 | 0.75 | 0.00 | 4.64 | 13.85 | 24.90 | 36.34 | 47.33 |
| 5 | 19 | [SiH3]C1CC1 | 4.63 | 0.75 | 0.00 | 4.63 | 13.85 | 24.90 | 36.33 | 47.33 |
| 5 | 19 | [SiH3]C1CCC1 | 4.65 | 0.76 | 0.00 | 4.63 | 13.82 | 24.85 | 36.28 | 47.29 |
| 5 | 19 | [SiH3]C1CCC1 | 4.66 | 0.76 | 0.00 | 4.59 | 13.75 | 24.75 | 36.13 | 47.08 |
| 5 | 19 | [SiH3]C1CCC1 | 4.67 | 0.77 | 0.00 | 4.58 | 13.73 | 24.73 | 36.11 | 47.06 |
| 5 | 19 | CO[SiH2]OC | 4.59 | 0.70 | 0.00 | 4.82 | 14.25 | 25.51 | 37.17 | 48.39 |
| 5 | 19 | CO[SiH2]OC | 4.60 | 0.71 | 0.00 | 4.82 | 14.24 | 25.50 | 37.16 | 48.38 |
| 5 | 19 | [SiH3]C1CC1 | 4.61 | 0.74 | 0.00 | 4.67 | 13.91 | 24.99 | 36.47 | 47.54 |
| 5 | 19 | [SiH3]C1CC1 | 4.63 | 0.75 | 0.00 | 4.63 | 13.85 | 24.90 | 36.34 | 47.33 |
| 5 | 19 | [SiH3]C1CC1 | 4.62 | 0.74 | 0.00 | 4.64 | 13.86 | 24.91 | 36.34 | 47.34 |
| 5 | 19 | O[SiH3] | 4.63 | 0.74 | 0.00 | 4.68 | 13.95 | 25.04 | 36.54 | 47.60 |
| 5 | 19 | O[SiH3] | 4.63 | 0.74 | 0.00 | 4.68 | 13.94 | 25.04 | 36.54 | 47.60 |
| 5 | 19 | O[SiH3] | 4.15 | 0.49 | 0.00 | 5.22 | 15.05 | 26.68 | 38.67 | 50.14 |
| 5 | 22 | O=N(=O)C1CC1 | 11.23 | 2.16 | 0.00 | 7.58 | 22.56 | 39.52 | 55.92 | 70.58 |
| 5 | 22 | O=N(=O)C1CC1 | 11.25 | 2.18 | 0.00 | 7.54 | 22.53 | 39.60 | 56.20 | 71.16 |
| 5 | 22 | O=N(=O)C1CC1 | 11.26 | 2.20 | 0.00 | 7.41 | 22.17 | 38.92 | 55.16 | 69.73 |
| 5 | 22 | O=N(=O)C1CC1 | 11.27 | 2.20 | 0.00 | 7.41 | 22.16 | 38.91 | 55.15 | 69.72 |
| 5 | 22 | O=N(=O)C1CC1 | 11.27 | 2.19 | 0.00 | 7.53 | 22.50 | 39.56 | 56.16 | 71.10 |
| 5 | 22 | O=N(=O)C1CC1 | 11.14 | 2.13 | 0.00 | 7.58 | 22.50 | 39.41 | 55.75 | 70.36 |
| 5 | 22 | O=N(=O)C1CC1 | 11.30 | 2.20 | 0.00 | 7.50 | 22.45 | 39.49 | 56.07 | 71.00 |
| 5 | 22 | O=N(=O)C1CC1 | 11.26 | 2.20 | 0.00 | 7.42 | 22.19 | 38.95 | 55.20 | 69.78 |
| 5 | 22 | O=N(=O)C1CC1 | 11.26 | 2.20 | 0.00 | 7.42 | 22.19 | 38.95 | 55.20 | 69.77 |
| 5 | 22 | O=N(=O)C1CC1 | 11.30 | 2.20 | 0.00 | 7.49 | 22.43 | 39.47 | 56.04 | 70.96 |
| 5 | 22 | [CH]C=C.[CH2].[C] | 11.84 | 2.66 | 0.00 | 5.77 | 18.24 | 32.39 | 45.99 | 58.11 |
| 5 | 22 | [CH]C=C.[CH2].[C] | 9.92 | 1.83 | 0.00 | 6.98 | 20.35 | 35.23 | 49.41 | 61.93 |
| 5 | 22 | CC1=CC1 | 12.30 | 2.76 | 0.00 | 6.01 | 18.97 | 33.65 | 47.65 | 60.05 |
| 5 | 22 | CC1=CC1 | 12.36 | 2.79 | 0.00 | 5.94 | 18.83 | 33.44 | 47.41 | 59.77 |
| 5 | 22 | CC1CC1 | 11.50 | 2.30 | 0.00 | 7.25 | 21.87 | 38.49 | 54.60 | 69.06 |
| 5 | 22 | CC1CC1 | 11.36 | 2.24 | 0.00 | 7.33 | 21.99 | 38.62 | 54.71 | 69.12 |
| 5 | 22 | CC1CC1 | 11.08 | 2.17 | 0.00 | 7.24 | 21.61 | 37.87 | 53.57 | 67.65 |
| 5 | 22 | CC1CC1 | 11.51 | 2.31 | 0.00 | 7.25 | 21.86 | 38.48 | 54.58 | 69.04 |
| 5 | 22 | CC1CC1 | 11.35 | 2.24 | 0.00 | 7.34 | 22.00 | 38.63 | 54.72 | 69.13 |
| 5 | 22 | O=C1CC1 | 11.69 | 2.39 | 0.00 | 7.13 | 21.68 | 38.26 | 54.32 | 68.68 |
| 5 | 22 | O=C1CC1 | 11.69 | 2.39 | 0.00 | 7.13 | 21.68 | 38.26 | 54.32 | 68.68 |
| 5 | 22 | O=C1CC1 | 11.70 | 2.39 | 0.00 | 7.13 | 21.67 | 38.26 | 54.32 | 68.67 |
| 5 | 22 | O=C1CC1 | 11.70 | 2.40 | 0.00 | 7.12 | 21.67 | 38.26 | 54.32 | 68.67 |
| 5 | 22 | O=CC1CC1 | 11.18 | 2.15 | 0.00 | 7.56 | 22.48 | 39.41 | 55.80 | 70.50 |
| 5 | 22 | O=CC1CC1 | 10.52 | 1.92 | 0.00 | 7.64 | 22.40 | 39.09 | 55.25 | 69.75 |
| 5 | 22 | O=CC1CC1 | 11.50 | 2.30 | 0.00 | 7.26 | 21.92 | 38.63 | 54.86 | 69.47 |
| 5 | 22 | O=CC1CC1 | 11.18 | 2.15 | 0.00 | 7.56 | 22.49 | 39.41 | 55.80 | 70.50 |
| 5 | 22 | O=CC1CC1 | 11.51 | 2.31 | 0.00 | 7.26 | 21.92 | 38.62 | 54.85 | 69.46 |
| 5 | 22 | C1CC(C1)C1CC1 | 11.45 | 2.29 | 0.00 | 7.27 | 21.90 | 38.53 | 54.66 | 69.15 |
| 5 | 22 | C1CC(C1)C1CC1 | 11.42 | 2.29 | 0.00 | 7.16 | 21.57 | 37.93 | 53.76 | 67.94 |
| 5 | 22 | C1CC(C1)C1CC1 | 11.69 | 2.38 | 0.00 | 7.18 | 21.79 | 38.45 | 54.62 | 69.14 |
| 5 | 22 | C1CC(C1)C1CC1 | 11.36 | 2.25 | 0.00 | 7.32 | 21.96 | 38.58 | 54.66 | 69.08 |
| 5 | 22 | C1CC(C1)C1CC1 | 11.40 | 2.27 | 0.00 | 7.26 | 21.82 | 38.34 | 54.31 | 68.63 |
| 5 | 22 | C=CC1CC1 | 11.37 | 2.25 | 0.00 | 7.34 | 22.04 | 38.77 | 55.01 | 69.61 |
| 5 | 22 | C=CC1CC1 | 11.38 | 2.25 | 0.00 | 7.30 | 21.91 | 38.49 | 54.53 | 68.91 |
| 5 | 22 | C=CC1CC1 | 11.35 | 2.24 | 0.00 | 7.35 | 22.04 | 38.71 | 54.85 | 69.32 |
| 5 | 22 | C=CC1CC1 | 11.45 | 2.27 | 0.00 | 7.40 | 22.24 | 39.13 | 55.53 | 70.26 |
| 5 | 22 | C=CC1CC1 | 10.63 | 1.98 | 0.00 | 7.45 | 21.90 | 38.14 | 53.73 | 67.60 |
| 5 | 22 | [GeH3]C1CC1 | 11.49 | 2.31 | 0.00 | 7.18 | 21.66 | 38.12 | 54.05 | 68.34 |
| 5 | 22 | [GeH3]C1CC1 | 11.33 | 2.23 | 0.00 | 7.36 | 22.04 | 38.69 | 54.80 | 69.24 |
| 5 | 22 | [GeH3]C1CC1 | 11.30 | 2.27 | 0.00 | 7.04 | 21.21 | 37.26 | 52.76 | 66.64 |
| 5 | 22 | [GeH3]C1CC1 | 11.51 | 2.32 | 0.00 | 7.17 | 21.65 | 38.11 | 54.04 | 68.34 |
| 5 | 22 | [GeH3]C1CC1 | 11.33 | 2.23 | 0.00 | 7.36 | 22.03 | 38.68 | 54.80 | 69.24 |
| 5 | 22 | [SiH3]C1CC1 | 11.46 | 2.30 | 0.00 | 7.21 | 21.74 | 38.24 | 54.23 | 68.57 |
| 5 | 22 | [SiH3]C1CC1 | 11.31 | 2.22 | 0.00 | 7.39 | 22.12 | 38.81 | 54.98 | 69.46 |
| 5 | 22 | [SiH3]C1CC1 | 11.02 | 2.17 | 0.00 | 7.15 | 21.38 | 37.47 | 53.00 | 66.90 |
| 5 | 22 | [SiH3]C1CC1 | 11.49 | 2.31 | 0.00 | 7.20 | 21.72 | 38.23 | 54.21 | 68.56 |
| 5 | 22 | [SiH3]C1CC1 | 11.26 | 2.20 | 0.00 | 7.42 | 22.15 | 38.84 | 55.00 | 69.48 |
| 5 | 22 | [SiH3]C1CC1 | 11.48 | 2.30 | 0.00 | 7.21 | 21.73 | 38.23 | 54.22 | 68.56 |
| 5 | 22 | [SiH3]C1CC1 | 11.30 | 2.21 | 0.00 | 7.40 | 22.12 | 38.82 | 54.98 | 69.47 |
| 5 | 22 | [SiH3]C1CC1 | 11.03 | 2.17 | 0.00 | 7.15 | 21.37 | 37.46 | 52.99 | 66.89 |
| 5 | 22 | [SiH3]C1CC1 | 11.48 | 2.30 | 0.00 | 7.21 | 21.73 | 38.23 | 54.22 | 68.56 |
| 5 | 22 | [SiH3]C1CC1 | 11.29 | 2.21 | 0.00 | 7.40 | 22.13 | 38.83 | 54.99 | 69.47 |
| 5 | 25 | OP | 6.72 | 1.44 | 0.00 | 4.25 | 13.48 | 24.55 | 35.59 | 45.79 |
| 5 | 25 | OP | 6.72 | 1.44 | 0.00 | 4.25 | 13.48 | 24.55 | 35.58 | 45.78 |
| 5 | 25 | CPC | 6.41 | 1.29 | 0.00 | 4.49 | 13.93 | 25.12 | 36.29 | 46.63 |
| 5 | 25 | PC=C | 6.55 | 1.35 | 0.00 | 4.47 | 13.94 | 25.14 | 36.22 | 46.37 |
| 5 | 25 | PC=C | 6.38 | 1.26 | 0.00 | 4.66 | 14.36 | 25.92 | 37.44 | 48.10 |
| 5 | 25 | C=CPC=C | 6.21 | 1.19 | 0.00 | 4.68 | 14.21 | 25.37 | 36.30 | 46.16 |
| 5 | 26 | BOC | 9.37 | 2.27 | 0.00 | 4.42 | 15.14 | 28.27 | 41.80 | 54.64 |
| 5 | 26 | BOC | 9.09 | 2.19 | 0.00 | 4.32 | 14.75 | 27.52 | 40.66 | 53.15 |
| 5 | 26 | COBC=C | 9.40 | 2.36 | 0.00 | 3.95 | 13.97 | 26.31 | 39.01 | 51.08 |
| 5 | 26 | COBC | 9.46 | 2.39 | 0.00 | 3.90 | 13.86 | 26.14 | 38.78 | 50.79 |
| 5 | 31 | [GeH3]C=C | 6.74 | 1.75 | 0.00 | 2.87 | 10.55 | 20.34 | 30.70 | 40.73 |
| 5 | 31 | [GeH3]C=C | 6.74 | 1.75 | 0.00 | 2.87 | 10.55 | 20.34 | 30.70 | 40.73 |
| 5 | 31 | [GeH3]C=C | 6.68 | 1.71 | 0.00 | 2.96 | 10.75 | 20.70 | 31.22 | 41.43 |
| 5 | 31 | [GeH3]C1CC1 | 6.45 | 1.66 | 0.00 | 2.89 | 10.52 | 20.29 | 30.69 | 40.86 |
| 5 | 31 | [GeH3]C1CC1 | 6.49 | 1.68 | 0.00 | 2.86 | 10.45 | 20.17 | 30.51 | 40.60 |
| 5 | 31 | [GeH3]C1CC1 | 6.51 | 1.68 | 0.00 | 2.85 | 10.45 | 20.16 | 30.51 | 40.60 |
| 5 | 31 | C[GeH3] | 6.77 | 1.76 | 0.00 | 2.86 | 10.55 | 20.38 | 30.81 | 40.94 |
| 5 | 31 | C[GeH3] | 6.77 | 1.76 | 0.00 | 2.86 | 10.55 | 20.38 | 30.81 | 40.94 |
| 5 | 31 | C[GeH3] | 6.77 | 1.76 | 0.00 | 2.86 | 10.55 | 20.38 | 30.81 | 40.94 |
| 5 | 31 | [GeH2][GeH2] | 6.11 | 1.53 | 0.00 | 2.92 | 10.35 | 19.71 | 29.54 | 38.99 |
| 5 | 31 | [GeH2][GeH2] | 6.11 | 1.53 | 0.00 | 2.92 | 10.35 | 19.71 | 29.54 | 38.99 |
| 5 | 31 | [GeH2][GeH2] | 6.11 | 1.53 | 0.00 | 2.92 | 10.35 | 19.71 | 29.54 | 39.00 |
| 5 | 31 | [GeH2][GeH2] | 6.11 | 1.53 | 0.00 | 2.92 | 10.35 | 19.71 | 29.54 | 38.99 |
| 5 | 32 | CC[SnH3] | 5.14 | 1.34 | 0.00 | 2.36 | 8.75 | 17.06 | 26.05 | 34.96 |
| 5 | 32 | CC[SnH3] | 5.14 | 1.34 | 0.00 | 2.36 | 8.75 | 17.06 | 26.05 | 34.96 |
| 5 | 32 | CC[SnH3] | 5.16 | 1.34 | 0.00 | 2.35 | 8.72 | 17.02 | 26.00 | 34.89 |
| 5 | 32 | C[SnH2]C | 5.19 | 1.36 | 0.00 | 2.30 | 8.62 | 16.88 | 25.83 | 34.70 |
| 5 | 32 | C[SnH2]C | 5.19 | 1.36 | 0.00 | 2.30 | 8.62 | 16.88 | 25.83 | 34.70 |
| 5 | 32 | C[SnH3] | 5.07 | 1.30 | 0.00 | 2.44 | 8.90 | 17.29 | 26.35 | 35.31 |
| 5 | 32 | C[SnH3] | 5.07 | 1.30 | 0.00 | 2.44 | 8.90 | 17.29 | 26.35 | 35.31 |
| 5 | 32 | C[SnH3] | 5.07 | 1.30 | 0.00 | 2.44 | 8.90 | 17.29 | 26.35 | 35.31 |
| 5 | 33 | CC[PbH3] | 1.39 | -0.19 | 0.00 | 3.97 | 10.72 | 18.70 | 26.96 | 34.94 |
| 5 | 33 | CC[PbH3] | 1.39 | -0.19 | 0.00 | 3.97 | 10.72 | 18.70 | 26.96 | 34.94 |
| 5 | 33 | CC[PbH3] | 1.33 | -0.22 | 0.00 | 4.03 | 10.83 | 18.86 | 27.17 | 35.18 |
| 5 | 33 | C[PbH2]C | 1.46 | -0.16 | 0.00 | 3.89 | 10.57 | 18.49 | 26.71 | 34.66 |
| 5 | 33 | C[PbH2]C | 1.46 | -0.16 | 0.00 | 3.89 | 10.57 | 18.49 | 26.71 | 34.66 |
| 5 | 33 | [PbH2].[PbH3].[CH3] | 3.70 | 1.00 | 0.00 | 1.52 | 5.92 | 11.80 | 18.32 | 24.94 |
| 5 | 33 | [PbH2].[PbH3].[CH3] | 3.70 | 1.00 | 0.00 | 1.52 | 5.92 | 11.81 | 18.32 | 24.95 |
| 5 | 33 | [PbH2].[PbH3].[CH3] | 4.47 | 1.32 | 0.00 | 1.16 | 5.52 | 11.51 | 18.01 | 24.15 |
| 5 | 33 | [PbH2].[PbH3].[CH3] | 4.47 | 1.32 | 0.00 | 1.17 | 5.53 | 11.52 | 18.01 | 24.16 |
| 5 | 33 | [PbH2].[PbH3].[CH3] | 4.20 | 1.18 | 0.00 | 1.50 | 6.19 | 12.48 | 19.21 | 25.57 |
| 5 | 34 | CC[SeH] | 6.76 | 1.37 | 0.00 | 4.73 | 14.51 | 26.07 | 37.68 | 48.34 |
| 5 | 34 | C[SeH] | 6.69 | 1.33 | 0.00 | 4.81 | 14.60 | 26.17 | 37.78 | 48.43 |
| 5 | 35 | CC[TeH] | 3.98 | 0.57 | 0.00 | 4.37 | 12.73 | 22.60 | 32.56 | 41.80 |
| 5 | 38 | CC1=CC1 | 14.16 | 3.24 | 0.00 | 6.79 | 21.90 | 39.32 | 56.20 | 71.25 |
| 5 | 50 | C=Cc1ccccc1 | 8.76 | 1.23 | 0.00 | 8.38 | 23.44 | 40.23 | 56.41 | 70.93 |
| 5 | 50 | C=Cc1ccccc1 | 8.67 | 1.19 | 0.00 | 8.43 | 23.52 | 40.32 | 56.52 | 71.03 |
| 5 | 50 | C=Cc1ccccc1 | 8.95 | 1.35 | 0.00 | 8.07 | 22.78 | 39.23 | 55.12 | 69.40 |
| 5 | 50 | C=Cc1ccccc1 | 9.05 | 1.37 | 0.00 | 8.14 | 23.01 | 39.65 | 55.72 | 70.16 |
| 5 | 50 | C=Cc1ccccc1 | 8.68 | 1.20 | 0.00 | 8.41 | 23.47 | 40.24 | 56.40 | 70.89 |
| 5 | 50 | Cc1ccccc1 | 8.96 | 1.35 | 0.00 | 8.09 | 22.82 | 39.27 | 55.15 | 69.41 |
| 5 | 50 | Cc1ccccc1 | 8.71 | 1.21 | 0.00 | 8.39 | 23.43 | 40.19 | 56.33 | 70.80 |
| 5 | 50 | Cc1ccccc1 | 8.75 | 1.23 | 0.00 | 8.37 | 23.40 | 40.15 | 56.30 | 70.78 |
| 5 | 50 | Cc1ccccc1 | 8.70 | 1.21 | 0.00 | 8.39 | 23.44 | 40.20 | 56.35 | 70.82 |
| 5 | 50 | Cc1ccccc1 | 8.98 | 1.36 | 0.00 | 8.04 | 22.72 | 39.13 | 54.97 | 69.22 |
| 5 | 50 | Nc1ccccc1 | 9.03 | 1.39 | 0.00 | 7.94 | 22.50 | 38.79 | 54.53 | 68.70 |
| 5 | 50 | Nc1ccccc1 | 8.66 | 1.19 | 0.00 | 8.45 | 23.56 | 40.37 | 56.58 | 71.10 |
| 5 | 50 | Nc1ccccc1 | 8.73 | 1.22 | 0.00 | 8.40 | 23.47 | 40.26 | 56.44 | 70.96 |
| 5 | 50 | Nc1ccccc1 | 8.66 | 1.18 | 0.00 | 8.45 | 23.56 | 40.38 | 56.58 | 71.10 |
| 5 | 50 | Nc1ccccc1 | 9.03 | 1.39 | 0.00 | 7.94 | 22.50 | 38.79 | 54.54 | 68.70 |
| 5 | 50 | c1ccccc1 | 8.73 | 1.22 | 0.00 | 8.37 | 23.40 | 40.14 | 56.29 | 70.76 |
| 5 | 50 | c1ccccc1 | 8.73 | 1.22 | 0.00 | 8.37 | 23.40 | 40.14 | 56.29 | 70.76 |
| 5 | 50 | c1ccccc1 | 8.73 | 1.22 | 0.00 | 8.37 | 23.40 | 40.15 | 56.29 | 70.76 |
| 5 | 50 | c1ccccc1 | 8.73 | 1.22 | 0.00 | 8.37 | 23.40 | 40.14 | 56.29 | 70.76 |
| 5 | 50 | c1ccccc1 | 8.73 | 1.22 | 0.00 | 8.37 | 23.40 | 40.14 | 56.29 | 70.76 |
| 5 | 50 | c1ccccc1 | 8.73 | 1.22 | 0.00 | 8.37 | 23.40 | 40.15 | 56.29 | 70.76 |
| 5 | 50 | Nc1ccccc1 | 8.66 | 1.18 | 0.00 | 8.45 | 23.56 | 40.38 | 56.58 | 71.10 |
| 5 | 50 | Nc1ccccc1 | 9.03 | 1.39 | 0.00 | 7.94 | 22.50 | 38.79 | 54.54 | 68.70 |
| 5 | 50 | Nc1ccccc1 | 9.03 | 1.39 | 0.00 | 7.94 | 22.50 | 38.79 | 54.53 | 68.70 |
| 5 | 50 | Nc1ccccc1 | 8.66 | 1.19 | 0.00 | 8.45 | 23.56 | 40.37 | 56.58 | 71.09 |
| 5 | 50 | Nc1ccccc1 | 8.73 | 1.22 | 0.00 | 8.40 | 23.47 | 40.26 | 56.44 | 70.96 |
| 5 | 56 | NC1CCC1 | 8.10 | 1.08 | 0.00 | 7.97 | 22.10 | 37.75 | 52.78 | 66.22 |
| 5 | 56 | NC1CCC1 | 8.47 | 1.25 | 0.00 | 7.70 | 21.57 | 36.96 | 51.72 | 64.94 |
| 5 | 56 | NC1CCC1 | 7.68 | 0.88 | 0.00 | 8.31 | 22.73 | 38.61 | 53.81 | 67.39 |
| 5 | 56 | NC1CCC1 | 8.05 | 1.06 | 0.00 | 7.99 | 22.07 | 37.63 | 52.51 | 65.79 |
| 5 | 56 | NC1CCC1 | 7.82 | 0.93 | 0.00 | 8.29 | 22.76 | 38.77 | 54.13 | 67.87 |
| 5 | 56 | NC1CCC1 | 8.48 | 1.24 | 0.00 | 7.75 | 21.72 | 37.25 | 52.17 | 65.54 |
| 5 | 56 | NC1CCC1 | 8.50 | 1.25 | 0.00 | 7.69 | 21.48 | 36.71 | 51.22 | 64.14 |
| 5 | 56 | [CH2]/C=C\1/[CH]CC1 | 8.13 | 1.10 | 0.00 | 7.89 | 21.87 | 37.29 | 52.00 | 65.04 |
| 5 | 56 | [CH2]/C=C\1/[CH]CC1 | 8.10 | 1.09 | 0.00 | 7.90 | 21.88 | 37.30 | 52.00 | 65.04 |
| 5 | 56 | [CH2]/C=C\1/[CH]CC1 | 8.14 | 1.11 | 0.00 | 7.89 | 21.91 | 37.43 | 52.28 | 65.52 |
| 5 | 56 | [CH2]/C=C\1/[CH]CC1 | 8.13 | 1.10 | 0.00 | 7.90 | 21.92 | 37.44 | 52.29 | 65.53 |
| 5 | 56 | C1CC=C1C=O | 7.94 | 1.00 | 0.00 | 8.16 | 22.51 | 38.37 | 53.57 | 67.13 |
| 5 | 56 | C1CC=C1C=O | 7.93 | 1.00 | 0.00 | 8.16 | 22.50 | 38.36 | 53.56 | 67.11 |
| 5 | 56 | C1CC=C1C=O | 8.08 | 1.07 | 0.00 | 8.02 | 22.20 | 37.88 | 52.90 | 66.27 |
| 5 | 56 | C1CC=C1C=O | 8.09 | 1.07 | 0.00 | 8.02 | 22.21 | 37.91 | 52.93 | 66.32 |
| 5 | 56 | C1CC(C1)C1CC1 | 7.49 | 0.93 | 0.00 | 7.64 | 20.96 | 35.60 | 49.59 | 62.08 |
| 5 | 56 | C1CC(C1)C1CC1 | 7.82 | 0.96 | 0.00 | 8.18 | 22.50 | 38.37 | 53.62 | 67.29 |
| 5 | 56 | C1CC(C1)C1CC1 | 8.36 | 1.22 | 0.00 | 7.61 | 21.28 | 36.41 | 50.90 | 63.85 |
| 5 | 56 | C1CC(C1)C1CC1 | 8.05 | 1.05 | 0.00 | 8.01 | 22.14 | 37.74 | 52.67 | 65.99 |
| 5 | 56 | C1CC(C1)C1CC1 | 7.57 | 0.83 | 0.00 | 8.42 | 22.97 | 39.00 | 54.38 | 68.13 |
| 5 | 56 | C1CC(C1)C1CC1 | 8.16 | 1.11 | 0.00 | 7.94 | 22.09 | 37.80 | 52.92 | 66.49 |
| 5 | 56 | C1CC(C1)C1CC1 | 8.24 | 1.16 | 0.00 | 7.75 | 21.56 | 36.83 | 51.44 | 64.48 |
| 5 | 56 | NC1CCC1 | 8.08 | 1.07 | 0.00 | 7.94 | 21.98 | 37.47 | 52.28 | 65.50 |
| 5 | 56 | NC1CCC1 | 7.57 | 0.83 | 0.00 | 8.40 | 22.92 | 38.93 | 54.27 | 67.99 |
| 5 | 56 | NC1CCC1 | 7.93 | 0.99 | 0.00 | 8.22 | 22.66 | 38.66 | 54.04 | 67.84 |
| 5 | 56 | NC1CCC1 | 8.39 | 1.20 | 0.00 | 7.79 | 21.76 | 37.25 | 52.10 | 65.36 |
| 5 | 56 | NC1CCC1 | 8.04 | 1.03 | 0.00 | 8.11 | 22.40 | 38.16 | 53.22 | 66.64 |
| 5 | 56 | NC1CCC1 | 7.84 | 0.94 | 0.00 | 8.28 | 22.76 | 38.77 | 54.15 | 67.93 |
| 5 | 56 | NC1CCC1 | 8.61 | 1.32 | 0.00 | 7.54 | 21.24 | 36.45 | 51.05 | 64.11 |
| 5 | 56 | O=CC1CCC1 | 8.08 | 1.07 | 0.00 | 7.98 | 22.08 | 37.67 | 52.59 | 65.92 |
| 5 | 56 | O=CC1CCC1 | 7.61 | 0.85 | 0.00 | 8.40 | 22.93 | 38.97 | 54.36 | 68.14 |
| 5 | 56 | O=CC1CCC1 | 7.81 | 0.95 | 0.00 | 8.20 | 22.57 | 38.48 | 53.79 | 67.53 |
| 5 | 56 | O=CC1CCC1 | 8.16 | 1.12 | 0.00 | 7.87 | 21.86 | 37.32 | 52.14 | 65.38 |
| 5 | 56 | O=CC1CCC1 | 7.79 | 0.96 | 0.00 | 8.12 | 22.34 | 38.04 | 53.08 | 66.49 |
| 5 | 56 | O=CC1CCC1 | 7.64 | 0.86 | 0.00 | 8.43 | 23.05 | 39.22 | 54.77 | 68.70 |
| 5 | 56 | O=CC1CCC1 | 8.03 | 1.04 | 0.00 | 8.05 | 22.23 | 37.90 | 52.90 | 66.29 |
| 5 | 56 | OC1CCC1 | 7.59 | 0.84 | 0.00 | 8.37 | 22.83 | 38.76 | 54.02 | 67.66 |
| 5 | 56 | OC1CCC1 | 7.54 | 0.80 | 0.00 | 8.54 | 23.29 | 39.59 | 55.24 | 69.25 |
| 5 | 56 | OC1CCC1 | 8.18 | 1.13 | 0.00 | 7.85 | 21.82 | 37.27 | 52.08 | 65.31 |
| 5 | 56 | OC1CCC1 | 8.27 | 1.18 | 0.00 | 7.65 | 21.28 | 36.28 | 50.58 | 63.29 |
| 5 | 56 | OC1CCC1 | 7.55 | 0.81 | 0.00 | 8.50 | 23.18 | 39.39 | 54.95 | 68.87 |
| 5 | 56 | OC1CCC1 | 8.37 | 1.22 | 0.00 | 7.63 | 21.37 | 36.59 | 51.18 | 64.24 |
| 5 | 56 | OC1CCC1 | 7.93 | 1.00 | 0.00 | 8.12 | 22.33 | 38.01 | 53.00 | 66.36 |
| 5 | 56 | [SiH3]C1CCC1 | 7.79 | 0.95 | 0.00 | 8.16 | 22.43 | 38.21 | 53.37 | 66.94 |
| 5 | 56 | [SiH3]C1CCC1 | 7.60 | 0.84 | 0.00 | 8.40 | 22.94 | 38.98 | 54.37 | 68.14 |
| 5 | 56 | [SiH3]C1CCC1 | 8.25 | 1.16 | 0.00 | 7.76 | 21.61 | 36.94 | 51.61 | 64.71 |
| 5 | 56 | [SiH3]C1CCC1 | 8.22 | 1.14 | 0.00 | 7.83 | 21.77 | 37.19 | 51.96 | 65.15 |
| 5 | 56 | [SiH3]C1CCC1 | 7.80 | 0.95 | 0.00 | 8.15 | 22.42 | 38.20 | 53.35 | 66.92 |
| 5 | 56 | [SiH3]C1CCC1 | 7.77 | 0.98 | 0.00 | 7.96 | 21.89 | 37.25 | 51.95 | 65.08 |
| 5 | 56 | [SiH3]C1CCC1 | 8.22 | 1.14 | 0.00 | 7.83 | 21.77 | 37.19 | 51.95 | 65.14 |
| 5 | 56 | CC1=CCC1 | 8.12 | 1.10 | 0.00 | 7.89 | 21.86 | 37.27 | 51.99 | 65.08 |
| 5 | 56 | CC1=CCC1 | 8.21 | 1.14 | 0.00 | 7.83 | 21.79 | 37.23 | 52.00 | 65.16 |
| 5 | 56 | CC1=CCC1 | 8.20 | 1.13 | 0.00 | 7.84 | 21.81 | 37.25 | 52.03 | 65.20 |
| 5 | 56 | CC1=CCC1 | 8.10 | 1.09 | 0.00 | 7.90 | 21.88 | 37.30 | 52.03 | 65.12 |
| 5 | 56 | CC1CCC1 | 8.30 | 1.18 | 0.00 | 7.73 | 21.56 | 36.86 | 51.52 | 64.62 |
| 5 | 56 | CC1CCC1 | 8.14 | 1.10 | 0.00 | 7.89 | 21.88 | 37.33 | 52.12 | 65.32 |
| 5 | 56 | CC1CCC1 | 7.60 | 0.85 | 0.00 | 8.39 | 22.89 | 38.89 | 54.22 | 67.94 |
| 5 | 56 | CC1CCC1 | 7.85 | 0.97 | 0.00 | 8.18 | 22.52 | 38.39 | 53.63 | 67.30 |
| 5 | 56 | CC1CCC1 | 8.27 | 1.17 | 0.00 | 7.74 | 21.57 | 36.87 | 51.53 | 64.63 |
| 5 | 56 | CC1CCC1 | 7.55 | 0.89 | 0.00 | 8.06 | 22.05 | 37.46 | 52.21 | 65.40 |
| 5 | 56 | CC1CCC1 | 7.84 | 0.96 | 0.00 | 8.19 | 22.52 | 38.39 | 53.64 | 67.31 |
| 5 | 56 | C1CC=C1C#C | 10.87 | 2.19 | 0.00 | 6.74 | 20.35 | 35.82 | 50.78 | 64.20 |
| 5 | 56 | C1CC=C1C#C | 11.04 | 2.27 | 0.00 | 6.60 | 20.09 | 35.45 | 50.32 | 63.69 |
| 5 | 56 | C1CC=C1C#C | 10.91 | 2.21 | 0.00 | 6.69 | 20.24 | 35.62 | 50.49 | 63.81 |
| 5 | 56 | C1CC=C1C#C | 11.07 | 2.29 | 0.00 | 6.56 | 20.00 | 35.29 | 50.09 | 63.36 |
| 5 | 56 | OC1=CCC1 | 10.93 | 2.23 | 0.00 | 6.65 | 20.16 | 35.51 | 50.36 | 63.68 |
| 5 | 56 | OC1=CCC1 | 11.12 | 2.32 | 0.00 | 6.49 | 19.86 | 35.08 | 49.83 | 63.06 |
| 5 | 56 | OC1=CCC1 | 10.94 | 2.24 | 0.00 | 6.58 | 19.92 | 35.02 | 49.56 | 62.55 |
| 5 | 56 | OC1=CCC1 | 11.10 | 2.31 | 0.00 | 6.47 | 19.75 | 34.80 | 49.33 | 62.33 |
| 5 | 56 | NC1CCC1 | 7.96 | 1.00 | 0.00 | 8.21 | 22.64 | 38.64 | 54.03 | 67.83 |
| 5 | 56 | NC1CCC1 | 8.04 | 1.03 | 0.00 | 8.11 | 22.39 | 38.15 | 53.21 | 66.63 |
| 5 | 56 | NC1CCC1 | 7.85 | 0.95 | 0.00 | 8.28 | 22.75 | 38.76 | 54.14 | 67.92 |
| 5 | 56 | NC1CCC1 | 8.61 | 1.32 | 0.00 | 7.54 | 21.24 | 36.46 | 51.06 | 64.12 |
| 5 | 56 | NC1CCC1 | 7.58 | 0.84 | 0.00 | 8.40 | 22.91 | 38.92 | 54.26 | 67.97 |
| 5 | 56 | NC1CCC1 | 8.10 | 1.08 | 0.00 | 7.94 | 21.97 | 37.46 | 52.28 | 65.50 |
| 5 | 56 | NC1CCC1 | 8.36 | 1.19 | 0.00 | 7.80 | 21.78 | 37.27 | 52.11 | 65.38 |
| 5 | 56 | O=C1CCC1 | 9.06 | 1.45 | 0.00 | 7.67 | 21.84 | 37.68 | 52.92 | 66.54 |
| 5 | 56 | O=C1CCC1 | 9.08 | 1.46 | 0.00 | 7.67 | 21.83 | 37.67 | 52.91 | 66.53 |
| 5 | 56 | O=C1CCC1 | 9.09 | 1.46 | 0.00 | 7.66 | 21.82 | 37.66 | 52.90 | 66.52 |
| 5 | 56 | O=C1CCC1 | 8.38 | 1.15 | 0.00 | 8.10 | 22.53 | 38.56 | 53.96 | 67.74 |
| 5 | 56 | O=C1CCC1 | 8.38 | 1.15 | 0.00 | 8.10 | 22.53 | 38.56 | 53.96 | 67.74 |
| 5 | 56 | O=C1CCC1 | 9.07 | 1.45 | 0.00 | 7.67 | 21.83 | 37.67 | 52.92 | 66.54 |
| 5 | 56 | C=CC1CCC1 | 7.68 | 0.90 | 0.00 | 8.24 | 22.58 | 38.40 | 53.56 | 67.11 |
| 5 | 56 | C=CC1CCC1 | 7.68 | 0.90 | 0.00 | 8.24 | 22.57 | 38.39 | 53.55 | 67.10 |
| 5 | 56 | C=CC1CCC1 | 8.45 | 1.27 | 0.00 | 7.50 | 21.06 | 36.09 | 50.49 | 63.38 |
| 5 | 56 | C=CC1CCC1 | 7.97 | 1.01 | 0.00 | 8.12 | 22.37 | 38.10 | 53.15 | 66.58 |
| 5 | 56 | C=CC1CCC1 | 7.64 | 0.85 | 0.00 | 8.44 | 23.04 | 39.15 | 54.58 | 68.37 |
| 5 | 56 | C=CC1CCC1 | 7.76 | 0.93 | 0.00 | 8.20 | 22.47 | 38.17 | 53.18 | 66.57 |
| 5 | 56 | C=CC1CCC1 | 8.44 | 1.27 | 0.00 | 7.50 | 21.06 | 36.09 | 50.49 | 63.37 |
| 5 | 57 | [CH2]/C=C\1/[CH]CC1 | 10.38 | 1.90 | 0.00 | 7.54 | 22.09 | 38.50 | 54.36 | 68.59 |
| 5 | 57 | C1CC=C1C=O | 10.37 | 1.89 | 0.00 | 7.59 | 22.21 | 38.72 | 54.69 | 69.03 |
| 5 | 57 | CC1=CCC1 | 10.32 | 1.86 | 0.00 | 7.64 | 22.30 | 38.81 | 54.75 | 69.04 |
| 5 | 57 | C1CC=C1C#C | 11.86 | 2.45 | 0.00 | 7.15 | 21.82 | 38.60 | 54.92 | 69.62 |
| 5 | 57 | OC1=CCC1 | 12.01 | 2.54 | 0.00 | 6.88 | 21.22 | 37.63 | 53.60 | 67.99 |
| 5 | 153 | O=P | 9.40 | 2.45 | 0.00 | 3.58 | 13.10 | 24.83 | 36.54 | 47.22 |
| 5 | 153 | O=P | 9.40 | 2.45 | 0.00 | 3.58 | 13.10 | 24.83 | 36.54 | 47.22 |
| 5 | 153 | OP(=O)O | 11.98 | 3.43 | 0.00 | 2.88 | 12.74 | 25.43 | 38.59 | 50.87 |
| 5 | 153 | COP(=O)OC | 10.40 | 2.83 | 0.00 | 3.35 | 13.08 | 25.38 | 38.06 | 49.96 |
| 5 | 153 | OP=O | 9.40 | 2.43 | 0.00 | 3.74 | 13.59 | 25.87 | 38.44 | 50.19 |
| 5 | 153 | OP=O | 9.62 | 2.55 | 0.00 | 3.44 | 12.95 | 24.91 | 37.18 | 48.69 |
| 5 | 153 | O=P | 9.40 | 2.45 | 0.00 | 3.58 | 13.10 | 24.83 | 36.54 | 47.22 |
| 5 | 153 | O=P | 9.40 | 2.45 | 0.00 | 3.58 | 13.10 | 24.83 | 36.54 | 47.22 |
| 5 | 153 | O=P | 9.40 | 2.45 | 0.00 | 3.58 | 13.10 | 24.83 | 36.54 | 47.22 |
| 6 | 6 | COOC | 6.07 | 0.65 | 0.00 | 6.28 | 16.39 | 26.34 | 34.45 | 40.32 |
| 6 | 19 | CO[SiH2]OC | 9.53 | 2.14 | 0.00 | 6.95 | 22.25 | 38.81 | 55.11 | 70.49 |
| 6 | 19 | CO[SiH2]OC | 8.22 | 1.61 | 0.00 | 7.71 | 23.19 | 39.71 | 55.95 | 71.23 |
| 6 | 19 | O[SiH3] | 7.25 | 1.23 | 0.00 | 8.21 | 23.63 | 40.11 | 56.67 | 72.36 |
| 6 | 21 | OCC(O)(C)C | 22.85 | 5.31 | 0.00 | 8.33 | 26.82 | 47.09 | 65.80 | 81.72 |
| 6 | 21 | OCC(O)(C)C | 23.50 | 5.50 | 0.00 | 8.36 | 27.04 | 47.42 | 66.02 | 81.53 |
| 6 | 21 | CCO | 23.05 | 5.30 | 0.00 | 8.61 | 27.41 | 47.76 | 66.16 | 81.33 |
| 6 | 21 | CO | 22.81 | 5.19 | 0.00 | 8.80 | 27.76 | 48.23 | 66.70 | 81.88 |
| 6 | 21 | OP | 23.55 | 5.50 | 0.00 | 8.55 | 27.58 | 48.25 | 66.84 | 81.81 |
| 6 | 21 | O | 23.53 | 5.45 | 0.00 | 8.77 | 28.21 | 49.60 | 69.35 | 86.06 |
| 6 | 21 | O | 23.53 | 5.45 | 0.00 | 8.77 | 28.22 | 49.60 | 69.35 | 86.06 |
| 6 | 21 | OC1CCC1 | 23.21 | 5.39 | 0.00 | 8.41 | 26.95 | 46.95 | 64.95 | 79.72 |
| 6 | 21 | OCC(O)(C)C | 22.85 | 5.31 | 0.00 | 8.33 | 26.82 | 47.09 | 65.80 | 81.72 |
| 6 | 21 | OCC(O)(C)C | 23.50 | 5.50 | 0.00 | 8.36 | 27.04 | 47.42 | 66.02 | 81.53 |
| 6 | 21 | O[SiH3] | 22.99 | 5.20 | 0.00 | 9.15 | 28.89 | 50.59 | 70.67 | 87.72 |
| 6 | 24 | OC=O | 17.10 | 3.23 | 0.00 | 10.01 | 28.77 | 48.93 | 67.57 | 83.58 |
| 6 | 24 | CC(=O)O | 16.94 | 3.16 | 0.00 | 10.10 | 28.92 | 49.10 | 67.72 | 83.68 |
| 6 | 25 | OP | 16.47 | 4.87 | 0.00 | 1.51 | 12.04 | 26.19 | 39.84 | 52.60 |
| 6 | 26 | BOC | -5.32 | -4.85 | 0.00 | 16.64 | 36.73 | 56.73 | 75.06 | 91.08 |
| 6 | 26 | COBC=C | -1.42 | -2.94 | 0.00 | 13.15 | 30.27 | 47.83 | 64.25 | 78.85 |
| 6 | 26 | COBC | -1.37 | -2.90 | 0.00 | 13.08 | 30.13 | 47.65 | 64.05 | 78.69 |
| 6 | 56 | OC1CCC1 | 13.49 | 3.16 | 0.00 | 5.96 | 19.47 | 35.06 | 50.23 | 63.93 |
| 6 | 57 | OC1=CCC1 | 36.28 | 11.50 | 0.00 | -0.34 | 13.45 | 31.59 | 49.82 | 66.37 |
| 6 | 73 | Oc1ccccc1 | 15.69 | 2.53 | 0.00 | 11.24 | 30.84 | 51.35 | 69.74 | 84.92 |
| 6 | 73 | OC1=CCC1 | 15.76 | 2.65 | 0.00 | 10.73 | 29.63 | 49.31 | 66.81 | 81.09 |
| 6 | 153 | COP(=O)OC | -10.47 | -8.37 | 0.00 | 24.73 | 52.50 | 79.65 | 102.85 | 122.00 |
| 7 | 17 | CS(=O)C | 15.86 | 4.09 | 0.00 | 6.95 | 26.22 | 46.05 | 63.55 | 79.36 |
| 7 | 17 | CS=O | 15.21 | 3.71 | 0.00 | 7.79 | 27.48 | 47.18 | 64.27 | 79.51 |
| 7 | 17 | CC[S@@](=O)C | 16.19 | 4.25 | 0.00 | 6.77 | 25.88 | 45.60 | 63.08 | 78.88 |
| 7 | 17 | CS(=O)C | 15.86 | 4.09 | 0.00 | 6.95 | 26.22 | 46.05 | 63.55 | 79.36 |
| 7 | 18 | CS(=O)(=O)C | 18.12 | 3.30 | 0.00 | 10.09 | 34.03 | 60.02 | 82.20 | 101.33 |
| 7 | 18 | CS(=O)(=O)C | 18.13 | 3.30 | 0.00 | 10.09 | 34.03 | 60.01 | 82.20 | 101.32 |
| 7 | 18 | C=CS(=O)(=O)C=C | 18.60 | 3.50 | 0.00 | 10.40 | 34.88 | 60.85 | 82.83 | 101.96 |
| 7 | 18 | C=CS(=O)(=O)C=C | 18.58 | 3.49 | 0.00 | 10.41 | 34.89 | 60.86 | 82.84 | 101.96 |
| 7 | 18 | CCS(=O)(=O)C | 18.73 | 3.61 | 0.00 | 9.66 | 33.37 | 59.20 | 81.34 | 100.48 |
| 7 | 18 | CCS(=O)(=O)C | 18.33 | 3.40 | 0.00 | 9.79 | 33.48 | 59.25 | 81.20 | 100.19 |
| 7 | 18 | C=CS(=O)(=O)C | 18.48 | 3.47 | 0.00 | 10.33 | 34.79 | 60.86 | 82.99 | 102.20 |
| 7 | 18 | C=CS(=O)(=O)C | 18.27 | 3.35 | 0.00 | 10.07 | 34.00 | 59.91 | 81.94 | 100.93 |
| 7 | 18 | CS(=O)(=O)C | 18.13 | 3.30 | 0.00 | 10.09 | 34.03 | 60.01 | 82.19 | 101.32 |
| 7 | 18 | CS(=O)(=O)C | 18.12 | 3.30 | 0.00 | 10.09 | 34.04 | 60.02 | 82.20 | 101.33 |
| 7 | 18 | CS(=O)=O | 17.97 | 3.16 | 0.00 | 10.85 | 35.76 | 62.39 | 84.90 | 104.25 |
| 7 | 18 | CS(=O)=O | 17.97 | 3.16 | 0.00 | 10.85 | 35.76 | 62.40 | 84.90 | 104.25 |
| 7 | 18 | C=CNS(=O)(=O)N | 17.01 | 2.59 | 0.00 | 11.86 | 37.18 | 63.71 | 86.04 | 105.29 |
| 7 | 18 | CS(=O)(=O)C | 18.12 | 3.30 | 0.00 | 10.09 | 34.03 | 60.02 | 82.20 | 101.33 |
| 7 | 18 | CS(=O)(=O)C | 18.13 | 3.30 | 0.00 | 10.09 | 34.03 | 60.01 | 82.19 | 101.32 |
| 7 | 46 | O=N(=O)C1CC1 | 23.00 | 4.34 | 0.00 | 12.95 | 36.98 | 61.89 | 84.00 | 102.31 |
| 7 | 46 | O=N(=O)C1CC1 | 22.99 | 4.33 | 0.00 | 12.94 | 36.97 | 61.88 | 83.99 | 102.29 |
| 7 | 46 | C=CN(=O)=O | 22.52 | 4.13 | 0.00 | 13.10 | 36.80 | 60.89 | 81.94 | 99.19 |
| 7 | 46 | C=CN(=O)=O | 23.98 | 4.72 | 0.00 | 12.56 | 36.49 | 61.53 | 83.92 | 102.52 |
| 7 | 46 | CN(=O)=O | 22.44 | 4.08 | 0.00 | 13.32 | 37.64 | 62.81 | 85.11 | 103.58 |
| 7 | 46 | CN(=O)=O | 22.31 | 4.00 | 0.00 | 13.52 | 37.95 | 63.05 | 85.13 | 103.28 |
| 7 | 46 | O=N(=O)C1CC1 | 23.04 | 4.34 | 0.00 | 12.96 | 37.01 | 61.94 | 84.06 | 102.39 |
| 7 | 46 | O=N(=O)C1CC1 | 23.01 | 4.34 | 0.00 | 12.96 | 37.00 | 61.92 | 84.05 | 102.37 |
| 7 | 46 | CC(CN(=O)=O)C | 22.43 | 4.08 | 0.00 | 13.31 | 37.45 | 62.19 | 84.00 | 102.05 |
| 7 | 46 | CC(CN(=O)=O)C | 22.85 | 4.29 | 0.00 | 12.97 | 37.05 | 62.02 | 84.24 | 102.81 |
| 7 | 46 | CCN(=O)=O | 22.60 | 4.16 | 0.00 | 13.20 | 37.38 | 62.31 | 84.35 | 102.60 |
| 7 | 46 | CCN(=O)=O | 22.58 | 4.15 | 0.00 | 13.21 | 37.39 | 62.32 | 84.35 | 102.60 |
| 7 | 58 | O=C1CCC1 | 29.01 | 5.68 | 0.00 | 16.53 | 47.86 | 81.34 | 112.21 | 138.88 |
| 7 | 67 | O=C1CC1 | 32.22 | 6.86 | 0.00 | 15.52 | 46.54 | 79.89 | 110.68 | 137.41 |
| 7 | 106 | C=CC=C=O | 35.14 | 6.96 | 0.00 | 19.07 | 55.00 | 92.73 | 126.65 | 155.01 |
| 7 | 153 | OP=O | 20.87 | 4.11 | 0.00 | 14.30 | 39.86 | 67.81 | 94.23 | 116.14 |
| 7 | 154 | CS(=O)(=O)N | 11.19 | 0.79 | 0.00 | 13.92 | 40.10 | 66.30 | 88.29 | 107.25 |
| 7 | 154 | CS(=O)(=O)N | 11.18 | 0.79 | 0.00 | 13.92 | 40.10 | 66.30 | 88.29 | 107.26 |
| 7 | 154 | CNS(=O)(=O)C | 13.49 | 0.87 | 0.00 | 16.71 | 42.22 | 67.73 | 90.14 | 108.92 |
| 7 | 154 | CNS(=O)(=O)C | 13.03 | 0.69 | 0.00 | 16.83 | 42.31 | 67.73 | 90.00 | 108.62 |
| 7 | 154 | CS(=O)(=O)N | 11.28 | 0.83 | 0.00 | 13.88 | 40.05 | 66.23 | 88.22 | 107.21 |
| 7 | 154 | CS(=O)(=O)N | 11.31 | 0.84 | 0.00 | 13.86 | 40.01 | 66.19 | 88.18 | 107.18 |
| 7 | 154 | CNS(=O)(=O)C | 11.59 | 0.98 | 0.00 | 13.61 | 39.41 | 65.23 | 86.98 | 105.84 |
| 7 | 154 | CNS(=O)(=O)C | 10.97 | 0.77 | 0.00 | 13.76 | 39.67 | 65.52 | 87.13 | 105.77 |
| 8 | 23 | NC1CCC1 | 12.61 | 2.03 | 0.00 | 9.73 | 27.11 | 45.74 | 62.79 | 77.04 |
| 8 | 23 | NC1CCC1 | 12.51 | 1.99 | 0.00 | 9.73 | 27.05 | 45.57 | 62.46 | 76.55 |
| 8 | 23 | NC=C | 13.80 | 2.27 | 0.00 | 10.51 | 29.64 | 50.58 | 70.34 | 87.68 |
| 8 | 23 | NC=C | 13.27 | 2.02 | 0.00 | 10.98 | 30.56 | 51.90 | 72.02 | 89.64 |
| 8 | 23 | CNC=C | 14.48 | 2.70 | 0.00 | 9.24 | 26.64 | 45.44 | 62.56 | 76.62 |
| 8 | 23 | CCN | 15.02 | 2.96 | 0.00 | 8.80 | 25.87 | 44.43 | 61.51 | 75.83 |
| 8 | 23 | CCN | 14.82 | 2.86 | 0.00 | 9.00 | 26.26 | 45.02 | 62.27 | 76.74 |
| 8 | 23 | CN | 14.75 | 2.83 | 0.00 | 9.05 | 26.33 | 45.07 | 62.26 | 76.62 |
| 8 | 23 | CN | 14.74 | 2.82 | 0.00 | 9.05 | 26.33 | 45.07 | 62.26 | 76.62 |
| 8 | 23 | CNC | 14.35 | 2.69 | 0.00 | 9.00 | 25.91 | 44.04 | 60.45 | 73.94 |
| 8 | 23 | NC1CCC1 | 12.62 | 2.05 | 0.00 | 9.59 | 26.76 | 45.15 | 61.96 | 76.03 |
| 8 | 23 | NC1CCC1 | 12.40 | 1.93 | 0.00 | 9.87 | 27.31 | 45.93 | 62.87 | 76.93 |
| 8 | 23 | C1CCCN1 | 14.20 | 2.62 | 0.00 | 9.13 | 26.09 | 44.16 | 60.36 | 73.53 |
| 8 | 23 | C1CCCNC1 | 14.38 | 2.71 | 0.00 | 8.96 | 25.83 | 43.93 | 60.33 | 73.88 |
| 8 | 23 | Nc1ccccc1 | 14.91 | 2.87 | 0.00 | 9.16 | 26.76 | 45.95 | 63.62 | 78.33 |
| 8 | 23 | Nc1ccccc1 | 14.92 | 2.87 | 0.00 | 9.16 | 26.75 | 45.95 | 63.62 | 78.35 |
| 8 | 23 | NC1CCC1 | 12.42 | 1.94 | 0.00 | 9.86 | 27.30 | 45.92 | 62.87 | 76.93 |
| 8 | 23 | NC1CCC1 | 12.66 | 2.07 | 0.00 | 9.58 | 26.74 | 45.13 | 61.94 | 76.02 |
| 8 | 23 | Nc1ccccc1 | 14.91 | 2.86 | 0.00 | 9.16 | 26.76 | 45.95 | 63.62 | 78.33 |
| 8 | 23 | Nc1ccccc1 | 14.92 | 2.87 | 0.00 | 9.16 | 26.75 | 45.95 | 63.62 | 78.34 |
| 8 | 50 | Nc1ccccc1 | 15.84 | 3.79 | 0.00 | 6.86 | 22.80 | 41.30 | 59.25 | 75.17 |
| 8 | 50 | Nc1ccccc1 | 15.84 | 3.79 | 0.00 | 6.86 | 22.80 | 41.29 | 59.25 | 75.16 |
| 8 | 56 | NC1CCC1 | 9.50 | 1.63 | 0.00 | 7.73 | 22.33 | 38.73 | 54.47 | 68.38 |
| 8 | 56 | NC1CCC1 | 10.84 | 2.23 | 0.00 | 6.74 | 20.53 | 36.23 | 51.40 | 64.89 |
| 8 | 56 | NC1CCC1 | 10.83 | 2.23 | 0.00 | 6.74 | 20.54 | 36.24 | 51.41 | 64.90 |
| 9 | 18 | C=CNS(=O)(=O)N | 9.37 | 1.71 | 0.00 | 7.88 | 21.79 | 37.67 | 53.55 | 67.92 |
| 9 | 28 | NC=O | 11.28 | 1.36 | 0.00 | 11.17 | 30.23 | 50.84 | 70.23 | 87.23 |
| 9 | 28 | NC=O | 10.58 | 1.03 | 0.00 | 11.73 | 31.21 | 52.10 | 71.61 | 88.57 |
| 9 | 28 | CC(=O)N | 11.05 | 1.25 | 0.00 | 11.39 | 30.66 | 51.49 | 71.10 | 88.33 |
| 9 | 28 | CC(=O)N | 10.62 | 1.05 | 0.00 | 11.71 | 31.17 | 52.01 | 71.43 | 88.24 |
| 9 | 28 | CNC=O | 11.07 | 1.30 | 0.00 | 11.09 | 29.87 | 50.09 | 69.01 | 85.50 |
| 9 | 28 | NC=C | 10.36 | 0.91 | 0.00 | 11.98 | 31.71 | 52.85 | 72.64 | 89.92 |
| 9 | 28 | NC=C | 9.77 | 0.64 | 0.00 | 12.49 | 32.70 | 54.28 | 74.44 | 92.00 |
| 9 | 28 | CNC(=O)C | 10.95 | 1.26 | 0.00 | 11.14 | 29.96 | 50.23 | 69.24 | 85.85 |
| 9 | 28 | CC(=O)N | 11.05 | 1.25 | 0.00 | 11.39 | 30.67 | 51.50 | 71.10 | 88.33 |
| 9 | 28 | CC(=O)N | 10.63 | 1.05 | 0.00 | 11.71 | 31.17 | 52.01 | 71.43 | 88.24 |
| 9 | 28 | CCC(=O)N | 11.07 | 1.26 | 0.00 | 11.36 | 30.59 | 51.37 | 70.91 | 88.06 |
| 9 | 28 | CCC(=O)N | 10.68 | 1.07 | 0.00 | 11.67 | 31.09 | 51.90 | 71.27 | 88.03 |
| 15 | 15 | SS | 207.77 | 188.37 | 0.00 | 141.20 | 118.55 | 71.54 | 80.38 | 81.85 |
| 15 | 15 | CSSC | 5.62 | 1.34 | 0.00 | 2.54 | 10.65 | 21.01 | 30.71 | 39.68 |
| 15 | 15 | CCSSC | 5.60 | 1.35 | 0.00 | 2.55 | 10.73 | 21.04 | 30.66 | 39.63 |
| 15 | 44 | SS | 178.71 | 172.33 | 0.00 | 176.26 | 184.78 | 203.20 | 216.22 | 228.23 |
| 15 | 44 | SS | 178.71 | 172.33 | 0.00 | 176.26 | 184.78 | 203.20 | 216.22 | 228.23 |
| 15 | 44 | CCS | 8.31 | 1.72 | 0.00 | 5.04 | 16.68 | 30.36 | 43.56 | 55.49 |
| 15 | 44 | CS | 8.11 | 1.64 | 0.00 | 5.19 | 16.98 | 30.77 | 44.03 | 55.97 |
| 15 | 56 | C1CCS1 | 12.29 | 3.76 | 0.00 | 1.86 | 11.40 | 25.10 | 40.01 | 54.95 |
| 19 | 19 | [SiH3][SiH2][SiH3] | 3.16 | 0.60 | 0.00 | 2.90 | 9.22 | 17.19 | 25.69 | 34.23 |
| 19 | 19 | [SiH3][SiH2][SiH3] | 3.06 | 0.56 | 0.00 | 2.96 | 9.31 | 17.30 | 25.80 | 34.33 |
| 19 | 19 | [SiH3][SiH3] | 3.15 | 0.60 | 0.00 | 2.96 | 9.43 | 17.59 | 26.24 | 34.91 |
| 19 | 22 | [SiH3]C1CC1 | 3.69 | 0.16 | 0.00 | 5.99 | 16.21 | 28.31 | 40.77 | 52.60 |
| 19 | 22 | [SiH3]C1CC1 | 3.70 | 0.16 | 0.00 | 5.99 | 16.20 | 28.30 | 40.76 | 52.59 |
| 19 | 56 | [SiH3]C1CCC1 | 0.68 | -0.95 | 0.00 | 6.66 | 16.62 | 27.81 | 38.99 | 49.55 |
| 21 | 145 | CNO | 15.09 | 2.35 | 0.00 | 11.18 | 30.34 | 49.96 | 66.92 | 80.08 |
| 21 | 145 | NO | 15.13 | 2.35 | 0.00 | 11.19 | 30.32 | 49.82 | 66.60 | 79.59 |
| 21 | 159 | OP(=O)O | 24.37 | 5.89 | 0.00 | 7.97 | 26.91 | 48.19 | 68.30 | 85.95 |
| 21 | 159 | OP(=O)O | 23.82 | 5.63 | 0.00 | 8.45 | 27.78 | 49.36 | 69.65 | 87.27 |
| 21 | 159 | OP=O | 24.01 | 5.71 | 0.00 | 8.30 | 27.46 | 48.85 | 68.99 | 86.54 |
| 22 | 22 | O=N(=O)C1CC1 | 18.11 | 5.47 | 0.00 | 3.21 | 17.41 | 36.83 | 58.04 | 78.98 |
| 22 | 22 | O=N(=O)C1CC1 | 16.83 | 4.78 | 0.00 | 4.56 | 19.80 | 39.81 | 61.18 | 81.90 |
| 22 | 22 | O=N(=O)C1CC1 | 18.18 | 5.50 | 0.00 | 3.18 | 17.39 | 36.83 | 58.08 | 79.06 |
| 22 | 22 | O=N(=O)C1CC1 | 16.39 | 4.57 | 0.00 | 4.89 | 20.36 | 40.54 | 62.03 | 82.82 |
| 22 | 22 | CC1CC1 | 10.57 | 2.28 | 0.00 | 7.03 | 22.75 | 42.26 | 62.80 | 82.77 |
| 22 | 22 | CC1CC1 | 10.58 | 2.29 | 0.00 | 7.02 | 22.73 | 42.23 | 62.76 | 82.73 |
| 22 | 22 | O=CC1CC1 | 10.55 | 2.33 | 0.00 | 6.83 | 22.32 | 41.62 | 61.93 | 81.61 |
| 22 | 22 | O=CC1CC1 | 10.52 | 2.31 | 0.00 | 6.85 | 22.36 | 41.68 | 62.00 | 81.69 |
| 22 | 22 | O=C1CC1 | 7.76 | 1.46 | 0.00 | 6.75 | 20.92 | 38.29 | 56.27 | 73.37 |
| 22 | 22 | C1CC(C1)C1CC1 | 11.29 | 2.59 | 0.00 | 6.63 | 22.14 | 41.50 | 61.80 | 81.37 |
| 22 | 22 | C1CC(C1)C1CC1 | 11.03 | 2.46 | 0.00 | 6.91 | 22.65 | 42.21 | 62.73 | 82.55 |
| 22 | 22 | C=CC1CC1 | 10.19 | 2.12 | 0.00 | 7.25 | 23.03 | 42.40 | 62.52 | 81.81 |
| 22 | 22 | [GeH3]C1CC1 | 11.13 | 2.54 | 0.00 | 6.67 | 22.25 | 41.78 | 62.40 | 82.45 |
| 22 | 22 | [GeH3]C1CC1 | 11.12 | 2.54 | 0.00 | 6.67 | 22.25 | 41.78 | 62.40 | 82.45 |
| 22 | 22 | [SiH3]C1CC1 | 11.37 | 2.64 | 0.00 | 6.54 | 22.03 | 41.51 | 62.09 | 82.11 |
| 22 | 22 | [SiH3]C1CC1 | 11.36 | 2.63 | 0.00 | 6.56 | 22.06 | 41.54 | 62.13 | 82.16 |
| 22 | 22 | [SiH3]C1CC1 | 11.37 | 2.64 | 0.00 | 6.54 | 22.03 | 41.51 | 62.10 | 82.12 |
| 22 | 22 | [SiH3]C1CC1 | 11.37 | 2.64 | 0.00 | 6.55 | 22.04 | 41.52 | 62.11 | 82.13 |
| 22 | 31 | [GeH3]C1CC1 | 7.22 | 1.89 | 0.00 | 2.98 | 11.15 | 21.62 | 32.70 | 43.50 |
| 22 | 38 | [CH]C=C.[CH2].[C] | 28.83 | 10.47 | 0.00 | -5.32 | 1.55 | 14.41 | 29.76 | 45.66 |
| 22 | 38 | C1=CC1 | -2.31 | -3.89 | 0.00 | 18.52 | 45.10 | 74.76 | 104.02 | 129.62 |
| 22 | 46 | O=N(=O)C1CC1 | 10.14 | 2.30 | 0.00 | 4.87 | 15.44 | 27.42 | 38.91 | 49.05 |
| 22 | 46 | O=N(=O)C1CC1 | 9.66 | 2.09 | 0.00 | 5.16 | 15.90 | 28.00 | 39.55 | 49.73 |
| 22 | 56 | C1CC(C1)C1CC1 | 9.39 | 1.89 | 0.00 | 6.37 | 19.46 | 34.66 | 49.68 | 63.45 |
| 22 | 67 | O=C1CC1 | 9.78 | 1.84 | 0.00 | 7.44 | 22.58 | 40.80 | 59.83 | 78.47 |
| 22 | 67 | O=C1CC1 | 9.77 | 1.84 | 0.00 | 7.44 | 22.58 | 40.81 | 59.83 | 78.48 |
| 23 | 40 | c1ccc[nH]1 | 9.88 | 0.82 | 0.00 | 11.66 | 30.77 | 51.18 | 70.23 | 86.81 |
| 23 | 43 | [NH]N[O] | 10.95 | 1.70 | 0.00 | 8.36 | 22.54 | 36.82 | 49.01 | 58.58 |
| 23 | 43 | [CH]/C(=C\C=C\[CH])/[N]N[O] | 21.97 | 6.02 | 0.00 | 3.78 | 16.03 | 29.45 | 41.20 | 50.49 |
| 23 | 72 | CC=N | 15.43 | 3.29 | 0.00 | 7.62 | 23.00 | 39.54 | 54.45 | 66.72 |
| 23 | 72 | CC=N | 15.44 | 3.29 | 0.00 | 7.62 | 23.00 | 39.54 | 54.44 | 66.72 |
| 23 | 109 | [NH]N[O] | 13.05 | 2.36 | 0.00 | 8.64 | 24.57 | 41.56 | 56.95 | 69.71 |
| 23 | 109 | C=CON=N | 23.61 | 6.27 | 0.00 | 5.34 | 20.57 | 37.25 | 51.64 | 62.45 |
| 23 | 146 | CNO | 13.72 | 2.55 | 0.00 | 8.57 | 24.40 | 41.02 | 55.68 | 67.46 |
| 23 | 146 | CONC | 13.91 | 2.63 | 0.00 | 8.56 | 24.54 | 41.48 | 56.59 | 68.89 |
| 23 | 146 | NO | 14.23 | 2.70 | 0.00 | 8.78 | 25.33 | 43.08 | 59.14 | 72.41 |
| 23 | 146 | NO | 14.19 | 2.69 | 0.00 | 8.79 | 25.35 | 43.09 | 59.14 | 72.40 |
| 23 | 150 | CNN | 13.05 | 2.20 | 0.00 | 9.43 | 26.29 | 44.08 | 59.97 | 72.91 |
| 23 | 150 | CNN | 13.61 | 2.42 | 0.00 | 9.25 | 26.22 | 44.31 | 60.65 | 74.12 |
| 23 | 150 | CNN | 14.27 | 2.81 | 0.00 | 8.17 | 23.78 | 40.41 | 55.43 | 67.96 |
| 23 | 150 | CNNC | 13.91 | 2.67 | 0.00 | 8.37 | 24.15 | 40.89 | 55.77 | 67.75 |
| 23 | 150 | CNNC | 13.92 | 2.67 | 0.00 | 8.37 | 24.14 | 40.88 | 55.76 | 67.73 |
| 23 | 150 | NN | 14.30 | 2.79 | 0.00 | 8.46 | 24.66 | 42.06 | 57.76 | 70.56 |
| 23 | 150 | NN | 14.30 | 2.79 | 0.00 | 8.46 | 24.66 | 42.06 | 57.76 | 70.56 |
| 23 | 150 | NN | 14.30 | 2.79 | 0.00 | 8.46 | 24.66 | 42.06 | 57.76 | 70.56 |
| 23 | 150 | NN | 14.30 | 2.79 | 0.00 | 8.46 | 24.66 | 42.06 | 57.76 | 70.56 |
| 23 | 155 | CS(=O)(=O)N | 14.33 | 2.67 | 0.00 | 9.31 | 27.07 | 46.67 | 65.19 | 81.38 |
| 23 | 155 | CS(=O)(=O)N | 14.35 | 2.68 | 0.00 | 9.30 | 27.06 | 46.65 | 65.17 | 81.36 |
| 23 | 155 | CNS(=O)(=O)C | 13.87 | 2.49 | 0.00 | 9.41 | 26.99 | 46.16 | 63.99 | 79.17 |
| 23 | 155 | CS(=O)(=O)N | 14.36 | 2.68 | 0.00 | 9.31 | 27.07 | 46.69 | 65.25 | 81.48 |
| 23 | 155 | CS(=O)(=O)N | 14.34 | 2.68 | 0.00 | 9.31 | 27.08 | 46.70 | 65.25 | 81.47 |
| 23 | 155 | CNS(=O)(=O)C | 13.89 | 2.51 | 0.00 | 9.37 | 26.92 | 46.06 | 63.86 | 79.04 |
| 24 | 75 | OC=O | 15.73 | 2.61 | 0.00 | 10.94 | 30.27 | 50.64 | 69.09 | 84.52 |
| 24 | 75 | CC(=O)O | 16.33 | 2.92 | 0.00 | 10.35 | 29.26 | 49.50 | 68.15 | 84.11 |
| 31 | 31 | [GeH2][GeH2] | 0.03 | -0.79 | 0.00 | 4.75 | 11.85 | 19.85 | 27.80 | 35.02 |
| 33 | 33 | C[Pb]=[Pb] | 95.59 | 42.82 | 0.00 | -61.48 | -99.05 | -115.33 | -131.64 | -136.09 |
| 33 | 33 | [PbH2].[PbH3].[CH3] | 124.69 | 57.13 | 0.00 | -87.80 | -147.58 | -186.68 | -210.90 | -224.68 |
| 38 | 38 | [CH]C=C.[CH2].[C] | 8.74 | 2.12 | 0.00 | 3.90 | 12.99 | 23.64 | 34.50 | 44.46 |
| 39 | 40 | [NH3]n1cccc1 | 73.95 | 27.55 | 0.00 | -21.16 | -19.06 | -8.00 | 5.38 | 17.76 |
| 39 | 48 | [NH3]C=C | 9.28 | 1.10 | 0.00 | 9.06 | 24.18 | 39.95 | 54.01 | 65.37 |
| 39 | 48 | [NH3]C=C | 8.10 | 0.46 | 0.00 | 10.61 | 27.65 | 45.82 | 62.89 | 78.01 |
| 39 | 48 | [NH3]C=C | 9.58 | 1.26 | 0.00 | 8.70 | 23.37 | 38.63 | 52.07 | 62.75 |
| 39 | 48 | C[NH3] | 7.97 | 0.38 | 0.00 | 10.86 | 28.20 | 46.70 | 64.07 | 79.41 |
| 39 | 48 | C[NH3] | 7.97 | 0.38 | 0.00 | 10.86 | 28.20 | 46.70 | 64.07 | 79.41 |
| 39 | 48 | C[NH3] | 7.97 | 0.38 | 0.00 | 10.86 | 28.20 | 46.70 | 64.07 | 79.41 |
| 39 | 48 | [NH4] | 8.40 | 0.54 | 0.00 | 10.80 | 28.31 | 47.18 | 65.08 | 81.13 |
| 39 | 48 | [NH4] | 8.40 | 0.54 | 0.00 | 10.80 | 28.31 | 47.18 | 65.08 | 81.13 |
| 39 | 48 | [NH4] | 8.40 | 0.54 | 0.00 | 10.79 | 28.31 | 47.18 | 65.08 | 81.13 |
| 39 | 48 | [NH4] | 8.40 | 0.54 | 0.00 | 10.80 | 28.31 | 47.18 | 65.08 | 81.13 |
| 39 | 48 | [NH3]n1cccc1 | 8.41 | 0.65 | 0.00 | 10.14 | 26.53 | 43.84 | 59.64 | 72.60 |
| 39 | 48 | [NH3]n1cccc1 | 11.92 | 2.51 | 0.00 | 5.97 | 17.60 | 29.11 | 37.62 | 42.34 |
| 39 | 48 | [NH3]n1cccc1 | 8.41 | 0.65 | 0.00 | 10.14 | 26.54 | 43.85 | 59.65 | 72.61 |
| 41 | 73 | OC=C | 18.24 | 3.46 | 0.00 | 10.52 | 30.05 | 50.57 | 68.77 | 83.40 |
| 41 | 73 | C/C=C/O | 18.15 | 3.41 | 0.00 | 10.59 | 30.17 | 50.71 | 68.90 | 83.51 |
| 41 | 73 | ON=C(C)C | 10.42 | 0.59 | 0.00 | 12.83 | 32.57 | 52.49 | 69.82 | 83.50 |
| 41 | 108 | ON=C(C)C | 10.12 | 2.00 | 0.00 | 5.94 | 17.21 | 29.24 | 40.04 | 48.73 |
| 43 | 69 | C[N]([O])(C)C | 34.58 | 11.33 | 0.00 | -2.61 | 7.35 | 21.44 | 36.26 | 50.29 |
| 43 | 69 | C[N]N([O])C | 13.44 | 1.23 | 0.00 | 14.11 | 35.72 | 56.68 | 74.80 | 89.75 |
| 43 | 69 | [NH]N[O] | 10.97 | -0.05 | 0.00 | 16.67 | 40.67 | 63.50 | 82.62 | 97.67 |
| 43 | 69 | C[N]N([O])C | 13.44 | 1.23 | 0.00 | 14.11 | 35.72 | 56.68 | 74.80 | 89.75 |
| 43 | 69 | C[N]N[O] | 12.72 | 0.76 | 0.00 | 15.33 | 38.22 | 60.15 | 78.67 | 93.42 |
| 43 | 69 | [CH]/C(=C\C=C\[CH])/[N]N[O] | 7.63 | -0.72 | 0.00 | 14.92 | 35.42 | 55.03 | 72.12 | 86.39 |
| 43 | 69 | C[N]N([O])C | 13.49 | 1.25 | 0.00 | 14.08 | 35.69 | 56.65 | 74.77 | 89.73 |
| 43 | 69 | C=CON=N | 22.27 | 6.29 | 0.00 | 1.81 | 9.87 | 17.98 | 24.88 | 30.41 |
| 43 | 109 | C[N]N([O])C | 19.27 | 3.47 | 0.00 | 12.01 | 33.53 | 56.42 | 77.78 | 94.24 |
| 43 | 109 | [NH]N[O] | 22.39 | 4.62 | 0.00 | 11.38 | 33.77 | 58.32 | 81.56 | 100.88 |
| 43 | 109 | C[N]N([O])C | 19.26 | 3.47 | 0.00 | 12.01 | 33.53 | 56.42 | 77.78 | 94.24 |
| 43 | 109 | C[N]N[O] | 19.47 | 3.45 | 0.00 | 12.37 | 34.32 | 57.57 | 79.99 | 98.65 |
| 43 | 109 | C[N]N([O])C | 19.28 | 3.47 | 0.00 | 12.01 | 33.53 | 56.42 | 77.78 | 94.24 |
| 43 | 109 | C=CON=N | 22.80 | 3.60 | 0.00 | 16.64 | 44.56 | 72.00 | 94.56 | 113.40 |
| 50 | 50 | C=Cc1ccccc1 | 17.26 | 3.13 | 0.00 | 14.93 | 45.51 | 82.52 | 120.66 | 156.89 |
| 50 | 50 | C=Cc1ccccc1 | 31.67 | 6.94 | 0.00 | 17.66 | 55.60 | 99.59 | 142.82 | 182.05 |
| 50 | 50 | C=Cc1ccccc1 | 20.85 | 5.00 | 0.00 | 11.10 | 38.37 | 73.03 | 109.94 | 145.95 |
| 50 | 50 | C=Cc1ccccc1 | 18.33 | 4.29 | 0.00 | 11.35 | 40.07 | 78.87 | 121.69 | 163.20 |
| 50 | 50 | C=Cc1ccccc1 | 15.86 | 3.26 | 0.00 | 11.67 | 37.36 | 69.11 | 101.71 | 132.37 |
| 50 | 50 | Cc1ccccc1 | 19.02 | 4.13 | 0.00 | 12.65 | 41.08 | 76.36 | 113.33 | 148.93 |
| 50 | 50 | Cc1ccccc1 | 30.29 | 6.25 | 0.00 | 18.99 | 58.18 | 103.29 | 147.45 | 187.56 |
| 50 | 50 | Cc1ccccc1 | 18.54 | 3.81 | 0.00 | 13.50 | 42.84 | 78.92 | 116.49 | 152.45 |
| 50 | 50 | Cc1ccccc1 | 17.32 | 3.82 | 0.00 | 11.53 | 38.47 | 72.57 | 107.62 | 139.38 |
| 50 | 50 | Cc1ccccc1 | 17.84 | 4.18 | 0.00 | 10.76 | 37.49 | 72.31 | 108.64 | 141.14 |
| 50 | 50 | Nc1ccccc1 | 18.27 | 3.66 | 0.00 | 13.89 | 43.63 | 80.07 | 117.93 | 154.12 |
| 50 | 50 | Nc1ccccc1 | 29.34 | 5.82 | 0.00 | 19.73 | 59.59 | 105.32 | 150.05 | 190.63 |
| 50 | 50 | Nc1ccccc1 | 19.51 | 4.38 | 0.00 | 12.18 | 40.28 | 75.42 | 112.42 | 148.21 |
| 50 | 50 | Nc1ccccc1 | 18.30 | 4.41 | 0.00 | 10.34 | 36.81 | 71.53 | 107.63 | 139.36 |
| 50 | 50 | Nc1ccccc1 | 17.00 | 3.65 | 0.00 | 11.92 | 39.29 | 73.90 | 109.60 | 142.12 |
| 50 | 50 | c1ccccc1 | 18.99 | 4.09 | 0.00 | 12.76 | 41.38 | 76.92 | 114.17 | 150.02 |
| 50 | 50 | c1ccccc1 | 30.91 | 6.52 | 0.00 | 18.61 | 57.56 | 102.51 | 146.56 | 186.58 |
| 50 | 50 | c1ccccc1 | 18.98 | 4.09 | 0.00 | 12.77 | 41.38 | 76.92 | 114.17 | 150.03 |
| 50 | 50 | c1ccccc1 | 17.15 | 3.77 | 0.00 | 11.46 | 38.24 | 72.17 | 107.04 | 138.58 |
| 50 | 50 | c1ccccc1 | 17.15 | 3.77 | 0.00 | 11.46 | 38.24 | 72.16 | 107.05 | 138.58 |
| 50 | 50 | Nc1ccccc1 | 16.95 | 2.95 | 0.00 | 15.46 | 46.78 | 84.60 | 123.47 | 160.21 |
| 50 | 50 | Nc1ccccc1 | 30.96 | 6.64 | 0.00 | 18.06 | 56.28 | 100.50 | 143.96 | 183.50 |
| 50 | 50 | Nc1ccccc1 | 21.17 | 5.19 | 0.00 | 10.54 | 37.08 | 70.97 | 107.18 | 142.63 |
| 50 | 50 | Nc1ccccc1 | 18.87 | 4.51 | 0.00 | 11.30 | 40.68 | 81.00 | 126.37 | 171.72 |
| 50 | 50 | Nc1ccccc1 | 14.55 | 2.64 | 0.00 | 12.75 | 39.39 | 72.07 | 105.82 | 138.02 |
| 56 | 56 | NC1CCC1 | 15.60 | 3.40 | 0.00 | 9.65 | 30.61 | 55.37 | 79.97 | 102.33 |
| 56 | 56 | NC1CCC1 | 8.41 | 1.64 | 0.00 | 6.63 | 20.80 | 38.59 | 57.86 | 77.37 |
| 56 | 56 | NC1CCC1 | 9.02 | 1.89 | 0.00 | 6.36 | 20.45 | 38.21 | 57.40 | 76.66 |
| 56 | 56 | C[C@H]1CC(=O)O1 | 13.31 | 1.59 | 0.00 | 13.68 | 37.09 | 62.48 | 86.44 | 107.61 |
| 56 | 56 | O=C1CCC1 | 16.65 | 3.61 | 0.00 | 9.70 | 30.42 | 54.51 | 78.26 | 99.96 |
| 56 | 56 | NC1CCC1 | 15.60 | 3.40 | 0.00 | 9.65 | 30.61 | 55.37 | 79.97 | 102.33 |
| 56 | 56 | NC1CCC1 | 8.41 | 1.64 | 0.00 | 6.63 | 20.80 | 38.59 | 57.86 | 77.37 |
| 56 | 56 | C1CC=C1 | 28.45 | 6.23 | 0.00 | 15.88 | 49.94 | 89.23 | 127.43 | 161.47 |
| 56 | 56 | C1CCS1 | 17.66 | 4.40 | 0.00 | 7.80 | 27.27 | 51.42 | 75.99 | 98.39 |
| 56 | 56 | C1CC(C1)C1CC1 | 15.13 | 3.25 | 0.00 | 9.79 | 30.98 | 56.22 | 81.55 | 104.80 |
| 56 | 56 | C1CC(C1)C1CC1 | 7.98 | 1.34 | 0.00 | 7.57 | 22.94 | 42.03 | 62.60 | 83.25 |
| 56 | 56 | C1CC(C1)C1CC1 | 9.10 | 2.06 | 0.00 | 5.57 | 18.61 | 35.40 | 53.95 | 73.06 |
| 56 | 56 | NC1CCC1 | 15.76 | 3.63 | 0.00 | 8.74 | 28.60 | 52.35 | 76.18 | 98.13 |
| 56 | 56 | NC1CCC1 | 8.94 | 1.88 | 0.00 | 6.29 | 20.32 | 38.12 | 57.54 | 77.26 |
| 56 | 56 | NC1CCC1 | 7.71 | 1.28 | 0.00 | 7.34 | 22.19 | 40.63 | 60.50 | 80.52 |
| 56 | 56 | O=CC1CCC1 | 15.59 | 3.42 | 0.00 | 9.72 | 31.09 | 56.65 | 82.36 | 106.02 |
| 56 | 56 | O=CC1CCC1 | 8.53 | 1.64 | 0.00 | 6.88 | 21.52 | 39.92 | 59.84 | 79.93 |
| 56 | 56 | O=CC1CCC1 | 8.84 | 1.84 | 0.00 | 6.31 | 20.27 | 37.99 | 57.33 | 77.02 |
| 56 | 56 | OC1CCC1 | 15.19 | 3.22 | 0.00 | 9.90 | 31.05 | 56.06 | 81.03 | 103.90 |
| 56 | 56 | OC1CCC1 | 8.86 | 1.79 | 0.00 | 6.63 | 21.06 | 39.26 | 59.00 | 78.96 |
| 56 | 56 | OC1CCC1 | 7.76 | 1.33 | 0.00 | 7.19 | 21.81 | 39.95 | 59.49 | 79.19 |
| 56 | 56 | [SiH3]C1CCC1 | 16.13 | 3.76 | 0.00 | 8.90 | 29.48 | 54.37 | 79.53 | 102.74 |
| 56 | 56 | [SiH3]C1CCC1 | 8.53 | 1.66 | 0.00 | 6.77 | 21.21 | 39.32 | 58.87 | 78.56 |
| 56 | 56 | [SiH3]C1CCC1 | 9.01 | 1.93 | 0.00 | 6.11 | 19.91 | 37.47 | 56.66 | 76.16 |
| 56 | 56 | CC1CCC1 | 15.34 | 3.34 | 0.00 | 9.69 | 30.87 | 56.15 | 81.54 | 104.91 |
| 56 | 56 | CC1CCC1 | 8.64 | 1.70 | 0.00 | 6.76 | 21.26 | 39.50 | 59.27 | 79.19 |
| 56 | 56 | CC1CCC1 | 8.35 | 1.62 | 0.00 | 6.64 | 20.81 | 38.63 | 58.00 | 77.65 |
| 56 | 56 | C1CC=C1C#C | 31.35 | 7.47 | 0.00 | 14.18 | 47.10 | 85.52 | 123.02 | 155.43 |
| 56 | 56 | OC1=CCC1 | 31.62 | 7.65 | 0.00 | 13.73 | 46.27 | 84.57 | 122.41 | 156.99 |
| 56 | 56 | NC1CCC1 | 15.34 | 3.37 | 0.00 | 9.46 | 30.22 | 54.94 | 79.70 | 102.44 |
| 56 | 56 | NC1CCC1 | 8.65 | 1.70 | 0.00 | 6.83 | 21.59 | 40.26 | 60.63 | 81.30 |
| 56 | 56 | NC1CCC1 | 8.74 | 1.80 | 0.00 | 6.31 | 20.14 | 37.55 | 56.44 | 75.54 |
| 56 | 56 | O=C1CCC1 | 8.63 | 1.75 | 0.00 | 6.51 | 20.71 | 38.68 | 58.28 | 78.20 |
| 56 | 56 | C=CC1CCC1 | 7.42 | 1.02 | 0.00 | 8.32 | 24.46 | 44.23 | 65.30 | 86.24 |
| 56 | 56 | C=CC1CCC1 | 15.99 | 3.65 | 0.00 | 9.08 | 29.64 | 54.29 | 79.03 | 101.70 |
| 56 | 56 | C=CC1CCC1 | 10.21 | 2.65 | 0.00 | 4.23 | 15.77 | 30.92 | 47.78 | 65.25 |
| 56 | 57 | [CH2]/C=C\1/[CH]CC1 | 9.61 | 1.93 | 0.00 | 7.25 | 22.97 | 42.70 | 63.98 | 85.29 |
| 56 | 57 | [CH2]/C=C\1/[CH]CC1 | 9.77 | 2.11 | 0.00 | 6.50 | 21.18 | 39.83 | 60.14 | 80.74 |
| 56 | 57 | C1CC=C1C=O | 9.09 | 1.70 | 0.00 | 7.58 | 23.49 | 43.32 | 64.62 | 85.90 |
| 56 | 57 | C1CC=C1C=O | 9.77 | 2.06 | 0.00 | 6.85 | 22.09 | 41.39 | 62.33 | 83.44 |
| 56 | 57 | C1CC=C1 | 10.29 | 2.18 | 0.00 | 7.14 | 23.20 | 44.16 | 68.24 | 94.33 |
| 56 | 57 | CC1=CCC1 | 10.14 | 2.20 | 0.00 | 6.71 | 21.88 | 41.06 | 61.82 | 82.66 |
| 56 | 57 | CC1=CCC1 | 9.55 | 1.96 | 0.00 | 6.92 | 22.09 | 41.18 | 61.85 | 82.68 |
| 56 | 57 | C1CC=C1C#C | 11.60 | 2.70 | 0.00 | 6.68 | 22.85 | 44.31 | 69.21 | 96.35 |
| 56 | 57 | C1CC=C1C#C | 11.53 | 2.62 | 0.00 | 6.90 | 23.12 | 44.36 | 68.62 | 94.71 |
| 56 | 57 | OC1=CCC1 | 13.57 | 3.52 | 0.00 | 5.81 | 21.88 | 43.76 | 69.34 | 97.27 |
| 56 | 57 | OC1=CCC1 | 8.56 | 1.29 | 0.00 | 9.14 | 27.56 | 51.20 | 78.04 | 106.59 |
| 56 | 58 | C[C@H]1CC(=O)O1 | 9.98 | 1.97 | 0.00 | 8.18 | 26.35 | 50.20 | 77.12 | 104.90 |
| 56 | 58 | O=C1CCC1 | 10.48 | 2.46 | 0.00 | 5.73 | 19.52 | 37.26 | 56.96 | 77.52 |
| 56 | 58 | O=C1CCC1 | 16.65 | 3.61 | 0.00 | 9.69 | 30.40 | 54.48 | 78.20 | 99.86 |
| 56 | 58 | O=C1CCC1 | 10.14 | 2.39 | 0.00 | 5.35 | 18.29 | 34.92 | 53.34 | 72.44 |
| 57 | 57 | [CH2]/C=C\1/[CH]CC1 | 30.22 | 7.20 | 0.00 | 13.68 | 45.38 | 82.39 | 118.53 | 149.78 |
| 57 | 57 | C1CC=C1C=O | 29.72 | 6.96 | 0.00 | 14.13 | 46.31 | 83.77 | 119.94 | 151.24 |
| 57 | 57 | CC1=CCC1 | 28.07 | 6.11 | 0.00 | 15.84 | 49.63 | 88.56 | 126.33 | 158.66 |
| 69 | 143 | C=C[N]N(C=C)[O] | 30.44 | 9.57 | 0.00 | -0.64 | 9.35 | 22.56 | 36.09 | 48.54 |
| 145 | 146 | CNO | 9.07 | 1.60 | 0.00 | 6.64 | 18.87 | 31.94 | 43.57 | 52.75 |
| 145 | 146 | CONC | 6.27 | 0.52 | 0.00 | 7.46 | 19.42 | 31.61 | 42.04 | 49.92 |
| 145 | 146 | NO | 8.70 | 1.38 | 0.00 | 7.31 | 20.35 | 34.13 | 46.21 | 55.56 |
| 150 | 150 | CNN | 8.85 | 1.25 | 0.00 | 8.35 | 22.84 | 38.08 | 51.50 | 61.93 |
| 150 | 150 | CNNC | 7.76 | 1.10 | 0.00 | 7.24 | 19.76 | 32.96 | 44.62 | 53.80 |
| 150 | 150 | NN | 10.19 | 2.08 | 0.00 | 6.39 | 19.05 | 32.90 | 45.40 | 55.37 |
| 153 | 159 | O=P | -6.44 | -6.33 | 0.00 | 20.44 | 44.40 | 68.03 | 88.02 | 104.71 |
| 153 | 159 | OP(=O)O | 27.07 | 8.10 | 0.00 | 2.79 | 16.36 | 34.48 | 52.92 | 69.05 |
| 153 | 159 | OP(=O)O | -7.14 | -7.14 | 0.00 | 23.79 | 51.56 | 79.59 | 103.96 | 123.99 |
| 153 | 159 | OP(=O)O | 20.71 | 5.61 | 0.00 | 4.51 | 18.15 | 35.90 | 53.52 | 69.04 |
| 153 | 159 | COP(=O)OC | 20.66 | 5.54 | 0.00 | 4.78 | 18.56 | 35.97 | 53.13 | 68.17 |
| 153 | 159 | COP(=O)OC | 17.71 | 4.26 | 0.00 | 6.46 | 21.41 | 39.59 | 57.07 | 72.12 |
| 153 | 159 | OP=O | 20.05 | 5.46 | 0.00 | 4.10 | 17.30 | 34.71 | 51.97 | 67.38 |
| 153 | 159 | O=P | -6.44 | -6.33 | 0.00 | 20.44 | 44.40 | 68.03 | 88.02 | 104.71 |
| 154 | 155 | CS(=O)(=O)N | 8.12 | 1.32 | 0.00 | 7.58 | 20.08 | 33.81 | 47.32 | 59.04 |
| 154 | 155 | CNS(=O)(=O)C | 4.25 | -0.17 | 0.00 | 8.08 | 20.75 | 34.25 | 46.73 | 57.34 |
| 154 | 155 | CS(=O)(=O)N | 7.72 | 1.13 | 0.00 | 7.86 | 20.51 | 34.31 | 47.83 | 59.51 |
| 154 | 155 | CNS(=O)(=O)C | 5.06 | 0.01 | 0.00 | 8.43 | 20.81 | 34.25 | 46.88 | 57.47 |
| 154 | 155 | CNS(=O)(=O)C | 5.06 | 0.01 | 0.00 | 8.43 | 20.81 | 34.25 | 46.88 | 57.47 |

Table S1. Full list of CCSD(T)(F12*) energies for each bond for each molecule in the test set. Type 1 and Type 2 refer to the MM3 types for the bond and the next column gives the SMILES code for the particular chemical species. All energies for the different bond displacements are given relative to the energy at the harmonic equilibrium bond distance $r_h$ in kJ $mol^{-1}$.

## S2. EVB Results

Having obtained the *ab initio* grid of energies, the EVB procedure utilises a matrix representation of the reactive PES of the following form:

$$\mathbf{H} = \begin{bmatrix} V_1 + \varepsilon_1 & H_{21} \\ H_{12} & V_2 + \varepsilon_2 \end{bmatrix}$$

(Eq. S1)

Here two adiabatic states are considered, corresponding to the $CNCH_3$ and $NCCH_3$ reactant and product connectivity's. $V_1$ and $V_2$ are the MM3 energies of the reactant and product state respectively at a given geometry, and $\varepsilon_1$ and $\varepsilon_2$ are diagonal energy offsets designed to reproduce the reaction energy. In this case respective values 0.0 kcal $mol^{-1}$ and 24.9 kcal $mol^{-1}$ were used. The off diagonal elements couple the two adiabatic states and take the form of a two dimensional Gaussian in the Jacobi coordinates:

$$H_{12} = A \exp[-b_{11}(r - r^*)^2 + b_{22}(\gamma - \gamma^*)^2 + b_{12}(r - r^*)(\gamma - \gamma^*)]$$

(Eq. S2)

Here the Gaussian is centred at the reference values of the coordinates $r^*$ and $\gamma^*$ and the b coefficients are functions of the spread of the Gaussian in the two Jacobi coordinates $\sigma_r$ and $\sigma_\gamma$ and the tilt angle of the Gaussian $\theta$ as follows:

$$b_{11} = \frac{cos^2\theta}{2\sigma_r^2} + \frac{sin^2\theta}{2\sigma_\gamma^2}$$
$$b_{22} = \frac{sin^2\theta}{2\sigma_r^2} + \frac{cos^2\theta}{2\sigma_\gamma^2}$$
$$b_{12} = \frac{sin\ 2\theta}{2\sigma_r^2} + \frac{sin\ 2\theta}{2\sigma_\gamma^2}$$

(Eq. S3)

The EVB matrix in Eq (S4) can then be diagonalised and the ground state energy is taken as the lowest eigenvalue $\lambda_0$.

Using both the harmonic and Morse MM3 force fields, the parameters $A, r^*$, $\gamma^*, \sigma_r \sigma_\gamma$, and $\theta$ were varied in order to fit the resulting $\lambda_0$ to the *ab initio* energies. The merit function used to determine the quality of the fit was the squared difference:

$$\Delta E = \sum_{i=1}^{N} \left[E_i^{ab} - E_i^{EVB}\right]^2$$

(Eq. S4)

where $E_i^{ab}$ and $E_i^{EVB}$ are the *ab initio* and EVB energies respectively at a given geometry indexed by $i$, which runs over all 400 *ab initio* points. The fitting was carried out using a genetic algorithm. The optimisation hyperparameters, such as number of genes, mutation rate, number of iterations (4000) were identical for fits carried out using both harmonic and anharmonic functions, in order to ensure a fair comparison. For each MM3 case, 21 separate fits were performed in order to quantify stochastic errors. It should be noted that in these fits two minor alterations were made to the MM3 parameters in order to better describe this system. Firstly the C(sp2)–N(sp2) equilibrium bond length had to be decreased by 0.3 Å and secondly the C(sp3)--N(sp2)--C(sp3) equilibrium angle had to be increased to 180°in order to quantitatively describe this specific reaction.

| Atom | MM3 Type |
|---|---|
| C in methyl in $CH_3CN$ | C($sp^3$) 1 |
| C in CN in $CH_3CN$ | C($sp^1$) 4 |
| C in CN in CN-$CH_3$ | C($sp^2$) 2 |
| N in CN in $CH_3CN$ | N($sp^1$) 10 |
| N in CN in (CN-$CH_3$) | N($sp^2$) 9 |
| H (all) | H(generic) 5 |

Table S2: MM3 types used in EVB fits

| *parameter* | unit | range | optimised Morse | optimised harmonic |
|---|---|---|---|---|
| $A$ | kcal / mol | 0–20000 | 205 ± 14 | 16000 ± 6000 |
| $r$ | Å | 0–25 | 1.36 ± 0.54 | 7.46 ± 1.36 |
| $\gamma$ | degrees | 0–180 | 2.31 ± 0.59 | 1.88 ± 0.33 |
| $\sigma_r$ | $Å^{-1}$ | 0.01–25 | 1.21 ± 0.07 | 1.3 ± 1.0 |
| $\sigma_\gamma$ | $degrees^{-1}$ | 0.01–25 | 0.65 ± 0.02 | 7.3 ± 2.9 |
| $\theta$ | degrees | 0–180 | 174.7 ± 3.4 | 1.7 ± 2.4 |

Table S3: Optimised parameters from EVB fits. Please note, due to the symmetry of the system the tilt and the spread of the Gaussians had bimodal distributions. The tilt is centred around 86 and 175 degrees. The widths are interchanged between the two maxima.

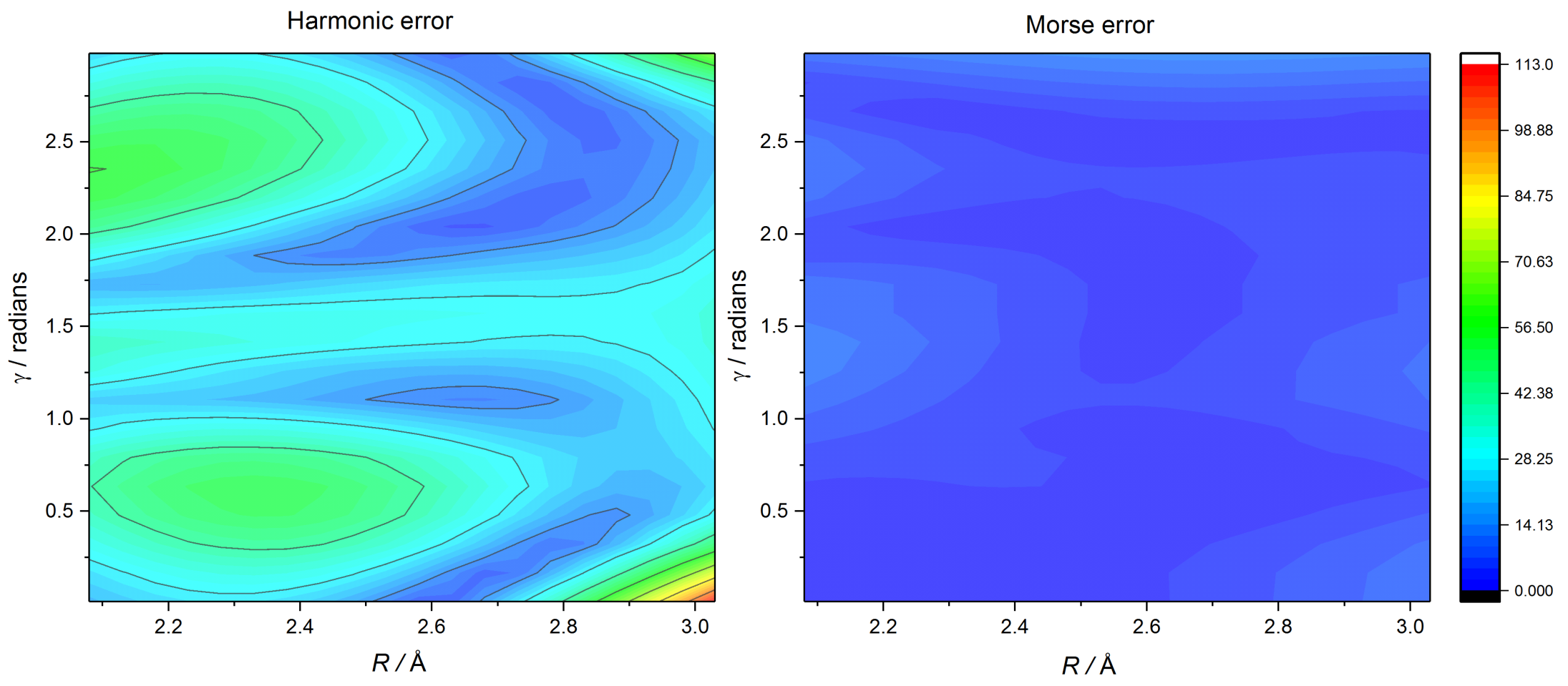

Figure S2: RMSE energy difference between EVB and *ab initio* $\Delta E$ on a point by point basis over the range of Jacobi coordinates for both the Harmonic and Morse force fields.